\documentclass{article}
\usepackage{graphicx}
\usepackage{amsfonts}
\usepackage{amsmath}
\usepackage{amssymb}
\usepackage{url}
\usepackage{wrapfig}
\usepackage{float}
\usepackage{fancyhdr}
\usepackage{indentfirst}
\usepackage{enumerate}
\usepackage{epstopdf}
\usepackage{dsfont}
 \usepackage[colorlinks=true,citecolor=blue]{hyperref}
\usepackage{amsthm}
\usepackage{multirow}
\usepackage{color}
\usepackage{natbib}
\usepackage{comment}
\usepackage{caption}
\usepackage{pifont}
\newcommand{\cmark}{\raisebox{0.6ex}{\scalebox{0.7}{$\sqrt{}$}}}%
\newcommand{\xmark}{\scalebox{0.85}[1]{$\times$}}%

\def\d{\mathrm{d}}
\def\laweq{\buildrel \mathrm{d} \over =}

\newcommand{\X}{\mathcal {X}}

\newcommand{\var}{\mathrm{var}}

\newcommand{\VaR}{\mathrm{VaR}}

\newcommand{\ES}{\mathrm{ES}}

\newcommand{\E}{\mathbb{E}}

\newcommand{\R}{\mathbb{R}}
\newcommand{\N}{\mathbb{N}}
\newcommand{\p}{\mathbb{P}}

\newcommand{\id}{\mathds{1}}

\renewcommand{\ge}{\geqslant}
\renewcommand{\le}{\leqslant}

\newcommand{\simeqm}{\buildrel \mathrm{m} \over \simeq }
\newcommand{\succeqm}{\buildrel \mathrm{m} \over \succeq }
\newcommand{\succm}{\buildrel \mathrm{m} \over \succ }
\renewcommand{\geq}{\geqslant}
\renewcommand{\leq}{\leqslant}
\renewcommand{\epsilon}{\varepsilon}
\newcommand{\esssup}{\mathrm{ess\mbox{-}sup}}
\newcommand{\essinf}{\mathrm{ess\mbox{-}inf}}

\theoremstyle{plain}
\newtheorem{theorem}{Theorem}

\newtheorem{lemma}{Lemma}
\newtheorem{proposition}{Proposition}
\theoremstyle{definition}
\newtheorem{definition}{Definition}
\newtheorem{example}{Example}
\usepackage{tkz-graph}
\usetikzlibrary{patterns, positioning, arrows}
\usetikzlibrary{decorations.markings}

\theoremstyle{remark}
\newtheorem{remark}{Remark}

\newcommand{\tbl}{\textcolor{blue}}
\newcommand{\trd}{\textcolor{red}}
\newcommand{\cet}{\begin{center}}
\newcommand{\ecet}{\end{center}}
\begin{document}

\captionsetup[figure]{labelfont={ },name={Figure},labelsep=period}
\captionsetup[table]{labelfont={},name={Table},labelsep=period}
\title{Diversification quotients: Quantifying diversification via risk measures}

\author{Xia Han \thanks{School of Mathematical Sciences and LPMC, Nankai University, China.
			\texttt{xiahan@nankai.edu.cn}}
\and Liyuan Lin\thanks{Department of Statistics and Actuarial Science, University of Waterloo, Canada.   \texttt{l89lin@uwaterloo.ca}}
\and Ruodu Wang\thanks{Department of Statistics and Actuarial Science, University of Waterloo, Canada.   \texttt{wang@uwaterloo.ca}}}
\date{\today}
\maketitle

\begin{abstract}

We establish the first axiomatic theory for diversification indices using six intuitive axioms: non-negativity, location invariance, scale invariance, rationality, normalization,  and continuity. The unique class of indices satisfying these axioms, called the diversification quotients (DQs), are defined based on a parametric family of risk measures. A further axiom of portfolio convexity pins down DQ based on coherent risk measures. DQ has many attractive properties, and it can address several theoretical and practical limitations of existing indices. In particular, for the popular risk measures Value-at-Risk and Expected Shortfall, the corresponding DQ admits simple formulas and it is efficient to optimize in portfolio selection. Moreover, it can properly capture tail heaviness and common shocks, which are neglected by traditional diversification indices. When illustrated with financial data, DQ is intuitive to interpret, and its performance is competitive against other diversification indices.  
\end{abstract}

\section{Introduction}

Portfolio diversification refers to   investment strategies that spread out among many assets, usually with the hope to reduce the volatility or risk of the resulting portfolio.
A mathematical formalization of diversification in a portfolio selection context was  made by \cite{M52}, and some early literature on diversification includes \cite{S64}, \cite{S67}, \cite{LS70} and \cite{FM72}, amongst others.  % However, this is not true in general. Negative diversification effects in heavy-tailed models with independent $\alpha$-stable asset returns are known at least since \cite{FM72}.

Although diversification is conceptually simple,  the question of how to measure diversification \emph{quantitatively} is never well settled. 
An intuitive, but non-quantitative, approach is to simply count the number of distinct stocks or industries of substantial weight in the portfolio;  see e.g., \cite{GH92}, \cite{DDY02} and \cite{DGU09} in different contexts. This approach is heuristic as it does not involve statistical or stochastic modeling. 
The second approach is to compute a quantitative index of the portfolio model, based on e.g., the volatility, variance, an expected utility, or a risk measure; this idea is certainly along the direction of \cite{M52}. In addition, one may empirically address diversification by combining both approaches; see e.g., \cite{TZ11} for the performance of different diversified portfolio strategies, \cite{DPR19}  in the context of robo-advising, and  \cite{BE21}  from the perspective of risk aversion and ambiguity aversion.
 \cite{GH92} studied conditions under which the two approaches are roughly in-line with each other.

In this paper, we take the second approach  by 
 assigning a quantifier, called a \emph{diversification index}, to each modeled portfolio.
Carrying the idea of \cite{M52}, we start our journey      with  a simple index, the diversification ratio (DR) based on the standard deviation (SD). For a random vector $\mathbf X=(X_1,\dots,X_n)$ representing future random losses and profits of individual components in a portfolio in one period,\footnote{We focus on the one-period losses to establish the theory. This is consistent with the vast majority of literature on risk measures and decision models.} 
DR based on SD is 
  defined as 
 \begin{equation}\label{eq:DR1}
{\rm DR}^{\mathrm{SD}}(\mathbf X)= \frac{ \mathrm{SD} \left(\sum_{i=1}^n X_i\right)}{ \sum_{i=1}^n \mathrm{SD}(X_i)};\end{equation}
see \cite{CC08}. One can also replace SD by   variance. Intuitively, with a smaller value indicating a stronger diversification, 
the index
$\mathrm{DR}^{\mathrm{SD}}$  quantifies the improvement of   the portfolio SD over the sum of SD of its components, and it
 has several convenient properties. 
Nevertheless, it is well-known that SD   is a coarse, non-monotone and symmetric measurement of risk, making it unsuitable for many risk management applications, especially in the presence of  heavy-tailed and skewed loss distributions; see   \cite{EMS02} for thorough discussions.

Risk measures, in particular the Value-at-Risk (VaR) and the Expected Shortfall (ES), are more flexible quantitative tools, widely  used in both financial institutions' internal risk management  and   banking and insurance regulatory frameworks, such as Basel III/IV (\cite{BASEL19}) and Solvency II (\cite{E11}).  ES has many nice theoretical properties and satisfies the four axioms of coherence (\cite{ADEH99}), whereas VaR is not subadditive in general, but it enjoys other practically useful properties; see  \cite{EPRWB14, ELW18}, \cite{EKT15} and the references therein for more discussions on the issues of VaR versus ES.   

Some indices of diversification based on various risk measures have been proposed in the literature.    
For a given risk measure $\phi$, an example of a diversification index is DR in \eqref{eq:DR1} with SD replaced by $\phi$; see  \cite{T07}. For a review of diversification indices, see \cite{K20}. 
We find several demerits of DR built on a general risk measure $\phi$ such as VaR or ES in Section \ref{sec:2}.  
A natural question is whether we can design a suitable index based on risk measures to quantify the magnitude of diversification, which avoids the deficiencies of DR. Answering this and related questions is the main purpose of this paper.

 We take an axiomatic approach to find our desirable diversification indices. 
Axiomatic approaches for risk and decision indices have been 
 prolific in economic and statistical decision theories; see e.g., the recent discussions of \cite{GPS19} and the monographs  
 \cite{G09} and  \cite{W10}. Closely related to diversification indices, risk measures (\cite{ADEH99,FR02,FS16}) and acceptability indices (\cite{CM09}) also admit sound axiomatic foundation;  the particular cases of VaR and ES are studied by \cite{C09} and \cite{WZ20}.

   In Section \ref{sec:axiom}, as our main contributions, we establish the first axiomatic foundation of diversification indices.\footnote{A different list of desirable axioms for diversification indices is studied by \cite{KD22}. Their framework is mathematically different from ours as their diversification indices are mappings of portfolio weights, instead of mappings of portfolio random vectors. They did not provide axiomatic characterization results.}
This axiomatic theory leads to the class of diversification quotients (DQs), the main object of this paper, which have an interpretation parallel to DR. 
Six simple axioms---non-negativity, location invariance, scale invariance, rationality, normalization, and continuity---are introduced and justified for their desirability in quantifying diversification. 
Their interpretations are self-evident and they describe the basic requirements for a diversification index. 
In Theorem \ref{th:ax-1}, these six axioms characterize DQ based on  monetary and positive homogeneous risk measures.
 A seventh axiom of portfolio convexity, planting an intuitive ordering over portfolio weights in the index, further pins down DQ based on coherent risk measures in Theorem \ref{thm:quasi-conv}. 
 Further, Proposition \ref{prop:convex} gives conditions for which such DQ has the range of a standard interval. Portfolio convexity means that, with a given list of assets, combining a portfolio with a better-diversified one does not lead to worse diversification than the original portfolio, reflecting a fundamental principle in economics (\cite{MWG95}).
The financial interpretation of DQ is that it quantifies the improvement of a risk-level parameter (such as the parameter in VaR or ES) caused by pooling assets, and this is discussed in Section \ref{sec:r1-34}.

% For the greatest generality, we call DQ the diversification index in \eqref{eq:intro-DQ}, without imposing any property on $\rho=(\rho_\alpha)_{\alpha \in I} $ other than  decreasing monotonicity in $\alpha$; this would include, for instance, DQ built on the standard deviation, the variance, or the family of entropic risk measures (we can build a one-parameter family from each of these choices).
A detailed analysis of
  the properties of DQ based on general risk measures is discussed
in Section \ref{sec:3}, which reveals that DQ has many appealing features,   both theoretically and practically. In addition to  standard operational properties (Proposition \ref{pro:loc-sca}),  DQ has intuitive behaviour for several benchmark portfolio scenarios (Theorem \ref{prop:div_ben}).  Moreover, DQ allows for consistency with stochastic dominance (Proposition \ref{pro:scx-consistent}) and a fair comparison across portfolio dimensions (Proposition \ref{prop:RI}).   
We proceed to focus on VaR and ES in Section \ref{VaR-ES}.  
It turns out that DQs based on VaR and ES have convenient alternative formulations (Theorem \ref{th:var}) and a natural range of $[0,n]$ and $[0,1]$,  respectively (Proposition \ref{th:var-01n}). 
Further, they report intuitive comparisons between normal and t-models and it has the nice feature that it can capture heavy tails and common shocks.

% \begin{table}[h]
% \begin{center}
% %\scriptsize
% \def\arraystretch{1} 
%   \caption{DQs based on VaR and ES,  $\alpha=0.05$ and $n=10$; numbers in bold indicate the most diversified among $\mathbf Z, \mathbf Y, \mathbf Y'$ according to the index $D$}
%   \label{tab:Ind01}
%     \begin{tabular}{c|cc}
%   $D$ & $\mathrm{DQ}^{\VaR}_\alpha $ &   $\mathrm{DQ}^{\ES}_{\alpha} $  \\ \hline
%   $\mathbf Z\sim \mathrm{N}(\boldsymbol 0,I_n)$ & $  \mathbf{2.0\times 10^{-6}}$  & $  \mathbf{1.9\times 10^{-9}}$ \\
%   %Ellip $\rm t(2)$   &0.1153    &0.0955  \\ \hline
%   % Ind $\rm t(2)$  &0.1369& 0.1111\\ \hline
%   $\mathbf Y\sim  {\rm it}_n(3) $   &0.0235  &  0.0124\\
%      $\mathbf Y'\sim \mathrm{t}(3,\boldsymbol 0,I_n)$  &0.0502 & 0.0340 \\ \hline
%       $D(\mathbf Z)<D(\mathbf Y)$ & Yes & Yes \\$D(\mathbf Y)<D(\mathbf Y')$ & Yes & Yes 
%     \\ \hline \hline
%     \end{tabular} 
%     \end{center}
% \end{table} 

   In Section \ref{sec:opt}, efficient algorithms for DQs based on VaR and ES in portfolio optimization based on empirical observations
  are obtained (Proposition \ref{thm:opt}).
Our new diversification index is applied to financial data in Section \ref{sec:emp}, where several empirical observations highlight   the advantages of DQ. 
%We find that while  DQ has the same tendency as DR in general,   DQ is more sensitive to the market changes than DR.  Moreover,   the portfolio optimization strategy based on minimizing DQ can 
%compete well with other peer strategies such as equivalent weight strategy and  those by minimizing DR. 
We conclude the paper in Section \ref{sec:conclusion}  by discussing a number of implications and promising future directions for DQ.  
Some additional results, proofs,  and  some omitted numerical results are relegated to the  E-Companion.

\textbf{Notation.} 
Throughout this paper,  $(\Omega,\mathcal F,\p)$ is an atomless probability space, on which almost surely equal random variables are treated as identical.  A risk measure  $\phi$ is a mapping from $\X$ to $\R$, where $\X$ is a convex cone of random variables on  $(\Omega,\mathcal F,\p)$ representing losses faced by a financial institution or an investor  {(i.e., a sign flip from \cite{ADEH99})}, and $\X$ is assumed to include all constants (i.e., degenerate random variables). For $p\in (0,\infty)$, denote by $L^p=L^p(\Omega,\mathcal F,\p)$ the set of all random variables $X$ with $\E[|X|^p]<\infty$ where $\E$ is the expectation under $\p$. Furthermore, $L^\infty=L^\infty(\Omega,\mathcal F,\p)$ is the space of all (essentially) bounded random variables, and $L^0=L^0(\Omega,\mathcal F,\p)$ is the space of all random variables.  Write $X\sim F$  if the random variable $X$ has the distribution function $F$  under $\mathbb{P}$, and 
$X \laweq  Y$ if two random variables $X$ and $Y$ have the same distribution. 
Further, denote by $\R_+=[0,\infty)$  and $\overline\R=[-\infty,\infty]$.  Terms such as increasing or decreasing functions are in the non-strict sense. For $X\in L^0$, $\esssup (X)$ and $\essinf(X)$ are  the essential supremum and the essential infimum of $X$, respectively. 
 Let $n$ be a fixed positive integer representing the number of assets in a portfolio, and write $[n]=\{1,\dots,n\}$.  It does not hurt to think about $n\ge 2$ although  our results hold also (trivially) for $n=1$.  The vector $\mathbf 0 $ represents the $n$-vector of zeros, and 
we always write $\mathbf X=(X_1,\dots,X_n)$ and $\mathbf Y=(Y_1,\dots,Y_n)$.  
%Below, we use the $L^\infty$-norm for both random variables taking values in $\R$ and random vectors taking values in $\R^n$ which is equipped with the Euclidean distance. 

\section{Preliminaries  and motivation}\label{sec:2}
The main object of the paper, a \emph{diversification index} $D$ is a mapping from  $\X^n$ to $\overline\R$, which is used to  quantify the magnitude of diversification of a risk vector $\mathbf X\in\X^n$ representing portfolio losses.
Our convention is that a smaller value of $D(\mathbf X)$   represents a stronger diversification in a sense specified by the design of $D$. 

As the evaluation of diversification is closely related to that of risk,  diversification indices in the literature are often defined through risk measures. An example of a diversification index is the diversification ratio (DR) mentioned in the Introduction based on measures of variability  such  as  the standard deviation ($\mathrm{SD}$) and   variance ($\var$):
$$
{\rm DR}^{\rm SD}(\mathbf X)= \frac{ \mathrm{SD} \left(\sum_{i=1}^n X_i\right)}{ \sum_{i=1}^n \mathrm{SD}(X_i)} \mbox{~~~and~~~}
{\rm DR}^{\rm var}(\mathbf X)= \frac{ \mathrm{var} \left(\sum_{i=1}^n X_i\right)}{ \sum_{i=1}^n \mathrm{var}(X_i)},$$
with the convention  $0/0=0$.  We refer to  \cite{RUZ06},  \cite{FWZ17} and \cite{BFWW22}  for  general measures of variability.    DRs based on   $\mathrm{SD}$ and $\var$ satisfy the three simple properties  below, which can be easily checked.
\begin{enumerate}[(i)]
\item[{[+]}] Non-negativity:  $D(\mathbf{X})\ge 0$ for all $\mathbf X\in \X^n$.
\item[{[LI]}] Location invariance: $D(\mathbf{X}+\mathbf{c})=D(\mathbf{X})$  for all $\mathbf{c}=(c_1,\dots,c_n) \in \R^n$ and all $ \mathbf X\in \X^n$.
\item[{[SI]}] Scale invariance: $D(\lambda \mathbf{X})=D(\mathbf{X})$  for all   $\lambda>0$ and all $ \mathbf X\in \X^n$.
\end{enumerate}
 The first property, [+], simply means that  diversification is measured in  non-negative values, where $0$ typically represents a fully diversified or hedged portfolio (in some sense).
%It is important to  deal with the situation where some component in the portfolio is used to hedge against other components,  possibly leading to a negative or zero risk value. %Another example is a  portfolio of credit default losses, where individual VaR are often $0$.
 %The second and third properties, [LI] and [SI], say that the measurement of diversification does not depend on the location or  scale of the risk vector.
The property [LI] means that injecting constant losses or gains to components of a portfolio, or changing the initial price of assets in the portfolio,\footnote{Recall that $X_i$ represents the loss from asset $i$. 
Suppose that two agents purchased the same 
portfolio of assets but at different prices of each asset.
Denote by $\mathbf X$ the  
portfolio loss vector  of agent $1$.
 The portfolio loss vector of agent $2$ is $\mathbf X+\mathbf c$, where $\mathbf c$ is the vector of differences between their purchase prices. 
The two agents should have the same level of diversification regardless of their purchase prices, as they hold the same portfolio.
} does not affect its diversification index. 
The property [SI] means that rescaling a portfolio  does not affect its diversification index.
%With this interpretation, either a location change or a scale change does not affect the Sharpe ratio of a portfolio.
The latter two properties are arguably natural, although they are not satisfied by some diversification indices used in the literature (see \eqref{eq:DR} below).
 A diversification index satisfying both [LI] and [SI] is called location-scale invariant.
% \begin{remark}
% The properties
% [LI] and [SI] are considered as desirable axioms for coherent diversification measures of \cite{KD19}, although their setting is different  from ours, as they mainly consider mappings from a subset of $\R^n $,  representing portfolio weights, to $\R$.
% \end{remark}
 %Compactness $\mathrm{[C]_\phi}$ means that the  range of the diversification index is continuous and bounded. The index constructed should  be sensitive and stationary to the risk, and thus  we  could  be able to identify it and  have some confidence of its persistence. In this sense, compactness can be regarded as an important property.

%Obviously, diversification cannot be measured without  measuring risks. Therefore,   diversification indices in the literature are often defined through risk measures.
Next, we define the two popular risk measures in banking and insurance practice. The VaR at level $\alpha \in [0,1)$ is defined as$$
\VaR_\alpha(X)=\inf\{x\in \R: \p(X\le x) \ge 1-\alpha\},~~~X\in L^0,
$$
and the ES (also called CVaR, TVaR or AVaR) at level $\alpha \in (0,1)$ is defined as
$$
\ES_{\alpha}(X) = \frac 1 \alpha \int_{0}^\alpha \VaR_\beta(X) \d \beta,~~~X\in L^1,
$$
and  $\ES_0(X)=\esssup(X)=\VaR_0(X)$, which may be $\infty$.
 The probability level $\alpha$ above is typically very small, e.g., $0.01$ or $0.025$ in \cite{BASEL19}; note that we use the  ``small $\alpha$" convention. 
 \cite{ADEH99} introduced \emph{coherent} risk measures  $\phi:\X\to \R$ as those satisfying the following four properties. 
\begin{itemize}
\item[{[M]}] Monotonicity: $\phi(X)\le \phi(Y)$ for all $X,Y \in \mathcal X$ with $X\le Y$.\footnote{The inequality $X\le Y$ between two random variables $X$ and $Y$ is pointwise.}
\item[{$[\mathrm{CA}]$}] Constant additivity:  $\phi(X+c)=\phi(X)+ c$ for all  $c\in \R$ and $X\in \X$.
\item[{$[\mathrm{PH}]$}] Positive homogeneity:  $\phi(\lambda X)=\lambda \phi(X)$ for all $\lambda \in (0, \infty)$ and $X\in \X$.
%{\color{red}\item[{[N$_{\pm}$]}] Non-positive or non-negative: $\phi(X)\ge 0$ or $\phi(X)\le 0$ for all $X \in
%\X$.}
\item[{[SA]}] Subadditivity:  $\phi(X+Y)\le \phi(X)+\phi(Y)$ for all $X,Y\in \X$.
\end{itemize}
 ES satisfies all four properties above, whereas VaR does not satisfy [SA]. 
%On the other hand,   many measures of variability, such as $\var$ and  $\mathrm{SD}$, satisfy $\mathrm{[CA]}_0$ and[PH] or $\mathrm{[PH]}_2$. 
 We say that a risk measure is \emph{monetary}  if it satisfies $[\mathrm{CA}]$ and [M],
and it is \emph{MCP} if it satisfies [M], [CA] and [PH].   For discussions and interpretations of these properties, we refer to \cite{FS16}.

Some diversification indices are defined via risk measures, such as DR (e.g.,  \cite{BDI08}, \cite{MR10} and 
   \cite{EWW15})  and the diversification benefit (DB, e.g., \cite{EFK09} and \cite{MFE15}).
For  a  risk measure $\phi$,    DR and DB    based on $\phi$ are defined as\footnote{If the denominator in the definition of ${\rm DR}^{\phi}(\mathbf X)$ is $0$, then we use the convention $0/0=0$ and $1/0=\infty$.}  
\begin{equation}\label{eq:DR}
{\rm DR}^{\phi}(\mathbf X)= \frac{ \phi \left(\sum_{i=1}^n X_i\right)}{ \sum_{i=1}^n \phi(X_i)} 
\mbox{~~~~and~~~~}   \mathrm{DB}^\phi(\mathbf X)=\sum_{i=1}^n \phi(X_i)-\phi\left(\sum_{i=1}^n X_i\right).
\end{equation}
In contrast to DR, a larger value of DB represents a stronger diversification, but this convention can be easily modified by flipping the sign to consider $-\mathrm{DB}^\phi$. 
% (thus, $c/0=\mathrm{sgn}(c)\times \infty$ with $0\times \infty=0$).}
By definition, DR is the ratio of the pooled risk to the sum of the individual risks, and thus a measurement of how substantially pooling reduces risk; similarly, DB measures the difference instead of the ratio.

DR has a number of deficiencies.  
First, the value of ${\rm DR}^{\phi}$ is not necessarily non-negative, violating $[+]$. Since the risk measure $\phi$ may take negative values,\footnote{A negative value of a risk measure has a concrete meaning as the amount of capital to be withdrawn from a portfolio position while keeping it acceptable; see \cite{ADEH99}.} it would be difficult to interpret the case where either the numerator or denominator in DR is negative, and this makes optimization of DR troublesome.  %This is particularly relevant if some components in the portfolio are used to hedge against other components, possibly leading to a negative or zero risk value. 
An example is a portfolio of credit default losses, where VaR of individual losses is often $0$ or negative but VaR of the portfolio loss is positive; see \citet[Example 2.25]{MFE15}.   
Second, for common risk measures,
 DR violates [LI], meaning that adding a risk-free asset changes the value of DR.
Third, DR is not necessarily quasi-convex in portfolio weights; this point is more subtle and will be explained later.
In addition to the above drawbacks, we also find that DR has wrong incentives for some simple models; for instance, it suggests that an iid portfolio of $t$-distributed risks is less diversified than a portfolio with a common shock and the same marginals; see Section \ref{sec:normal-t} for details.  
Similarly to DR,
the index  DB satisfies [LI] for $\phi$ satisfying [CA], but it does not satisfy [SI] for common risk measures, and it may take both positive and negative values.

%\begin{example}\label{ex:div-index1}
%  One may also consider using a centered   version of risk measures to construct DR; see Example \ref{ex:div-index2} in Appendix \ref{App:A}. 
%\end{example}

% All monetary risk measures (\cite{FS16}), such as VaR and ES, satisfy  $\mathrm{[CA]}$. % Hence, Theorem \ref{thm:div-index} puts a strong restriction on diversification indices based on these risk measures.
 In financial applications, the risk measures VaR and ES are specified in regulatory documents such as \cite{BASEL19} and \cite{E11}, and therefore it is beneficial to stick to VaR or ES as the risk measure when assessing diversification.  
Both  $\mathrm{DR}^{\VaR_{\alpha}}$ and  $\mathrm{DR}^{\ES_{\alpha}}$ satisfy {\rm[SI]}, but they do not satisfy [+] or {\rm[LI]}.% In contrast to ${\rm DR}^{\VaR_\alpha}$, which is unbounded, ${\rm DR}^{\ES_\alpha} $ is bounded above by $1$ for   $\mathbf X$ with positive components, because $\ES_{\alpha}$ satisfies {\rm [SA]}. 
\footnote{An impossibility result (Proposition \ref{thm:div-index}) is presented in Appendix \ref{App:A}, which suggests that it is not possible to construct non-trivial diversification indices like DR and DB satisfying  [+], [LI] and [SI].} It remains unclear how one can define a diversification index based on VaR or ES satisfying these properties. In the remainder of the paper, we will introduce and study a new index of diversification to bridge this gap. 

%\begin{remark}
% It turns out that it is impossible to construct a non-trivial diversification index only based on the values of $\VaR_\alpha$ or $\ES_\alpha$ for $X_1, \dots, X_n$ and $\sum_{i=1}^n X_i$ to satisfy all three properties [+], [LI] and [SI]; see Proposition \ref{thm:div-index} in Appendix \ref{App:A} for a precise statement of this claim.
%\end{remark}
%  Moreover, by Proposition \ref{thm:div-index} in Appendix \ref{App:A}, we show that  it is not possible to  construct a diversification index only based on the values of $\VaR_\alpha$ or $\ES_\alpha$ for $X_1, \dots, X_n$ and $\sum_{i=1}^n X_i$ to satisfy all three properties [+], [LI] and [SI]  unless it is in the form of \eqref{eq:D}, i.e., it takes at most $3$ different values. Such a diversification index can only provide the information about whether the diversification   reduces the risk value or not, which is of limited value in applications.

%If $\phi$ is a  variability risk measure (see \cite{BFWW22}), such as $\var$ and $\mathrm{SD}$,  $\mathrm{DR}^\phi$ may satisfy [+], [LI] and [SI], 
% and we show in  Proposition \ref{prop:DRgood} that all such DRs   belong  to $\mathrm{DQ}$.  
 %For  $\phi$ which satisfies $\mathrm{[CA]}_0$, such as $\var$ and $\mathrm{SD}$, $\mathrm{DR}^\phi$ may satisfy [+], [LI] and [SI], and we will see that all such DRs   belong  to the new class of diversification indices.

\section{Diversification indices: An axiomatic theory} \label{sec:axiom}
In this section, 
we fix $\X=L^\infty$ as the standard choice in the literature of axiomatic theory of risk measures. 
In addition to  $\mathrm{[+]}$, $\mathrm{[LI]}$ and $\mathrm{[SI]}$  introduced in  Section \ref{sec:2}, we propose four new axioms.
The first six axioms together characterize a new class of diversification indices, that is, diversification quotients (DQ) based on MCP risk measures.
With the seventh axiom of portfolio convexity, we further pin down the class of DQ based on coherent risk measures.

\subsection{Axioms of rationality, normalization, and continuity}

We first present three axioms, which depend on a risk measure $\phi$. These three axioms are  standard and weak in the sense that they do not impose a specific functional structure on $D$ other than some forms of monotonicity, normalization, and continuity.
  
For a risk measure $\phi,$ we say that two vectors $\mathbf X,\mathbf Y\in \X^n$ are \emph{$\phi$-marginally equivalent} if 
$\phi(X_i) = \phi(Y_i)$ for each $i\in [n]$, and we denote this by 
$
\mathbf X \simeqm   \mathbf Y$. 
In other words, if an agent evaluates risks using the risk measure $\phi$, then she would view the individual components of $\mathbf X$ and those of $\mathbf Y$ as equally risky. 
Similarly,  denote   by 
$
\mathbf X  \succeqm  \mathbf Y$   if 
$\phi(X_i) \le \phi(Y_i)$ for each $i\in [n]$, and by $\mathbf X   \succm\mathbf Y$ if  $\phi(X_i)  <\phi(Y_i)$ for each $i\in [n]$.
% Furthermore,  we say $\mathbf X$ is  \emph{$\phi$-marginally better than}  $\mathbf Y$  if 
% $\phi(X_i) \le \phi(Y_i)$ for each $i\in [n]$, and we denote this by 
% $
% \mathbf X  \succeqm  \mathbf Y$. We also write $\mathbf X   \succm\mathbf Y$ if  $\phi(X_i)  <\phi(Y_i)$ for each $i$. 
The other three desirable axioms are presented below, and they are built on a given risk measure $\phi$, such as VaR or ES, typically specified exogenously by  financial regulation. 

\begin{enumerate}[(i)]
 \item[{$\mathrm{[R]}_\phi$}] Rationality:  $D(\mathbf{X})\le D(\mathbf Y) $ for $\mathbf X,\mathbf Y\in \X^n$ satisfying  $
\mathbf X  \simeqm   \mathbf Y
$ and $\sum_{i=1}^n X_i \le \sum_{i=1}^n Y_i$.  
\end{enumerate} 

To interpret the axiom $\mathrm{[R]}_\phi$,
consider two portfolios $\mathbf X$ and $\mathbf Y$
satisfying  $
\mathbf X  \simeqm   \mathbf Y
$.  %,  meaning that the individual components $X_i$ and $Y_i$ are  equally risky for each $i\in[n]$. 
If further $\sum_{i=1}^n X_i \le \sum_{i=1}^n Y_i$ holds, then the total loss from portfolio $\mathbf X$
is always less or equal to that from portfolio $\mathbf Y$, making the portfolio $\mathbf X$  safer than  $\mathbf Y$. 
Since the individual components in $\mathbf X$ and those in $\mathbf Y$ are equally risky, the fact that $\mathbf X$ is safer in aggregation is a result of the different diversification effects in $\mathbf X$ and $\mathbf Y$, leading to the inequality $D(\mathbf{X})\le D(\mathbf Y) $.
This axiom is called rationality because a rational agent always prefers to have smaller losses. 

% means that if two portfolios $\mathbf X$ and $\mathbf Y$  are marginally equivalent, then the one with a smaller  total loss  has a better diversification. 
% Also, for most commonly used   risk measures  in practice,  the risk is usually measured  by the loss distribution, but not by the realized loss outcome. Thus, for  a law-invariant risk measure such as VaR or ES,    $\mathrm{[R]}_\phi$  implies that   the portfolios   whose components incur large losses simultaneously has a worse diversification if the components  $X_i$ and $Y_i$  for $i\in [n]$ have the same distribution. 

% For  the law-invariant risk measure,  we could interpret  $\mathrm{[R]}_\succeq$  as with marginal distributions fixed,  the portfolio   whose components incur large losses simultaneously has a worse diversification, which is intuitive economic. 
 %The  property  $\mathrm {[C]_\phi}$   implies some specific continuity satisfied by the  sum of the individual risks, which will be used in the axiomatic representation result.

 % if $\sum_{i=1}^n X_i -\sum_{i=1}^n Y_i\le \delta$ for a small value $\delta>0$, then $D(\mathbf X)-D(\mathbf Y)\le \epsilon$ for some small value $\epsilon>0$. 

 Next, we formulate our idea about normalizing representative values of the diversification index. 
 First, we assign the zero portfolio $\mathbf 0$ the value $D(\mathbf 0)=0$, as it carries no risk in every sense.\footnote{Indeed, the value of $D(\mathbf 0)$ may be rather arbitrary; this is   the case for DR where $0/0$ occurs.}
 A natural benchmark of a non-diversified portfolio is one in which all components are the same. Such a portfolio $\mathbf X^{\rm du}=(X,\dots, X)$ will be called a \emph{duplicate} portfolio, and we may, ideally, wish to assign the value $D(\mathbf X^{\rm du})=1$.
 However, since the zero portfolio $\mathbf 0$ is also duplicate but $D(\mathbf 0)=0$, we will require the weaker assumption $D(\mathbf X^{\rm du})\le 1$ for duplicate portfolios.\footnote{Theorem \ref{prop:div_ben} gives some mild conditions that yield $D(\mathbf X^{\rm du})= 1$ for the class $D$ characterized in this section.}
 Lastly, we should understand for what portfolios $D(\mathbf X)\ge 1$ needs to occur. 
 We say that a portfolio $\mathbf X^{\rm wd}=(X_1,\dots,X_n)$ is \emph{worse than duplicate}, if there exists a duplicate portfolio $\mathbf X^{\rm du}=(X,\dots,X)$ such that $\mathbf X^{\rm wd} \succm \mathbf X^{\rm du}$ and $\sum_{i=1}^n X_i \ge n X$.
 Intuitively, this means that each component of $\mathbf X^{\rm wd}$ is strictly less risky than $X$, but putting them together always incurs a larger loss than $nX$; in this case, diversification creates nothing but a penalty to the risk manager,  and we assign $D(\mathbf X^{\rm wd})\ge 1$.\footnote{Such situations may be regarded as diversification disasters; see \cite{IJW11}.}
{Existence of worse-than-duplicate portfolios is characterized in Appendix \ref{app:worse}.}
 Putting all of the considerations above, we propose the following normalization axiom.

 \begin{enumerate}[(i)]
\item [{$\mathrm{[N]}_\phi$}] Normalization: $D(\mathbf 0)=0$, $D(\mathbf X ) \le 1$ if  $\mathbf X $   is duplicate, 
and 
$ D(\mathbf X ) \ge 1$  if $\mathbf X  $ is worse than duplicate. 
\end{enumerate}

%  The axiom $\mathrm{[N]}_\phi$ implies that the diversification of the constant vector equals $0$ as  the risk-free assets portfolio has no systemic risk which corresponds to the most diversified case. 
%  The duplicated portfolios are considered as no diversification effect. We normalize the upper bound for the duplicated portfolios to 1, leading to $D(\mathbf X)\le 1$ for a duplicated portfolio $\mathbf X$. For a portfolio that is worse than a duplicated portfolio, it is more risky in aggregation. Hence,  we  normalize the lower bound for the portfolios that are worse than a duplicated portfolio to 1.

Finally, we propose a continuity axiom which is mainly for technical convenience. 
 
\begin{enumerate}[(i)]
  \item [{$[\mathrm{C}]_\phi$}] Continuity: 
For $\{\mathbf Y^k\}_{k\in\N}\subseteq \X^n $ and  $\mathbf X \in \X^n$ satisfying   $\mathbf Y^k \simeqm \mathbf X$ for each $k$, if $(\sum_{i=1}^n X_i-\sum_{i=1}^n Y^k_i)_+ \buildrel {L^\infty} \over \longrightarrow 0$ as $k\to\infty$, then $(D(\mathbf X)- D(\mathbf Y^k))_+\to 0$. %\com{Change $m$ to $k$ to avoid the confusion with marginal ``m".}
\end{enumerate}

 The axiom $[\mathrm{C}]_\phi$ is 
 a special form of semi-continuity. 
To interpret it, consider  portfolios $\mathbf X$ and $\mathbf Y$ that are marginally equivalent. If the sum of components of $\mathbf{X}$ is not much worse than that of $\mathbf{Y}$ in $L^\infty$, then the axiom says that the diversification of $\mathbf{X}$ is not much worse than the diversification of $\mathbf{Y}$.
This property allows for a special form of stability or robustness\footnote{In the literature of statistical robustness, often a different metric than the $L^\infty$ metric is used;  see \cite{HR09} for a general treatment. Our choice of formulating continuity via the $L^\infty$ metric is standard in the axiomatic theory of risk mappings on $L^\infty$.} with respect to 
statistical errors 
when estimating the distributions of portfolio losses.

%We say a random variable is \emph{doubly continuous} if both its   distribution function and  its quantile function are continuous. % We can take $\X^*$ to be the set of all doubly continuous random variables in $\X$.  \com{Where do we use this?}
 %We say that $\rho=(\rho_{\beta})_{\beta \in I}$ is \emph{mildly decreasing} if it is decreasing and
%$\beta \mapsto \rho_\beta(X)$ is strictly decreasing for any doubly continuous $X$,
%and 
%$\rho=(\rho_{\beta})_{\beta \in I}$ is \emph{mildly continuous} if 
%$\beta \mapsto \rho_\beta(X)$ is continuous for any doubly continuous $X$.
%These two conditions are satisfied by the classes of VaR and ES.
%We denote by $\mathcal R$ the risk measure class
%\begin{align*}
%\mathcal R&=\{\rho= (\rho_{\alpha})_{\alpha \in (0,\bar \alpha) }:  \mbox{$\bar \alpha>0$, each $\rho_\alpha$ is monetary scalable, and}\\&\qquad\qquad\mbox{$\rho$ is  mildly decreasing and mildly continuous}\}.
%\end{align*}
%The classes of VaR and ES belong to $\mathcal R$. 
%

One can check that the axioms $\mathrm{[R]}_\phi$, {$\mathrm{[N]}_\phi$} and  $[\mathrm{C}]_\phi$ are satisfied by $\mathrm{DR^{VaR_\alpha}}$ and $\mathrm{DR^{ES_\alpha}}$ if we only consider positive portfolio vectors. The axioms are not satisfied by  $\mathrm{DR^{SD}}$ because $\mathrm{SD}$ is not monotone and hence  the inequalities  $ \sum_{i=1}^n X_i \le \sum_{i=1}^n Y_i$ and 
$ \sum_{i=1}^n X_i \ge  nX$ 
used in $\mathrm{[R]}_\phi$ and $\mathrm{[N]}_\phi$ are not relevant for SD.  % As mentioned in Introduction,  even thought the index $\mathrm {DR^{SD}}$  has several convenient
%properties and  satisfies all three properties [+], [LI] and [SI],   SD is a coarse and symmetric measurement of
%risk and fails in capturing
%heavy-tailed and skewed loss distributions.  Therefore,  these three axioms   can help us  to rule out a class of  diversification  index  including $\mathrm {DR^{SD}}$.

\subsection{Portfolio convexity}

The next axiom, different from the   three above, 
imposes a natural form of convexity on the diversification index.  
Portfolio diversification is intrinsically connected to   convexity of ordering relations. 
Quoting \citet[p.~44]{MWG95}, ``Convexity can also be viewed as the formal expression of a basic inclination of economic agents for diversification."
For this purpose, we propose an axiom of portfolio convexity in this section.

Let a 
random vector  $\mathbf{X}\in\X^n$ represent losses from  $n$ assets and a vector $\mathbf w= (w_{1},  \dots, w_{n}) \in \Delta_{n}$ of portfolio weights, where
$\Delta_{n}$ is the standard $n$-simplex,  given by
$$  \Delta_{n}=\left\{\mathbf{x} \in[0,1]^{n}: x_{1} +\dots+x_{n}=1\right\}.$$ 
The total loss  of the portfolio is  $\mathbf{w}^\top \mathbf{X}$. 
We write $\mathbf w \odot \mathbf X=\left(w_1X_1,\dots,w_nX_n\right)$, which is the portfolio loss vector with the weight $\mathbf w$.
The \emph{portfolio convexity} axiom is formulated below.
\begin{enumerate}
\item[{[PC]}] Portfolio convexity: The set $\{\mathbf w\in \Delta_n: D(\mathbf w \odot \mathbf X)\le d)\}$ is convex for each $\mathbf X \in \X^n$ and $d \in \overline{\R}$.
\end{enumerate}

Intuitively, portfolio convexity means that, for a given vector $\mathbf X$ of assets, combining a portfolio strategy with a better-diversified one on the same set of assets  does not result in a portfolio that is less diversified than the original portfolio.
As convexity is the decision-theoretic counterpart
of diversification, 
[PC] is desirable  for diversification indices.

\begin{remark}\label{rem:qc}
Axiom [PC]   is equivalent to \emph{quasi-convexity} of $\mathbf w \mapsto  D(\mathbf w \odot \mathbf X)$ for each $\mathbf X\in \X^n$; that is, $D((\lambda \mathbf w+ (1-\lambda) \mathbf w') \odot \mathbf X)\le D(\mathbf w \odot \mathbf X) \vee  D(\mathbf w' \odot \mathbf X)$ for all $\lambda \in [0,1]$, $\mathbf w,\mathbf w'\in \Delta_n$ and $\mathbf X\in \X^n$. 
\end{remark}

\begin{remark}\label{rem:undesirable}
Convexity or quasi-convexity of $\mathbf X\mapsto D(\mathbf X)$ is not natural or desirable.
For instance, combining two diversified portfolios $(X,Y)$ and $(Y,X)$ may
result in a duplicate portfolio; 
see Example \ref{ex:non-convex} in Appendix \ref{app:convex}.  
 Convexity of $\mathbf w \mapsto  D(\mathbf w \odot \mathbf X)$,  which is stronger than [PC], is not desirable either; see Example \ref{ex:DRSDnon-convex}   in Appendix \ref{app:convex}. 
\end{remark}

The four axioms introduced above, together with the three  in Section \ref{sec:2},
lead to a class of diversification indices, which we define next.

\subsection{Characterization results}

We first formally introduce the diversification index DQ relying on a parametric class of risk measures,
which will be characterized in two results below.

\begin{definition}\label{def:DQ}Let $  \rho =(\rho_{\alpha})_{\alpha \in  I}$ be a class of risk measures indexed by $\alpha\in I=(0,\overline \alpha) $ with $\overline \alpha\in(0,\infty]$ such that $\rho_\alpha$ is decreasing in $\alpha$. 
  For $\mathbf X \in \X^n$, the \emph{diversification quotient} based on the class $ \rho$  at level $\alpha\in I$ is defined  by
 \begin{equation}\label{def:alpha*}
 {\rm DQ}^{\rho}_\alpha(\mathbf X)=\frac{\alpha^*}{\alpha}, \mbox{~~
where }
 \alpha^*= \inf\left\{\beta \in I :  \rho_{\beta} \left(\sum_{i=1}^n X_i\right) \le \sum_{i=1}^n \rho_{\alpha}(X_i) \right\}, 
 \end{equation} with the convention $\inf(\varnothing)=\overline \alpha$.
\end{definition}

We first characterize DQ based on MCP risk measures by six axioms without [PC]. 
 \begin{theorem}\label{th:ax-1}
 A diversification index $D:\X^n\to \overline \R$
satisfies $\mathrm{[+]}$, $\mathrm{[LI]}$, $\mathrm{[SI]}$, $\mathrm{[R]_\phi}$, $\mathrm{[N]}_\phi$ and $\mathrm{[C]_\phi}$ for some  MCP risk measure $\phi$
%for some preference $\succeq$ satisfying $\mathrm{[A1]}$-$\mathrm{[A4]}$
   if and only if $D$ is $\mathrm{DQ}_\alpha^\rho$  for some $\alpha$ and   decreasing  class  $\rho $ of  MCP risk measures. %such that $\rho$ is non-flat from left at $(\alpha, X)$ for any $X \in \X^*$
   Moreover, in both directions of the above equivalence, it can be required that $\rho_\alpha=\phi$. 
\end{theorem}

Theorem \ref{th:ax-1} gives the first axiomatic characterization of diversification indices, to the best of our knowledge. 
The proof techniques to show the important ``only if" statement of Theorem \ref{th:ax-1} are based on a sophisticated analysis of an auxiliary mapping
$$R:\X\to [0,\infty],~  R(X) = \inf\left\{D(\mathbf X): X\le \sum_{i=1}^n X_i,~\mathbf X \simeqm \mathbf 0\right\},$$ and this is explained in Appendix \ref{App:main}.

Next, we incorporate portfolio convexity into our axiomatic framework. For this purpose, it is natural to build the diversification indices based on risk measures with convexity. When formulated on monetary risk measures, convexity represents the idea that diversification reduces the risk; see \cite{FS16}. For risk measures that are not constant additive, \cite{CMMM11} argued that quasi-convexity is more suitable than convexity to reflect the consideration of diversification; moreover, convexity and quasi-convexity are equivalent if $\mathrm{[CA]}$ holds.  
A risk measure is \emph{linear} if it satisfies $\phi(aX+bY)=a\phi(X)+b\phi(Y)$ for all $X, Y\in \X$ and $a,b \in \R$.
Since linear risk measures correspond to expectations (under monotonicity), which do not reflect diversification, we will focus on non-linear ones. 
The next theorem characterizes DQ based on coherent risk measures.
\begin{theorem}\label{thm:quasi-conv}
Suppose $n \ge 4$ and $\phi$ is a non-linear coherent risk measure. A diversification index $D:\X^n\to \overline \R$ satisfies $\mathrm{[+]}$, $\mathrm{[LI]}$, $\mathrm{[SI]}$, $\mathrm{[R]_\phi}$, $\mathrm{[N]}_\phi$, $\mathrm{[C]_\phi}$ and $\mathrm{[PC]}$  
   if and only if $D$ is $\mathrm{DQ}_\alpha^\rho$  for some $\alpha$ and   decreasing  class  $\rho $ of  coherent risk measures with $\rho_\alpha=\phi$. 
\end{theorem}

The conditions $n\ge 4$ and non-linearity of $\phi$ are essential to the proof of Theorem \ref{thm:quasi-conv}. They are harmless for financial applications since typical portfolios have more than a few components, and common risk measures are not linear. 

Although portfolio convexity is crucial for diversification indices, making Theorem \ref{thm:quasi-conv} a central result,  
we present Theorem  \ref{th:ax-1} separately for the following reasons. 
First, Theorem \ref{th:ax-1} reveals the fundamental properties needed to pin down the form of DQ and this helps to clarify the role of [PC].  
Second,  the proof of 
Theorem \ref{thm:quasi-conv} is technically built on Theorem \ref{th:ax-1}.
Third, the class of DQ characterized by Theorem \ref{th:ax-1} allows for DQ based on VaR, which is  popular  in financial regulation.

In the next proposition, we show that for  sub-linear risk measures,
DQ satisfies [PC] (thus, the ``if" direction of Theorem \ref{thm:quasi-conv} does not need [M] and [CA] for $\rho$),
and its range is $[0,1]$ under mild conditions, avoiding non-degeneracy. 
A risk measure is sub-linear if it satisfies subadditivity and positive homogeneity (equivalently,  convexity and positive homogeneity).

\begin{proposition}\label{prop:convex} Let  $\rho=\left(\rho_{\beta}\right)_{\beta \in I}$ be a decreasing class of sub-linear risk measures and $\alpha\in I$.  Then $\mathrm{DQ}_{\alpha}^{\rho}$ satisfies $\mathrm{[PC]}$. If $n\ge 3$,
 $\rho_\alpha$ is non-linear and there exists $X \in \X$ such that $\beta \mapsto \rho_\beta(X)$ is  strictly decreasing,  then 
  $\{{\rm DQ}^\rho_\alpha (\mathbf X): \mathbf X \in \X^n\}=[0,1]$.
\end{proposition} 

Given a sub-linear risk measure $\rho_\alpha$, the conditions in  Proposition \ref{prop:convex} for $\{{\rm DQ}^\rho_\alpha (\mathbf X): \mathbf X \in \X^n\}=[0,1]$ are mild and satisfied by e.g., DQ based on the family of ES.
In contrast to DQ, DR based on sub-linear risk measures may not satisfy [PC] since the denominator in \eqref{eq:DR} may be negative.    
% The stronger condition of convexity in $\mathbf w$  
% generally fails to hold for any DQ or DR (this is welcome as discussed in Remark \ref{rem:undesirable}). 
% \begin{remark}
% Propositions \ref{prop:non_deg}  shows that the diversification indices built in Theorem \ref{thm:quasi-conv} are non-degenerate under some mild conditions. Proposition \ref{prop:uni_rho} shows that the choice of the risk measure family is unique up to strictly increasing transformation of the parameter
% if the ordering structure on portfolio diversification is specified by a given ordering relation $\succeq$ on $\X^n$ that can be numerically represented by a DQ. 
% \end{remark} 
For a clear comparison, 
we summarize  
in Table \ref{tab:axiom} the axioms satisfied by the diversification indices that appear in the paper.

\begin{table}[t]
\caption{Summary of  axioms satisfied by diversification indices $\mathrm{DR}^\phi$, $\mathrm{DB}^\phi$ and $\mathrm{DQ}^\rho_\alpha$ (with $\phi=\rho_\alpha$), where $\mathcal X_+$ is the set of non-negative elements in $\X$ and $\alpha \in (0,1)$}%, and for DB, [PC], $[\mathrm{R}]_\phi$ and $[\mathrm{C}]_\phi $ are adjusted so that a larger value represents better diversification}
%, and $\phi$ is the corresponding risk measure ($\VaR_\alpha$, $\ES_\alpha$, $\var$ or $\mathrm{SD}$)}
\label{tab:axiom}
\setlength\tabcolsep{5pt}
\centering
\begin{tabular}{c|c|cccccccc}
Index &Domain & $\mathrm{[+]}$ & $\mathrm{[LI]}$ & $\mathrm{[SI]}$ & $\mathrm{[R]}_\phi$ &$\mathrm{[N]}_\phi$&$[\mathrm{C}]_\phi$ &$\mathrm{[PC]}$\\
\hline
$\mathrm{DR}^{\VaR_\alpha}$ and $\mathrm{DR}^{\ES_\alpha}$ &$ \mathcal{X}^n$ & \xmark & \xmark & \cmark & \xmark &\xmark &\xmark&\xmark\\
%$\mathrm{DR}^{\ES}$& $\mathbf X \in \mathcal{X}^n$ & \xmark & \xmark & \cmark & \xmark &\xmark &\xmark&\xmark\\
%$\mathrm{DR}^{\ES}$, $\mathbf X \in \mathcal{X}^n$ & \xmark & \xmark & \cmark & \xmark &\xmark &\xmark\\
$\mathrm{DR}^{\VaR_\alpha}$& ${\mathcal{X}_+^n}$ & \cmark & \xmark & \cmark & \cmark &\cmark &\cmark&\xmark\\
$\mathrm{DR}^{\ES_\alpha}$& ${\mathcal{X}_+^n}$ & \cmark & \xmark & \cmark & \cmark &\cmark &\cmark&\cmark\\
%& $\mathbf X \in \trd{\mathcal{X}_+^n}$ & \cmark & \xmark & \cmark & \cmark &\cmark &\cmark&\cmark\\
%$\mathrm{DR}^{\ES}$, $\mathbf X \in \mathcal{X}_+^n$ & \cmark & \xmark & \cmark & \cmark &\cmark &\cmark\\
$\mathrm{DR}^{\mathrm{SD}}$& $\mathcal{X}^n$  & \cmark & \cmark & \cmark & \xmark &\xmark &\xmark&\cmark\\
$\mathrm{DR}^{\mathrm{\var}}$& $\mathcal{X}^n$  & \cmark & \cmark & \cmark & \xmark &\xmark &\xmark&\xmark\\
%  $\mathrm{DR}^{\mathrm{\var}}$, $\mathbf X \in \mathcal{X}^n$  & \cmark & \cmark & \cmark & \xmark &\xmark &\xmark\\
$-\mathrm{DB}^{\VaR_\alpha}$& $\mathcal{X}^n$  & \xmark & \cmark & \xmark & \cmark &\xmark &\cmark&\xmark\\
$-\mathrm{DB}^{\ES_\alpha}$& $\mathcal{X}^n$  & \xmark & \cmark & \xmark & \cmark &\xmark &\cmark&\cmark\\\hline\hline
$\mathrm{DQ}^{\VaR}_\alpha$& $\mathcal{X}^n$  & \cmark & \cmark & \cmark & \cmark &\cmark &\cmark&\xmark\\
$\mathrm{DQ}^{\ES}_\alpha$& $ \mathcal{X}^n$  & \cmark & \cmark & \cmark & \cmark &\cmark &\cmark&\cmark\\\hline
\end{tabular}
\end{table}

\subsection{Interpretation of DQ}
\label{sec:r1-34}
DQ based on MCP or coherent risk measures have been characterized axiomatically, but we have not interpreted the meaning of DQ in \eqref{def:alpha*}. 
For an interpretation, 
consider a decreasing class of risk measures $(\rho_\beta)_{\beta \in I}$.
The values of risk measures typically represent the capital requirement of a risky asset, and hence 
$\beta$ is interpreted as a parameter of risk level  (as in $\VaR_\beta$ or $\ES_\beta$), that is, a smaller $\beta$ means a larger capital requirement for the same risk. 
Notice from \eqref{def:alpha*} that, under mild conditions, $\alpha^*$ is uniquely determined by 
 $$\rho_{\alpha^*} \left(\sum_{i=1}^n X_i\right) = \sum_{i=1}^n \rho_{\alpha}(X_i). 
 $$
 Therefore, $\alpha^*$ is the parameter of risk level achieved by pooling, assuming that the portfolio maintains the same total capital requirement assessed by $\rho_\alpha$ when there is no pooling, that is, $\sum_{i=1}^n \rho_{\alpha}(X_i)$. %\footnote{In case of VaR or ES, $\sum_{i=1}^n \rho_{\alpha}(X_i)$ is equal to $\rho_{\alpha}(S)$ where $S$ is the sum of comonotonic random variables with marginal distributions of $(X_1,\dots,X_n)$, and thus $\rho_{\alpha}(S)$ represents the total risk with no diversification.} 
As  $\mathrm{DQ}^\rho_\alpha(\mathbf X)=\alpha^*/\alpha$, DQ is the ratio of the 
 risk-level parameters before and after pooling. 
 To summarize, 
 \begin{center}
 \emph{the index DQ quantifies the improvement of the risk-level parameter caused by pooling assets.} 
\end{center}

In the most relevant case  $\rho_{\alpha} \left(\sum_{i=1}^n X_i\right)  < \sum_{i=1}^n \rho_{\alpha}(X_i)$,  we present in Figure \ref{fig:DQDR} the conceptual symmetry between DQ, which measures the improvement by pooling in the horizontal direction, and DR, which measures an improvement in the vertical direction. 
In particular, in the case of VaR, DQ measures the probability improvement, whereas DR measures the quantile improvement; see Theorem \ref{th:var} and \eqref{eq:compare-DQDR} below.

\begin{figure}[t]
\centering
\caption{Conceptual symmetry between DQ and DR}\label{fig:DQDR}
\begin{tikzpicture}[scale=.5,description/.style=auto]
\draw[<->,thick] (0,10) -- (0,0) -- (11,0);
\draw[thick, red] (0.2,9.5) .. controls (1,4.5) and (8,1.8).. (9.5,1.4);
\draw[thick, blue] (0.2,6.5) .. controls (1,3) and (8,1).. (9.5,0.6);
\node[above] at (2.65,8.3) {\trd{\small$\sum_{i=1}^n \rho_\beta\left( X_i\right)$ }};
\node[above] at (11.65,0.1) {\tbl{\small $\rho_\beta\left(\sum_{i=1}^n X_i\right)$}};
\node[above] at (11,2.9) {\small$\displaystyle\mathrm{DR}^{\rho_\alpha}(\mathbf X)=\frac{\tbl{\rho_\alpha\left(\sum_{i=1}^n X_i\right)}}{\trd{\sum_{i=1}^n \rho_\alpha\left(X_i\right)}}$};
\node[below] at (11,0) {\small $\beta$};
\draw[dash pattern={on 1.84pt off 1.51pt}] (1.01,0)--(1.01,4.9);  
%{\color{red}\draw[thick] (0,4.7)--(9.5,4.7);}
\node[left] at (0,4.9) {\trd{\small$\sum_{i=1}^n \rho_\alpha(X_i)$}};
\node[below] at (1.16,0) {\tbl{\small$\alpha^*$}};
 %{\color{red}\draw[thick] (0,3.1)--(9.5,3.1);}
\draw[dash pattern={on 1.84pt off 1.51pt}] (3,0)--(3,4.9);
%\draw[dash pattern={on 1.84pt off 1.51pt}] (3,0)--(3,3.26);
\draw[dash pattern={on 0.84pt off 2.51pt}] (0,3.26)--(10,3.26); 
\draw[dash pattern={on 0.84pt off 2.51pt}] (0,4.9)--(10,4.9);
\node at (3,-0.6) {\trd{$\alpha$}};
\node at (0,-0.54) {$0$};
\node[left] at (0,3.26) {\tbl{\small $\rho_\alpha\left(\sum_{i=1}^n X_i\right)$}};
\end{tikzpicture}\\
\small $\mathrm{DQ}_\alpha^\rho(\mathbf X)=\tbl{\alpha^*}/\trd{\alpha}$~~~~~~~~~~~~~~~~~~~~~~~~~~~
\end{figure}

% \begin{figure}[t]
% \centering
% \caption{Conceptual symmetry between DQ and DR}\label{fig:DQDR}
% \begin{tikzpicture}[scale=.5,description/.style=auto]
% \draw[<->,thick] (0,10) -- (0,0) -- (10,0);
% \draw[thick] (0.2,9.5) .. controls (1,3) and (8,0.5).. (9.5,0.2);
% \node[above] at (2.5,8.5) {\small$\rho_\beta\left(\sum_{i=1}^n X_i\right)$};
% \node[above] at (9,4.2) {\small$\displaystyle\mathrm{DR}^{\rho_\alpha}(\mathbf X)=\frac{\rho_\alpha\left(\sum_{i=1}^n X_i\right)}{ \sum_{i=1}^n \rho_\alpha\left(X_i\right)}$};
% \node[below] at (10,0) {\small $\beta$};
% \draw[dash pattern={on 0.84pt off 2.51pt}] (1,0)--(1,6.7);
% {\color{blue}\draw[dashed, thick] (0,6.7)--(9.5,6.7);}
% \node[left] at (0,6.7) {\small$\sum_{i=1}^n \rho_\alpha(X_i)$};
% \node[below] at (1,0) {\small$\alpha^*$};
%  {\color{blue}\draw[dashed, thick] (0,3.9)--(9.5,3.9);}
% \draw[dash pattern={on 0.84pt off 2.51pt}] (3,0)--(3,3.9)--(0,3.9);
% \node at (3,-0.6) {$\alpha$};
% \node at (0,-0.54) {$0$};
% \node[left] at (0,3.9) {\small $\rho_\alpha\left(\sum_{i=1}^n X_i\right)$};
% \node[left] at (4.5,-1.5) {\small $\mathrm{DQ}_\alpha^\rho(\mathbf X)=\alpha^*/\alpha$};
% \end{tikzpicture}\\
% %\small $\mathrm{DQ}_\alpha^\rho(\mathbf X)=\alpha^*/\alpha$~~~~~~~~~~~~~~~~
% \end{figure} 

\begin{remark}\label{rem:literature}
The idea of improvement of risk level is closely related to  acceptability indices, proposed by
\cite{CM09}.
More precisely, an acceptability index for a loss $X\in \X$ is defined by  $\mathrm { AI}^\rho(X)=\sup \{ \gamma \in \R_+: \rho_{1/\gamma}(X)\le 0\}$ for a decreasing class of coherent risk measures $(\rho_\gamma)_{\gamma \in \R_+}$, which has visible  similarity to $\alpha^*$ in  \eqref{def:alpha*}; see \cite{KRC20} for optimization of acceptability indices.  If $\rho$ is a class of risk measures satisfying [CA], then
$$\mathrm{DQ}_\alpha^\rho(\mathbf X)=\frac{1}{\alpha}\left(\mathrm{AI}^\rho\left(\sum_{i=1}^n (X_i-\rho_\alpha(X_i))\right)\right)^{-1}.$$ 
\cite{DTVV12}  studied several    methods for  capital allocation, among which the quantile allocation principle
computes a capital allocation $(C_1,\dots,C_n)$ such that
$\sum_{i=1}^n C_i = \VaR_{\alpha}(\sum_{i=1}^n X_i)$
and $C_i=\VaR_{c\alpha}(X_i)$ for some $c>0$.
The constant $c$ appearing as a nuisance parameter in the above rule has a visible mathematical similarity to $\mathrm{DQ}^{\VaR}_{\alpha}$.  
 \cite{MU18} studied the so-called {buffered probability of exceedance}, which is the inverse of the ES curve $\beta\mapsto\mathrm{ES}_{\beta}(X)$ at a specific point $x\in\R$; note that 
$\alpha^*$ in \eqref{def:alpha*}  is obtained by inverting the ES curve $\beta\mapsto\mathrm{ES}_{\beta}(\sum_{i=1}^n X_i )$ at  $\sum_{i=1}^n \ES_\alpha(X_i)$.  
\end{remark}

We close the section with discussions on the construction of DQ. 
First, 
 DQ can be constructed from any monotonic parametric family of risk measures.
 All commonly used risk measures belong to a monotonic family,
as this includes  VaR, ES,  expectiles (e.g., \cite{BKMR14}), mean-variance (e.g., \cite{M52} and \cite{MMRT09}), and entropic risk measures (e.g., \cite{FS16}); some   choices do not guarantee all axioms to hold.
Our results imply that using ES or expectiles guarantees all axioms and non-degeneracy for DQ. 
In addition,
there are ways to construct DQ from a single risk measure; see Appendix \ref{sec:43}.  
DQ can also be axiomatized using preferences instead of risk measures; see Appendix \ref{app:Preference}.

  DQ can be used as a normative tool for measuring diversification.
In this context, the choice of  the parametric family of risk measures
is up to the user, and DQ serves as a versatile tool that accommodates various risk attitudes. The choice of risk measures (e.g., VaR, ES) and the determination of the confidence level ($\alpha$) should be aligned with the risk tolerance, objectives, and regulatory requirements of the decision maker.
For instance, conservative investors, prioritizing capital preservation, may gravitate towards the family of ES at a high level $\alpha$, which reflects an assessment of downside risk, whereas those with aggressive risk preferences may opt for VaR or ES at a lower level $\alpha$.  
Most generally, we would recommend the use of DQ based on ES, which has a natural and strong connection to financial regulation and tail risk management, and the parameter $\alpha$ allows for flexibility in the assessment of tail risk.

\section{Properties of DQ}\label{sec:3}
In this section, we study the properties of DQ defined in Definition \ref{def:DQ}.
 For the greatest generality, 
 we do not impose any properties of risk measures in the decreasing family $\rho=(\rho_\alpha)_{\alpha\in I}$, i.e., the family $\rho$ is not limited to  MCP or coherent risk measures, so that our results can be applied to more flexible contexts in which some of the seven axioms are relaxed. In this section, $\X$ is not restricted to $L^\infty$. 
 
 \subsection{Basic properties}
 \label{sec:41}
 We first make a few immediate observations by the definition of DQ. From  \eqref{def:alpha*}, we can see that  computing $\mathrm{DQ}^\rho_\alpha$ is to invert the decreasing function $\beta\mapsto \rho_\beta (\sum_{i=1}^n X_i)$ at $\sum_{i=1}^n \rho_{\alpha}(X_i)$.
For the cases of VaR and ES,   $I=(0,1)$,  $\alpha^* \in [0,1]$, and DQ  has simple formulas; see Theorem \ref{th:var}  in Section \ref{VaR-ES}.
For a fixed value of $\sum_{i=1}^n \rho_\alpha(X_i)$,
DQ is larger if the  curve $\beta\mapsto \rho_\beta (\sum_{i=1}^n X_i)$ is larger,
and DQ is smaller if the  curve $\beta\mapsto \rho_\beta (\sum_{i=1}^n X_i)$ is smaller.  This is consistent with our intuition that a diversification index is large if there is little or no diversification, thus a large value of the portfolio risk,
and a diversification index is small if there is strong diversification. 
%Consistency of DQ with respect to stochastic dominance  is discussed in Appendix \ref{app:extremal}.
%Furthermore, we will show in Proposition \ref{pro:loc-sca} that ${\rm DQ}^{\rho}_\alpha$ is  invariant for both location and scale transforms if $\rho$ is a class of  risk measures satisfying $\mathrm{[CA]}_m$ and $\mathrm{[PH]}_\gamma$ with any fixed $m \in \R$ and $\gamma \in \R$.

 In  Theorem \ref{th:ax-1}, we have seen that DQ satisfies [SI] and [LI] if $\rho$ is a class of MCP risk measures. These properties of DQ can be obtained based on  a more general version of   properties [CA] and [PH] of risk measures, allowing us to include SD and the variance. The results are summarized  in  Proposition \ref{pro:loc-sca}, which are straightforward to check by definition. 
  \begin{itemize}
\item[{$[\mathrm{CA}]_m$}] Constant additivity with $m\in \R$:  $\phi(X+c)=\phi(X)+m c$ for all  $c\in \R$ and $X\in \X$.
\item[{$[\mathrm{PH}]_\gamma$}] Positive homogeneity with $\gamma \in \R$:  $\phi(\lambda X)=\lambda^{\gamma} \phi(X)$ for all $\lambda \in (0, \infty)$ and $X\in \X$.
\end{itemize}

 \begin{proposition}\label{pro:loc-sca}
 Let $  \rho =(\rho_{\alpha})_{\alpha \in I}$  be a class of risk measures decreasing in $\alpha$.  For each $\alpha \in I$,
 \begin{enumerate}[(i)]
 \item  if  $\rho_\beta $ satisfies ${\mathrm{[PH]}_\gamma}$ with the same $\gamma$ across $\beta \in I$, then
  ${\rm DQ}^{\rho}_\alpha$ satisfies {\rm [SI]}.
   \item   if  $\rho_\beta $ satisfies ${\mathrm{[CA]}_m}$ with the same $m$ across $\beta \in I$, then
  ${\rm DQ}^{\rho}_\alpha$ satisfies {\rm [LI]}.
 
\item
if $\rho_\alpha$ satisfies {\rm[SA]}, then
  ${\rm DQ}^{\rho}_\alpha$ takes value in $[0,1]$.
  \end{enumerate}
\end{proposition}

It is clear that [CA]  is {$[\mathrm{CA}]_m$} with $m=1$ and    [PH]   is  {$[\mathrm{PH}]_\gamma$} with $\gamma=1$. 
 %Proposition \ref{pro:loc-sca} implies that if  $\rho$ is a class of risk measures satisfying [CA], [PH] and [SA] such as ES,  then ${\rm DQ}^{\rho}_{\alpha}$ is location-scale invariant and takes value in $[0,1]$.  
More properties of DQs on the important families of VaR and ES will be discussed in Section \ref{VaR-ES}. In particular, we will see that the ranges of ${\rm DQ}_\alpha^{\VaR}$ and ${\rm DQ}_\alpha^{\ES}$ are $[0,n]$ and $[0,1]$, respectively.
%Therefore, DQ  can provide non-trivial diversification indices that satisfy the three properties [+], [LI] and [SI] based on $\VaR$ or $\ES$.

\begin{example}[Liquidity and temporal consistency]
In risk management practice, liquidity and time-horizon of potential losses need to be taken into account; see \citet[p.89]{BASEL19}.
If liquidity risk is of concern, one may use a risk measure with $[\mathrm{PH}]_\gamma$ with $\gamma>1$ to penalize large exposures of losses. For such risk measures,  $\mathrm{DQ}^\rho_\alpha$ remains scale invariant, as shown by Proposition \ref{pro:loc-sca}.
On the other hand, if the risk associated to the loss $\mathbf X(t)$ at different time spots $t>0$ is scalable by a   function $f>0$ (usually of the order $f(t)=\sqrt{t}$ in standard models such as the Black-Scholes), then DQ is consistent across different  horizons in the sense that $\mathrm{DQ}^\rho_\alpha (\mathbf X(t)) = \mathrm{DQ}^\rho_\alpha (\mathbf X(s))$ for two time spots $s,t>0$, given that $\rho_\beta(  X_{i}(t)) = f(t) \rho_\beta(  X_i(1))$ for $i\in [n]$, $t>0$ and $\beta \in I$.
\end{example}

% {\com{HX:  Is this result a little repetitive with Propositions 2 and 6? }
% \color{blue}
% In next result,  we demonstrate that for a general class of $\rho$, DQ is upper-bounded by $1$ and exhibits continuous ranges.   Note that the continuous ranges   give more information on diversification. Moreover, similarly to the continuity axiom of preferences (e.g., \cite{FS16}), a bounded interval can also provide mathematical convenience for applications. 
% }

%\subsection{Interpreting DQ from portfolio models}\label{sec:compar}

Next, we explain that the values taken by DQ are consistent with 
%To use DQ as a diversification index, we need to make sure that it coincides with 
our usual perceptions of portfolio diversification.
For  a given risk measure $\phi $ and a portfolio risk vector $\mathbf X$, we consider the following three   situations which   yield intuitive values of DQ.
   \begin{enumerate}[(i)]
\item  There is no insolvency risk with pooled individual capital, i.e.,  $\sum_{i=1}^n X_i\le \sum_{i=1}^n \phi (X_i)$ a.s.;
\item  diversification benefit exists, i.e., $ \phi \left(\sum_{i=1}^n X_i\right) <  \sum_{i=1}^n \phi (X_i)$;
%\item $\mathbf X$ has comonotonic components;
\item the portfolio relies on a single asset, i.e., $\mathbf X=(\lambda_1 X,\dots,\lambda_n X)$ for some $X\in \X$ and $\lambda_1,\dots,\lambda_n\in \R_+$. 
  A duplicate portfolio relies on a single asset.
\end{enumerate}
The above three situations
receive special attention because they intuitively correspond to very strong diversification, some diversification, and no diversification, respectively.
Naturally, we would expect DQ to be very small for (i),
DQ to be smaller than $1$ for (ii),
and DQ to be $1$ for (iii). 
It turns out that the above intuitions all check out under very weak conditions that are  satisfied by commonly used classes of risk measures.

Before presenting this result, we fix some technical terms.
For a class $\rho$ of risk measures $\rho_\alpha$ decreasing in $\alpha$,  we say that  $\rho$ is \emph{non-flat from the left} at $(\alpha,X)$ if   $\rho_{\beta}(X)>\rho_{\alpha}(X)$ for all $\beta \in(0, \alpha)$, and 
$\rho$ is \emph{left continuous} at $(\alpha,X)$ if
$\alpha\mapsto \rho_\alpha(X)$ is left continuous.
A random vector $(X_1,\dots,X_n)$ is \emph{comonotonic}
if there exists a random variable $Z$ and increasing functions $f_1,\dots,f_n$ on $\R$ such that $X_i=f_i(Z)$ a.s.~for every $i\in [n]$. 
A risk measure is \emph{comonotonic-additive} if $\phi(X+Y)=\phi(X)+\phi(Y)$ for comonotonic $(X,Y)$. Each  of ES and VaR satisfies comonotonic-additivity, as well as any distortion risk measures (\cite{Y87}, \cite{K01}) and  signed Choquet integrals (\cite{WWW20}).
We denote by $\rho_0=\lim_{\alpha \downarrow 0}\rho_\alpha$.
Note that $\rho_0=\esssup $ for common classes $\rho$ such as VaR, ES, expectiles, and entropic risk measures.

\begin{theorem}\label{prop:div_ben}
 For  given $\mathbf X \in \X^n$ and $\alpha\in I$, if $\rho$ is left continuous and non-flat from the left at $(\alpha, \sum_{i=1}^n X_i)$,
 the following hold.
  \begin{enumerate}[(i)]
\item Suppose $\rho_0 \le \esssup $.
 If for $\rho_\alpha$ there is no insolvency risk with pooled individual capital,  then ${\rm DQ}^{\rho}_\alpha(\mathbf X)=0$.
 The converse holds true if $\rho_0=\esssup$.
 \item Diversification benefit exists if and only if  ${\rm DQ}^{\rho}_\alpha(\mathbf X)<1$.
\item  If $\rho_\alpha$ satisfies $\mathrm{[PH]}$ and $\mathbf X$ relies on a single asset, then  ${\rm DQ}^{\rho}_\alpha(\mathbf X)=1$.
\item  If $\rho_\alpha$ is comonotonic-additive   and  $\mathbf X$ is comonotonic, then  ${\rm DQ}^{\rho}_\alpha(\mathbf X)=1$.
\end{enumerate}
\end{theorem}

In (i), we see that if there is no insolvency risk with pooled individual capital, then $\mathrm{DQ}^\rho_\alpha(\mathbf X)=0$.
In typical models, such as some elliptical models in Section \ref{sec:normal-t}, $
\sum_{i=1}^n X_i$ is unbounded from above unless it is a constant.
Hence, for such models and $\rho$ satisfying $\rho_0 = \esssup $,
$\mathrm{DQ}^\rho_\alpha(\mathbf X)=0$
if and only if $\sum_{i=1}^n X_i$ is a constant, thus full hedging is achieved.
This is also consistent with our intuition of full hedging as the strongest form of diversification.
%Under the conditions of Theorem %\ref{prop:div_ben},
% for a given  $\alpha\in(0,1)$,  i%f  $\sum_{i=1}^n \rho_{\alpha}(X_i)$ and $\rho_{\alpha}(\sum_{i=1}^n X_i)$ are  positive, then
The  existence of diversification benefit is the main idea behind coherent risk measures of \cite{ADEH99}.
By (ii),
 DQ and DR agree on whether diversification benefit exists under mild conditions,
 and this intuition is consistent with \cite{ADEH99}.

% \begin{proposition}
% For  given $\mathbf X \in \X^n$ and $\alpha\in(0,1)$, if $\rho$ is left continuous and non-flat from the left at $(\alpha, \sum_{i=1}^n X_i)$,  the following statements are equivalent:
% \begin{enumerate}[(i)]
%\item Diversification benefit exists, i.e., $ \rho_\alpha \left(\sum_{i=1}^n X_i\right) <  \sum_{i=1}^n \rho_\alpha(X_i)$.
%\item ${\rm DQ}^{\rho}_\alpha(\mathbf X)<1$.
%\end{enumerate}
% \end{proposition}

\begin{remark}
We require $\rho$ to be left continuous and non-flat from the left to make the inequality in (ii)  holds strictly.
This requirement excludes, in particular, trivial cases like $\mathbf X=\mathbf c\in \R^n$ which gives $\mathrm{DQ}^\VaR_\alpha(\mathbf X)=0$ by definition.
%Risk measures such as $\VaR$ may not satisfy these conditions for all $\mathbf X$ since their quantiles may have a jump. 
In case the conditions fail to hold, ${\rm DQ}^{\rho}_\alpha(\mathbf X)<1$ may not guarantee  $ \rho_\alpha \left(\sum_{i=1}^n X_i\right) <  \sum_{i=1}^n \rho_\alpha(X_i)$, but it implies the non-strict inequality
$ \rho_\alpha \left(\sum_{i=1}^n X_i\right) \le  \sum_{i=1}^n \rho_\alpha(X_i)$, and thus the portfolio risk is not worse than the sum of the individual risks. 
\end{remark}

% With the help of Theorem \ref{prop:div_ben} (ii),
%  we can show that  DQ and DR are equivalent when it comes to the existence of the diversification benefit.
% Under the conditions of Theorem \ref{prop:div_ben},
% for a given  $\alpha\in(0,1)$,  if  $\sum_{i=1}^n \rho_{\alpha}(X_i)$ and $\rho_{\alpha}(\sum_{i=1}^n X_i)$ are  positive, then it is straightforward to check that the following three statements are equivalent: 
% (i) Diversification benefit exists, i.e., $ \rho_\alpha \left(\sum_{i=1}^n X_i\right) <  \sum_{i=1}^n \rho_\alpha(X_i)  $;
%  (ii) ${\rm DR}^{\rho_\alpha}(\mathbf X)<1$;
% (iii) ${\rm DQ}^{\rho}_\alpha(\mathbf X)<1$.
% The above equivalence only says that $\mathrm{DQ}^\rho_\alpha$ and
% $\mathrm{DR}^{\rho_\alpha}$ agree on whether they are smaller than $1$,
% but they do not have to agree in other situations, and they are not meant to be compared on the same scale. 

	\subsection{Stochastic dominance and dependence}\label{sec:44} 
	
% 	In the elliptical and MRV models, the dependence structures among components of the portfolio are restricted to certain distributions. 
	In this section, we  discuss
	the consistency of DQ with respect to stochastic dominance, 
	as well as 
	the best  and worst cases for DQ among all dependence structures with given marginal distributions of the risk vector.
	
	For a diversification index, monotonicity with respect to stochastic dominance yields   consistency  with common decision-making criteria such as the expected utility model and the rank-dependent utility model. 
	A random variable $X$ (representing random loss) is  dominated by a random variable $Y$   in second-order stochastic dominance (SSD) if  $\E[f(X)]\le \E[f(Y)]$ for all decreasing concave functions $f:\R\to \R$ provided that the expectations exist, and we denote this by $X\le_{\rm SSD} Y$.\footnote{If $X$ and $Y$ represent gains instead of losses, then SSD is typically defined via increasing concave functions.
	} 
	A risk measure $\phi$ is \emph{SSD-consistent} if $\phi(X)\ge \phi(Y)$ for all $X,Y\in \X$ whenever $X\le_{\rm SSD} Y$. 
	SSD consistency is known as strong risk aversion in the classic decision theory literature (\cite{RS70}).  SSD-consistent monetary risk measures, which include all law-invariant convex risk measures such as ES, admit an ES-based characterization (\cite{MW20}).

	\begin{proposition}\label{pro:scx-consistent}
		Assume that $\rho=(\rho_\alpha)_{\alpha \in I} $ is a decreasing class of   SSD-consistent  risk measures. For $\mathbf X ,\mathbf Y \in \X^n$ and $\alpha \in I$, if $\sum_{i=1}^n \rho_\alpha(X_i) \le  \sum_{i=1}^n \rho_\alpha(Y_i) $
		and $\sum_{i=1}^n X_i \le_{\rm SSD}  \sum_{i=1}^n Y_i$, then ${\rm DQ}^{\rho}_{\alpha}(\mathbf X) \ge {\rm DQ}^{\rho}_{\alpha}(\mathbf Y)$.
	\end{proposition} 
		%For  $\mathbf X, \mathbf Y \in \mathcal Y_\mathbf{F}$,  as $\rho_{\alpha}$ is law-invariant, we have $\sum_{i=1}^n \rho_{\alpha}(X_i)=\sum_{i=1}^n \rho_{\alpha}(Y_i)$.
		% Further,   
		Proposition \ref{pro:scx-consistent} follows from the simple observation that 
% 		Because $\sum_{i=1}^n X_i \le_{\rm{SSD}} \sum_{i=1}^n Y_i$  and  $\rho_\beta$ is  SSD-consistent for all $\beta \in I$, it follows that $\rho_{\beta}(\sum_{i=1}^n X_i) \ge \rho_{\beta}(\sum_{i=1}^n Y_i)$. Using $\sum_{i=1}^n \rho_\alpha(X_i) \le  \sum_{i=1}^n \rho_\alpha(Y_i) $, we get that 
$$\left\{\beta \in I:  \rho_{\beta} \left(\sum_{i=1}^n X_i\right) \le \sum_{i=1}^n \rho_{\alpha}(X_i) \right\} \subseteq \left\{\beta \in I:  \rho_{\beta} \left(\sum_{i=1}^n Y_i\right) \le \sum_{i=1}^n \rho_{\alpha}(Y_i) \right\},$$
and we omit the proof. 
% 		Taking an infimum on both sets above yields 
		%Then, for $\beta \in B$, we have
		%$$\rho_{\beta}\left(\sum_{i}^n X_i\right) \le \rho_{\beta}\left(\sum_{i}^n Y_i\right)\le \sum_{i=1}^n \rho_{\alpha}(Y_i)=\sum_{i=1}^n \rho_{\alpha}(X_i).$$
		%Thus, $\beta \in A$ and $B \subseteq A$. Therefore, we have $\inf B \ge \inf A$, which leads to 
% 		${\rm DQ}^{\rho}_{\alpha}(\mathbf X) \ge {\rm DQ}^{\rho}_{\alpha}(\mathbf Y)$. 

Assume $\rho$ is a class of SSD-consistent risk measures (e.g.,  law-invariant convex risk measures).	Proposition \ref{pro:scx-consistent} implies that if 
	the sum of marginal risks is the same for $\mathbf X$ and $\mathbf Y$ (this holds in particular if $\mathbf X$ and $\mathbf Y$ have the same marginal distributions), then DQ  is decreasing in SSD of the total risk. 
	The dependence structures which maximize or minimize DQ for $\mathbf X$ with specified marginal distributions are discussed in Appendix \ref{App:dep}.
	For instance, a comonotonic portfolio has the largest DQ (thus the smallest diversification) among all portfolios with the same marginal distributions; this observation is related to Proposition \ref{pro:loc-sca} (iii) and Theorem \ref{prop:div_ben} (iv). %Recall that DQ for a comonotonic portfolio can attain $1$ if $\rho$ is comonotonic-additive (Theorem \ref{prop:div_ben}), DQ does not exceed $1$ if $\rho$ is subadditive (Proposition \ref{pro:loc-sca}), and a comonotonic-additive risk measure is subadditive if and only if it is SSD-consistent (Theorem 3 of \cite{WWW20}).

 \subsection{Consistency across dimensions} \label{sec:45}

  All properties in the previous sections are discussed under the assumption that the dimension $n\in \N $ is fixed. 
Letting $n$ vary allows for a comparison of diversification between portfolios with different dimensions.
 In this section, we slightly generalize our framework by considering a diversification index $D$ as a mapping on $\bigcup_{n\in \N} \X^n$; note that  the input vector $\mathbf X$ of DQ and DR can naturally have any dimension $n$. 
 We present two more useful properties of DQ in this setting. 
 For $\mathbf X\in \X^n$ and $c\in \R$, $(\mathbf X,c)$ is the $(n+1)$-dimensional random vector obtained by pasting $\mathbf X$ and $c$, and $(\mathbf X,\mathbf X)$ is the $(2n)$-dimensional random vector obtained by pasting $\mathbf X$ and  $\mathbf X$.
\begin{itemize}
\item[{[RI]}] Riskless invariance: $D(\mathbf X,c)=D(\mathbf X)$ for all $n\in \N$, $\mathbf X\in \X^n$ and $c\in \R$.
\item[{[RC]}] Replication consistency: $D(\mathbf X,\mathbf X)=D(\mathbf X)$ for all $n\in \N$ and $\mathbf X\in \X^n$.
\end{itemize}
Riskless invariance means that adding a risk-free asset into the portfolio $\mathbf X$ does not affect its diversification. For instance, the Sharpe ratio of the portfolio does not change under such an operation.  Replication consistency means  that replicating the same portfolio composition does not affect $D$. Both properties   are arguably desirable in most applications
due to their natural interpretations. %of adding a risk-free asset or replicating the same portfolio does not affect diversification.

\begin{proposition}\label{prop:RI}
 Let $  \rho =(\rho_{\alpha})_{\alpha \in I}$  be a class of risk measures decreasing in $\alpha$. 
% and  {\color{red}$\phi:L^p\to \R $ be a  continuous and law-invariant risk measure.}  
 For $\alpha \in I$,
 %$\phi$ is a risk measure.
 \begin{enumerate}[(i)]
     \item If $\rho_\beta$ satisfies $\mathrm{[CA]_m}$ with $m\in\R$ for $\beta\in I$ and $\rho_{\alpha}(0)=0$   then
  ${\rm DQ}^{\rho}_\alpha$ satisfies {\rm[RI]}.
  %   \item Suppose that $\mathrm{DR}^\phi$ is not degenerate for some  input dimension. Then    ${\rm DR}^{\phi}$ satisfies {\rm[RI]} and {\rm [+]}  if and only if  $\phi$ satisfies  $\mathrm{[CA]}_0$, $[\pm]$ and $\phi(0)=0$.
       \item If $  \rho_\beta  $ satisfies  $\mathrm{[PH]}$  for $\beta\in I$, then  ${\rm DQ}^{\rho}_\alpha$ satisfies $\mathrm{[RC]}$.
 %  \item     If $\phi$ satisfies $\mathrm{[PH]}$, then  ${\rm DR}^{\phi}$ satisfies $\mathrm{[RC]}$.
 \end{enumerate}
\end{proposition}
We further show in Proposition \ref{prop:RI2} that if [RI] is assumed, then the only option for DR is to use a non-negative $\phi$ (which is a  subclass of DQ). Thus,  if [RI] is considered as desirable, then DQ becomes useful compared to DR as it offers more choices, and in particular, it works for any classes $\rho$ of monetary risk measures with $\rho_\alpha(0)=0$ including VaR and ES. Both   DQ and DR satisfy [RC] and [RI] for MCP risk measures.

 \begin{example}
 Let $\phi$ be  a risk measure satisfying $\mathrm{[CA]}$, such as $\ES_\alpha$ or $\VaR_\alpha$. 
 Suppose that $\phi (\sum_{i=1}^n X_i)=1$
 and $\sum_{i=1}^n  \phi (X_i)=2$, and thus $\mathrm{DR}^\phi(\mathbf X)=1/2$. If  a non-random payoff of $c>0$ is added to the portfolio, then 
 $\mathrm{DR}^\phi (\mathbf X,-c)=(1-c)/(2-c)$, which turns to $0$ as $c\uparrow 1$, and it becomes negative as soon as $c>1$.
 Hence, $\mathrm{DR}^\phi$ is improved or made negative by  including  constant payoffs (either as a new asset or added to an existing asset).  This creates problematic incentives in optimization. On the other hand, $\mathrm{DQ}$ does not suffer from this problem due to [LI] and [RI].
 \end{example}
%The properties of [RI] comes from the property of  [CA] on risk measures. It implies that adding a risk-free asset does not change the degree of portfolio diversification.  Proposition \ref{prop:RI} shows that adding the amount of cash to asset return can reduce portfolio risk, but does not affect the values of DQ, while the values of   DR are not affected only when the cash does not affect the degree of portfolio risk.

\section{DQ based on the classes of VaR and ES}\label{VaR-ES}

Since VaR and ES are the two most common classes of risk measures in practice,  we focus on the theoretical properties of ${\rm DQ}^{\VaR}_{\alpha}$ and ${\rm DQ}^{\ES}_{\alpha}$ in this section. 
% The  interval   in Definition \ref{def:DQ} has a natural range of $I=(0,1)$.
We fix the parameter range $I=(0,1)$,
and we choose $\X^n$ to be $(L^0)^n$ when we discuss ${\rm DQ}^{\VaR}_{\alpha}$ and $(L^1)^n$ when we discuss ${\rm DQ}^{\ES}_{\alpha}$, but all results hold true if we fix $\mathcal X=L^1$.
\subsection{General properties}

%\subsection{Alternative formulas}
We  first provide alternative formulations of ${\rm DQ}^{\VaR}_\alpha$ and ${\rm DQ}^{\ES}_\alpha$. The formulations offer clear interpretations and simple ways to compute the values of DQs. 
The formula  \eqref{eq:es-alter3} below can be derived from the  optimization formulation for the buffered probability of exceedance in Proposition 2.2 of \cite{MU18}. 

\begin{theorem}\label{th:var}
For a given $\alpha \in (0,1)$, ${\rm DQ}^{\VaR}_\alpha $ and ${\rm DQ}^{\ES}_\alpha$  have the alternative formulas
\begin{equation}\label{eq:var-alter}
{\rm DQ}^{\VaR}_\alpha (\mathbf X)  =\frac{1}{\alpha} \p\left(\sum_{i=1}^n X_i> \sum_{i=1}^n \VaR_{\alpha}(X_i)\right),~~~\mathbf X\in \X^n, 
\end{equation}
and 
\begin{equation}\label{eq:es-alter}
{\rm DQ}^{\ES}_\alpha (\mathbf X) =\frac{1}{\alpha} \p\left( Y> \sum_{i=1}^n \ES_{\alpha}(X_i)\right),~~~\mathbf X\in \X^n, 
\end{equation}
where 
$Y = \ES_U(\sum_{i=1}^n X_i)$ and $U\sim\mathrm{U}[0,1]$.
Furthermore, if $\p(\sum_{i=1}^n  X_i>\sum_{i=1}^n \ES_\alpha(X_i) )>0$, then 
\begin{equation}\label{eq:es-alter3}{\rm DQ}^{\ES}_\alpha(\mathbf X)= \frac{1}{\alpha}\min_{r\in (0,\infty)} \E\left[\left(r \sum_{i=1}^n (X_i-\ES_\alpha(X_i))+1\right)_+\right],\end{equation}
and otherwise ${\rm DQ}^{\ES}_\alpha(\mathbf X)=0.$
 \end{theorem}

As a first observation from Theorem \ref{th:var}, it is straightforward to compute $\mathrm{DQ}^\VaR_\alpha$ and $\mathrm{DQ}^\ES_\alpha$ on real or simulated data by applying \eqref{eq:var-alter} and \eqref{eq:es-alter} to the empirical distribution of the data.

 Theorem \ref{th:var}  also gives ${\rm DQ}^{\VaR}_{\alpha}$ a clear economic interpretation as the improvement of insolvency probability when risks are pooled,   making the discussion in Section \ref{sec:r1-34} more concrete.
 Suppose that $X_1,\dots,X_n$ are continuously distributed and they represent losses from $n$ assets.  The total pooled capital is $s_\alpha= \sum_{i=1}^n \VaR_{\alpha}(X_i)$, which is determined by the marginals of $\mathbf X$ but not the dependence structure.
An agent investing only in asset $X_i$ with capital computed by $\VaR_\alpha$ has an insolvency probability $\alpha=\p(X_i>\VaR_{\alpha}(X_i)).$
On the other hand, by  Theorem \ref{th:var}, $\alpha^*  $  is the probability that the pooled loss $\sum_{i=1}^nX_i$ exceeds the pooled capital $s_\alpha$.
The improvement from $\alpha$ to $\alpha^*$, computed by $\alpha^*/\alpha$, is precisely $\mathrm{DQ}^\VaR_\alpha(\mathbf X)$.
From here, it is also clear that ${\rm DQ}^{\VaR}_{\alpha}(\mathbf X)<1$ is equivalent to $\p\left(\sum_{i=1}^n X_i> s_\alpha\right)<\alpha$.  

To compare $\mathrm{DQ}^\VaR_\alpha$ with
$\mathrm {DR}^{\VaR_\alpha}$, recall that the two diversification indices can be rewritten as
\begin{equation}
    \label{eq:compare-DQDR}
\mathrm{DQ}^\VaR_\alpha(\mathbf X)= \frac{\p\left(\sum_{i=1}^n X_i >s_\alpha \right)}{\alpha} \mbox{~~~and~~} \mathrm{DR}^{\VaR_\alpha}(\mathbf X)= \frac{\VaR_{\alpha}\left(\sum_{i=1}^n  X_i\right) }{s_\alpha}.
\end{equation}
From \eqref{eq:compare-DQDR}, we can see a clear symmetry between DQ, which measures the probability improvement, and DR, which measures the quantile improvement.
DQ and DR based on ES have a similar comparison.

The range of DQ based on VaR is different from that based on ES, which is $[0,1]$ by Proposition \ref{prop:convex}. We summarize them below.

\begin{proposition}\label{th:var-01n}
 For $\alpha \in (0,1)$ and $n\ge 2$, $\{\mathrm{DQ}^{\VaR}_{\alpha}(\mathbf{X}): \mathbf{X} \in \X^n\}=[0,\min\{n,1/\alpha\}]$ and $\{\mathrm{DQ}^{\ES}_{\alpha}(\mathbf{X}): \mathbf{X} \in \X^n\}=[0,1]$.
\end{proposition}

Both ${\rm DQ}^{\VaR}_\alpha$ and ${\rm DQ}^{\ES}_\alpha$ take values on a bounded interval. In contrast, the diversification ratio ${\rm DR}^{\VaR_\alpha}$ is unbounded,  and  ${\rm DR}^{\ES_\alpha}$  is  bounded above by $1$ only  when the ES of the total risk is non-negative.  %Moreover, similarly to the continuity axiom of preferences (e.g., \cite{FS16}), a bounded interval can  provide mathematical convenience for applications.
%The values of DQs are simple to interpret.
%To be specific,
%for ${\rm DQ}^{\VaR}_\alpha$, its value is $0$ if there is a very good hedge in the sense of Proposition \ref{th:var-01n}  (ii);
%its value is $1$ if there is strong positive dependence such as comonotonicity, and
%its value is $n$ if there is strong negative dependence conditional on the tail event.   
%For ${\rm DQ}^{\ES}_\alpha$,
% its value is $0$ if there is a very good hedge in the sense of  Proposition \ref{th:var-01n} (ii) and
%its value is $1$ if there is strong positive dependence such as comonotonicity or $\alpha$-concentration.

\begin{remark}
It is a coincidence that  $\mathrm{DQ}^{\VaR}_\alpha$ for $\alpha <1/n$ and $\mathrm{DR}^{\var}$ both have a maximum value $n$. 
The latter maximum value is attained by a risk vector $(X/n,\dots,X/n)$ for any $X\in L^2$.
\end{remark}

\subsection{Capturing heavy tails and common shocks}
\label{sec:normal-t}
In this section,  we  analyze three simple normal and t-models to illustrate some features of DQ regarding heavy tails and common shocks in the portfolio models.   Here, we only present some key observations. 
A detailed study of DQs based on VaR and ES
for elliptical distributions and multivariate regularly varying  models, 
including explicit formulas to compute DQ for these models, can be found in  \cite{HLW22}. 
%featuring the advantages of DQ. 

Let  $\mathbf Z=(Z_1,\dots,Z_n)$ be an $n$-dimensional standard normal random vector, and let $\xi^2$ have an inverse gamma distribution independent of $\mathbf Z$. Denote by $ {\rm it}_n(\nu)$ the joint distribution with $n$ independent t-marginals $\mathrm t(\nu,0,1)$, where the parameter $\nu$ represents the degrees of freedom; see \cite{MFE15} for t-models. The model $\mathbf Y=(Y_1,\dots,Y_n)\sim  {\rm it}_n(\nu)$ can be stochastically represented by
\begin{equation}
Y_i=  \xi_i Z_i,~~~\mbox{for~}i\in [n], \label{eq:commonshock-2}
\end{equation}
where $\xi_1,\dots,\xi_n$ are iid following the same distribution as $\xi$, and independent of $\mathbf Z$.
In contrast, a joint t-distributed random vector $\mathbf Y'=(Y'_1,\dots,Y'_n)\sim \mathrm{t}(\nu,\mathbf 0,I_n)$ has a stochastic representation
 $ 
\mathbf Y' = \xi   \mathbf Z,
$
that is,
%where $A\in \R^{n\times n}$ is from the Cholesky decomposition  
% $AA^\top = \Sigma$  of $\Sigma$.
%Assume $\boldsymbol \mu=
%\mathbf 0$ and $\Sigma =I_n$.  In this case,
\begin{equation}
Y'_i=  \xi Z_i,~~~\mbox{for~}i\in [n]. \label{eq:commonshock}
\end{equation}
In other words, $\mathbf Y'$ is a standard normal random vector multiplied by a heavy-tailed common shock $\xi $.
% this also explains why $r=0$ does not imply independence of components in the t-model.
All three models $\mathbf Z,\mathbf Y,\mathbf Y'$ have the same correlation matrix, the identity matrix $I_n$.

 Because of the common shock $\xi$ in \eqref{eq:commonshock}, large losses from components of $\mathbf Y'$ are more likely to occur simultaneously, compared to  $\mathbf Y$ in \eqref{eq:commonshock-2}, which does not have a common shock. Indeed, $\mathbf Y'$ is tail dependent (Example 7.39 of \cite{MFE15}) whereas $\mathbf Y$ is tail independent.
As such,  at least intuitively (if not rigorously),    diversification for portfolio $\mathbf Y'$ should be considered as weaker than $\mathbf Y$, although both models are uncorrelated and have the same marginals.\footnote{On a related note, as discussed by \cite{EMS02},  correlation is not a good measure of diversification in the presence of heavy-tailed and skewed distributions.}
By the central limit theorem, for $\nu>2$, the  component-wise average of $\mathbf Y $ (scaled by its variance) is asymptotically normal as $n$ increases, whereas the component-wise average of  $\mathbf Y'$  is always t-distributed.
Hence, one may intuitively expect the order $D(\mathbf Z)<D(\mathbf Y)<D(\mathbf Y') $ to hold.

\begin{table}[t]
\def\arraystretch{1} \begin{center}  \caption{DQs/DRs based on VaR, ES, SD and var, where   $\alpha=0.05$,  $n=10$ and $\nu=3$; numbers in bold indicate the most diversified  among $\mathbf Z, \mathbf Y, \mathbf Y'$ according to the index $D$}  \label{tab:Ind1}  \begin{tabular}{c|cccccc}   $D$ & $\mathrm{DQ}^{\VaR}_\alpha $ &   $\mathrm{DQ}^{\ES}_{\alpha} $&$ \mathrm{DR}^{\VaR_\alpha}$&$ \mathrm{DR}^{\ES_\alpha}$ & 
$ \mathrm{DR}^{\mathrm{SD}}$& $\mathrm{DR}^{\mathrm{var}}$ \\ \hline
   %Ellip $\rm t(2)$   &0.1153    &0.0955& 0.3162& 0.3162 &1 \\ \hline
  % Ind $\rm t(2)$  &0.1369& 0.1111& 0.4190&  0.3617 &1\\ \hline
 $\mathbf Z\sim \mathrm{N}(\boldsymbol 0,I_n)$ &  $\mathbf{2.0\times 10^{-6}}$  & $\mathbf{1.9\times 10^{-9}}$  &\textbf{0.3162} &0.3162 & \textbf{0.3162}& \textbf{1}  \\ 
   $\mathbf Y\sim  {\rm it}_n(3) $   &0.0235  &  0.0124& 0.3569&  \textbf{0.2903}  & \textbf{0.3162} & \textbf{1}\\ 
   $\mathbf Y'\sim \mathrm{t}(3,\boldsymbol 0,I_n)$  &0.0502 & 0.0340& \textbf{0.3162}&  0.3162  & \textbf{0.3162}& \textbf{1}  \\ \hline  
    $D(\mathbf Z)<D(\mathbf Y)$       & Yes & Yes & Yes &  No & No&No \\  
  $D(\mathbf Y)<D(\mathbf Y')$    &Yes & Yes&No &   Yes  & No &No\\ \hline
  \hline  \end{tabular}  \end{center}\end{table}

\begin{table}[t]
\def\arraystretch{1} \begin{center}  \caption{DQs/DRs based on VaR, ES, SD and var, where   $\alpha=0.05$,  $n=10$ and $\nu=4$; numbers in bold indicate the most diversified  among $\mathbf Z, \mathbf Y, \mathbf Y'$ according to the index $D$}  \label{tab:Ind2}  \begin{tabular}{c|cccccc}   $D$ & $\mathrm{DQ}^{\VaR}_\alpha $ &   $\mathrm{DQ}^{\ES}_{\alpha} $&$ \mathrm{DR}^{\VaR_\alpha}$&$ \mathrm{DR}^{\ES_\alpha}$ & 
$ \mathrm{DR}^{\mathrm{SD}}$& $\mathrm{DR}^{\mathrm{var}}$ \\ \hline
  $\mathbf Z\sim \mathrm{N}(\boldsymbol 0,I_n)$ &  $  \mathbf{2.0\times 10^{-6}}$  & $ \mathbf{1.9\times 10^{-9}}$  & \textbf{0.3162} &0.3162 & \textbf{0.3162}& \textbf{1}  \\
 $\mathbf Y\sim  {\rm it}_n(4) $    &0.0050  & 0.0017&  0.3415&  \textbf{0.2828}  & \textbf{0.3162}& \textbf{1}\\
 $\mathbf Y'\sim \mathrm{t}(4,\boldsymbol 0,I_n)$    &0.0252 & 0.0138& \textbf{0.3162} &  0.3162  & \textbf{0.3162} & \textbf{1}\\  \hline
  $D(\mathbf Z)<D(\mathbf Y)$       & Yes & Yes & Yes&   No & No&No \\ 
 $D(\mathbf Y)<D(\mathbf Y')$       & Yes& Yes & No&   Yes & No&No \\  \hline 
 \hline  \end{tabular}  \end{center}\end{table}

% \begin{table}[t]
% \def\arraystretch{1} \begin{center}  \caption{DQs/DRs based on VaR, ES, SD and var, where   $\alpha=0.05$ and $n=10$}  \label{tab:Ind1}  \begin{tabular}{cc|cccccc}  &  $D$ & $\mathrm{DQ}^{\VaR}_\alpha $ &   $\mathrm{DQ}^{\ES}_{\alpha} $&$ \mathrm{DR}^{\VaR_\alpha}$&$ \mathrm{DR}^{\ES_\alpha}$ & 
% $ \mathrm{DR}^{\mathrm{SD}}$& $\mathrm{DR}^{\mathrm{var}}$ \\ \hline
%   %Ellip $\rm t(2)$   &0.1153    &0.0955& 0.3162& 0.3162 &1 \\ \hline
%   % Ind $\rm t(2)$  &0.1369& 0.1111& 0.4190&  0.3617 &1\\ \hline
%   \multirow{3}*{t(3)} &    $\mathbf Y'\sim  {\rm it}_n(3) $   &0.0235  &  0.0124& 0.3569&  0.2903  & 0.3162 &1\\ & $\mathbf Y\sim \mathrm{t}(3,\boldsymbol 0,I_n)$  &0.0502 & 0.0340& 0.3162&  0.3162  & 0.3162&1 \\ &   $D(\mathbf Y')/D(\mathbf Y)$    &0.4681 & 0.3647 &1.1287 &   0.9180  & 1 &1 \\ \hline \multirow{3}*{t(4)}   & $\mathbf Y'\sim  {\rm it}_n(4) $    &0.0050  & 0.0017&  0.3415&  0.2828  & 0.3162&1\\  &   $\mathbf Y\sim \mathrm{t}(4,\boldsymbol 0,I_n)$    &0.0252 & 0.0138& 0.3162&  0.3162  & 0.3162 &1\\ & $D(\mathbf Y')/D(\mathbf Y)$       & 0.1984 & 0.1231 & 1.0800&   0.8943 & 1&1 \\  \hline normal  & $\mathbf Z\sim \mathrm{N}(\boldsymbol 0,I_n)$ &  $  2.0\times 10^{-6}$  & $  1.9\times 10^{-9}$  &0.3162 &0.3162 & 0.3162&1  \\ \hline \hline  \end{tabular}  \end{center}\end{table}
In Tables \ref{tab:Ind1} and  \ref{tab:Ind2}, we present DQ and DR for a few different models based on $\mathrm{N}(\mathbf 0,I_n)$, $\mathrm{t}(\nu,\mathbf 0,I_n)$,   and $ {\rm it}_n(\nu)$.
We choose $n=10$ and $\nu=3$ or $4$,\footnote{Most financial asset log-loss data have a tail-index between $[3,5]$, which corresponds to $\nu\in[3,5]$; see e.g., \cite{JD91}.} and thus we have five models in total.
As  we see  from Tables \ref{tab:Ind1} and  \ref{tab:Ind2},
 DQs based on both VaR and ES report a lower value for $ {\rm it}_n(\nu)$
and a larger value for $\mathrm{t}(\nu,\mathbf 0,I_n)$,
meaning that diversification is weaker for the common shock t-model \eqref{eq:commonshock} than the iid t-model \eqref{eq:commonshock-2}.
For the iid normal model, the diversification is the strongest according to DQ.
In contrast, DR sometimes reports that the iid t-model has a larger diversification than the common shock t-model, which is counter-intuitive.
In the setting of both Tables \ref{tab:Ind1} and  \ref{tab:Ind2}, a risk manager governed by $\mathrm{DQ}^{\VaR}_{\alpha}$ would prefer the iid portfolio over the common shock portfolio, but the preference is flipped if the risk manager uses $\mathrm{DR}^{\VaR_{\alpha}}$. A more detailed analysis on this phenomenon for varying $\alpha \in (0,0.1]$  is presented in Figure \ref{fig:ratio} in   Appendix \ref{App:D}, and consistent results are observed.

%In the next sections, we study optimization and properties of DQ as functions of portfolio weights. %, as well as their behavior in specific models. 
%The methods and results will help us to understand  and use DQ in risk management applications.

% In particular,   both ${\rm DQ}^{\VaR}_\alpha$ and  ${\rm DQ}^{\ES}_\alpha$ have the nice property that the corresponding  range  is a bounded interval, which shows  the sensitivity  to  risk.

% A similar phenomenon can be observed if the individual losses have small probability of being positive and large probability to be zero.

\section{Portfolio selection with DQ}\label{sec:opt}

%\subsection{Portfolio optimization}
%Next, 
Next, we focus on the optimal diversification problem
\begin{equation}\label{eq:optimal_DQ}
\min _{\mathbf{w} \in \Delta_n}  {\rm DQ}^{\VaR}_{\alpha}(\mathbf w \odot \mathbf X) \mbox{~~~and~~~} \min _{\mathbf{w} \in \Delta_n}  {\rm DQ}^{\ES}_{\alpha}(\mathbf w \odot \mathbf X);
\end{equation}
recall that a smaller value of DQ means better diversification.\footnote{A possible alternative formulation to \eqref{eq:optimal_DQ} is to use DQ as a constraint instead of an objective in the optimization. This is mathematically similar to a risk measure constraint (e.g., \cite{BS01}, \cite{RU02} and \cite{MU18}), but with a different interpretation, as DQ is not designed to measure or control risk.}
Recall from Table \ref{tab:axiom} that the first optimization 
is not quasi-convex 
and the second one is quasi-convex (Proposition \ref{prop:convex}). 
We do not say that optimizing a diversification  index has a decision-theoretic benefit; here we simply illustrate the advantage of DQ in computation and optimization. Whether optimizing diversification is desirable for individual or institutional investors is an open-ended question which goes beyond the current paper; we refer to \cite{VV10},    \cite{BGUW12} and \cite{CFSS17} for relevant discussions.

For the portfolio weight $\mathbf w$, DQ based on VaR at level $\alpha \in(0,1)$ is given by
%$
% {\rm DQ}^{\VaR}_{\alpha}(\mathbf w \odot \mathbf X)=\alpha^*_\mathbf{w}/\alpha
%$ where
 $$ {\rm DQ}^{\VaR}_{\alpha}(\mathbf w \odot \mathbf X)= \frac{1}\alpha    \inf\left\{\beta \in (0,1) :  \VaR_{\beta}\left(\sum_{i=1}^n w_iX_i\right) \le \sum_{i=1}^n w_{i}\VaR_{\alpha}(X_i) \right\},$$
and DQ based on ES is similar. 
%and DQ based on ES at level $\alpha \in(0,1)$ is given by
%$
% {\rm DQ}^{\ES}_{\alpha}(\mathbf w \odot \mathbf X)=\alpha^*_\mathbf{w}/\alpha
%$ where $$ \alpha^*_\mathbf{w}= \inf\left\{\beta \in (0,1) :  \ES_{\beta}\left(\sum_{i=1}^n w_iX_i\right) \le \sum_{i=1}^n w_{i}\ES_{\alpha}(X_i) \right\}.$$ 
In what follows, we fix $\alpha\in (0,1)$ and $\mathbf X=(X_1,\dots,X_n)\in \X^n$, where $\X$ is $L^0$ for VaR and $L^1$ for ES, as in Section \ref{VaR-ES}. 
%We write  $\mathbf x^{\VaR}_{\alpha}=(\VaR_\alpha(X_1),\dots,\VaR_\alpha (X_n))$ and $\mathbf x^{\ES}_{\alpha}=(\ES_\alpha(X_1),\dots,\ES_\alpha (X_n))$ which are two known vectors that do not depend on the decision variable $\mathbf w$. 
Write $\mathbf 0=(0,\dots,0)\in \R^n$ and $\mathbf x^{\rho}_{\alpha}=(\rho_\alpha(X_1),\dots,\rho_\alpha (X_n))$ for a given risk measure $\rho$.

\begin{proposition}
\label{thm:opt}
%Assuming each component of $\mathbf{X}$ is non-constant.
Fix $\alpha\in(0,1)$ and $\mathbf X \in \X^n$.
The optimization of $\mathrm{DQ}^\VaR_\alpha(\mathbf w \odot\mathbf X)$ in \eqref{eq:optimal_DQ} can be solved by
\begin{align}\label{eq:opt_DQ_VaR}
    \min_{\mathbf w \in \Delta_n}\p\left(\mathbf w ^\top \left(\mathbf X-\mathbf  x^{\VaR}_{\alpha}\right)>0 \right).
\end{align} 
Assuming $\p(X_i>\ES_\alpha(X_i))>0$ for each $i\in [n]$,
 the optimization of $\mathrm{DQ}^\ES_\alpha(\mathbf w \odot\mathbf X)$ in \eqref{eq:optimal_DQ} can be solved by the convex program
\begin{align}\label{eq:opt_DQ_ES}
\min_{\mathbf v\in\R_+^n\setminus\{\mathbf 0\} }\mathbb{E}\left[\left(\mathbf v^\top \left(\mathbf X   - \mathbf x^{\ES}_{\alpha}\right)+1\right)_{+}\right],
\end{align}
 and the optimal $\mathbf w^*$ is given by $\mathbf v/\Vert \mathbf v \Vert_1$.

\end{proposition}

%\begin{remark}
%If there exists $\mathbf w$ such that $\p(\mathbf w^\top (\mathbf X-\mathbf x^{\ES}_\alpha)>0)=0$, then the optimal $\mathbf w^*$ satisfies $\mathrm{DQ}^{\ES}_\alpha(\mathbf w^* \odot \mathbf X)=0$. In this case, the optimal problem \eqref{eq:opt_DQ_ES} is the same as 
%$$\min_{\mathbf w \in \Delta_n} \p(\mathbf w(\mathbf X-\mathbf x^{\ES}_\alpha)=0).$$
%The optimal value of \eqref{eq:opt_DQ_ES} might not equal to 0 unless $\min_{\mathbf w \in \Delta_n} \p(\mathbf w(\mathbf X-\mathbf x^{\ES}_\alpha)=0)=0$.
%In fact, all $\mathbf w \in \{\mathbf w \in \Delta_n: \p(\mathbf w(\mathbf X-\mathbf x^{\ES}_\alpha)>0)=0\}$ are the solutions to the optimal diversification problem $\min_{\mathbf w \in \Delta_n}\mathrm{DQ}^{\ES}_\alpha(\mathbf w \odot \mathbf X)$, while the optimal problem \eqref{eq:opt_DQ_ES} gives us the solution that maximize the probability $\p(\mathbf w(\mathbf X-\mathbf x^{\ES}_\alpha)<0)$.
%\end{remark}

 Proposition \ref{thm:opt} offers efficient algorithms to optimize $\mathrm{DQ}^\VaR_\alpha$ and $\mathrm{DQ}^\ES_\alpha$ 
 in real-data applications.
 The values of $\mathbf x^{\VaR}_\alpha$ and $\mathbf x^\ES_\alpha$ can be computed by many existing estimators  of the individual losses (see e.g., \cite{MFE15}).
 In particular, 
 a simple way to estimate these risk measures is to use  an empirical estimator.  
 More specifically, if we have data $\mathbf X^{(1)},\dots,\mathbf X^{(N)}$ sampled from $\mathbf X$ satisfying some ergodicity condition (being iid would be sufficient), then the empirical version of the problem \eqref{eq:opt_DQ_VaR}  is 
 %we can compute the empirical version of $\mathrm{DQ}^\VaR_\alpha$  by  
%\begin{align}\label{eq:general-opt2}
 %\widehat{\mathrm{DQ}}^\VaR_\alpha  =
 %\frac{1}{N\alpha}\sum_{j=1}^N \id_{\left\{\mathbf w ^\top  \left(\mathbf X^{(j)}- \widehat{\mathbf x}^{\VaR}_\alpha\right)  >   0 \right\}}, 
 %\end{align} 
 \begin{align}\label{eq:general-opt2}
  \mbox{minimize ~~~}  \sum_{j=1}^N \id_{\left\{\mathbf w ^\top  \left(\mathbf X^{(j)}- \widehat{\mathbf x}^{\VaR}_\alpha\right)  >   0 \right\}} ~~~& \mbox{over $\mathbf w\in \Delta_n$},
\end{align} 
 where $\widehat{\mathbf x}^{\VaR}_\alpha $ is the empirical estimator of $\mathbf x^{\VaR}_\alpha$ based on sample  $\mathbf X^{(1)},\dots,\mathbf X^{(N)}$;  see  \cite{MFE15}. 
Write $\mathbf y^{(j)}= \mathbf X^{(j)}- \widehat {\mathbf x}^{\VaR}_\alpha$ 
and $z_j=\id_{\{\mathbf w^\top \mathbf y^{(j)} > 0\}}$ for $j\in  [n]$. 
 Problem \eqref{eq:general-opt2} involves a chance constraint (see e.g., \cite{L14} and \cite{LKL16}).   By using the big-M method (see e.g., \cite{SSA10}) via choosing a sufficient large $M$ %such that $\mathbf w^\top \mathbf y^{(j)}-Mz_j\le0$ is redundant when $z_j=1$ 
 (e.g., it is sufficient if $M$ is larger than the components of $\mathbf y^{(j)}$ for all $j$),  \eqref{eq:general-opt2}  can be converted into the following linear integer program:
\begin{equation} \begin{array}{ll}
~~\mbox{minimize}&   \sum_{j=1}^N z_j \\
~\text { subject to }& \mathbf w^\top \mathbf y^{(j)}-Mz_j\le0,~~~\sum_{i=1}^n w_i=1, \\
&z_j \in\{0,1\},~~~ w_i\geq0 \mbox{~~~for all $j\in [N]$ and $i\in [n]$}.
\end{array}\label{eq:integer-prog}
\end{equation} 
Similarly, the optimization problem 
 \eqref{eq:opt_DQ_ES}
 for $\mathrm{DQ}^{\ES}_\alpha$ can be solved   the empirical version of the problem \eqref{eq:opt_DQ_ES}, which is a convex program:
\begin{align}\label{eq:general_opt_ES_1}
     \mbox{minimize ~~~}  \sum_{j=1}^N \max\left\{\mathbf v ^\top  \left(\mathbf X^{(j)}- \widehat{\mathbf x}^{\ES}_\alpha\right)+1,0\right\} ~~~& \mbox{over $\mathbf v\in \R_+$},
\end{align}
 where $\widehat{\mathbf x}^{\ES}_\alpha $ is the empirical estimator of $\mathbf x^{\ES}_\alpha$ based on sample  $\mathbf X^{(1)},\dots,\mathbf X^{(N)}$.
Both problems \eqref{eq:integer-prog} and \eqref{eq:general_opt_ES_1} can be efficiently solved by modern optimization programs, such as CVX programming (see e.g., \cite{MP13}).

Additional linear constraints, such as those on budget or expected return, can be easily included in  
\eqref{eq:opt_DQ_VaR}-\eqref{eq:general_opt_ES_1}, and the corresponding optimization problems can be solved similarly.

 Tie-breaking  needs to be addressed when working with  \eqref{eq:general-opt2}  since %$\widehat{\mathrm{DQ}}^\VaR_\alpha$
 its objective function 
 takes integer values. 
 %on multiples of $1/(N\alpha)$.
 In dynamic portfolio selection, it is desirable to avoid adjusting positions  too drastically or   frequently. 
 Therefore, in the real-data analysis in Section \ref{opt_dp},
 among tied optimizers, 
 we pick the closest one (in $L^1$-norm $\Vert\cdot \Vert_1$ on $\R^n$) to a given benchmark $\mathbf w_0$, the portfolio weight of the previous trading period. 
With this tie-breaking rule,   we 
solve 
 \begin{align}\label{eq:general-opt3}
     \mbox{minimize ~~~}
     \Vert\mathbf w-\mathbf w_0\Vert_1  
   ~~~  & \mbox{over $\mathbf w\in \Delta_n$}
  \mbox{~~~~subject to}~~~    \sum_{j=1}^N  \id_{\{\mathbf w^\top \mathbf y^{(j)} > 0\}}\le m^* ,
 \end{align}
where  $m^*$ is the optimum of \eqref{eq:general-opt2}. A tie-breaking for \eqref{eq:general_opt_ES_1}  may need to  be addressed similarly since \eqref{eq:general_opt_ES_1} is not strictly convex.

\section{Numerical illustrations}\label{sec:emp}
 To illustrate the performance of  DQ,   we   collect historical   asset prices from Yahoo Finance and conduct three sets of numerical  experiments based on the data. We use the period from January 3, 2012,  to December 31, 2021, with a total of 2518 observations of daily losses  and   500 trading days for the initial training.
 In Section \ref{com:DQ_DR}, we first   compare DQs and DRs based  on  $\VaR$ and $\ES$.  In Section \ref{per_DQ},  we calculate the values of ${\rm DQ}^{\VaR}_{\alpha}$ and  ${\rm DQ}^{\ES}_{\alpha}$ under different selections of stocks. Finally, we construct  portfolios  by minimizing ${\rm DQ}_\alpha^{\mathrm{ VaR}}$, ${\rm DQ}_\alpha^{\ES}$ and ${\rm DR}^{\rm SD}$ and by the mean-variance criterion in Section \ref{opt_dp}.  

\subsection{Comparing  DQ and DR}\label{com:DQ_DR}
We first identify the largest stock in each of the S\&P 500 sectors ranked by market cap in 2012.
Among these stocks, we select the 5 largest stocks\footnote{XOM from ENR, AAPL from IT, BRK/B from FINL,  WMT from  CONS, and GE from INDU.} to build our portfolio.
We compute  ${\rm DQ}^{\VaR}_{\alpha}$,  ${\rm DQ}^{\ES}_{\alpha}$, ${\rm DR}^{\VaR_\alpha}$ and ${\rm DR}^{\ES_\alpha}$ on each day using the empirical distribution in a rolling window of  500 days, where we set $\alpha=0.05$.

 \begin{figure}[t]
\caption{DQs  and DRs based on  $\VaR$   and $\ES$  with $\alpha=0.05$ }\label{fig:D_R_VaR}
\centering
 \includegraphics[width=14.5cm]{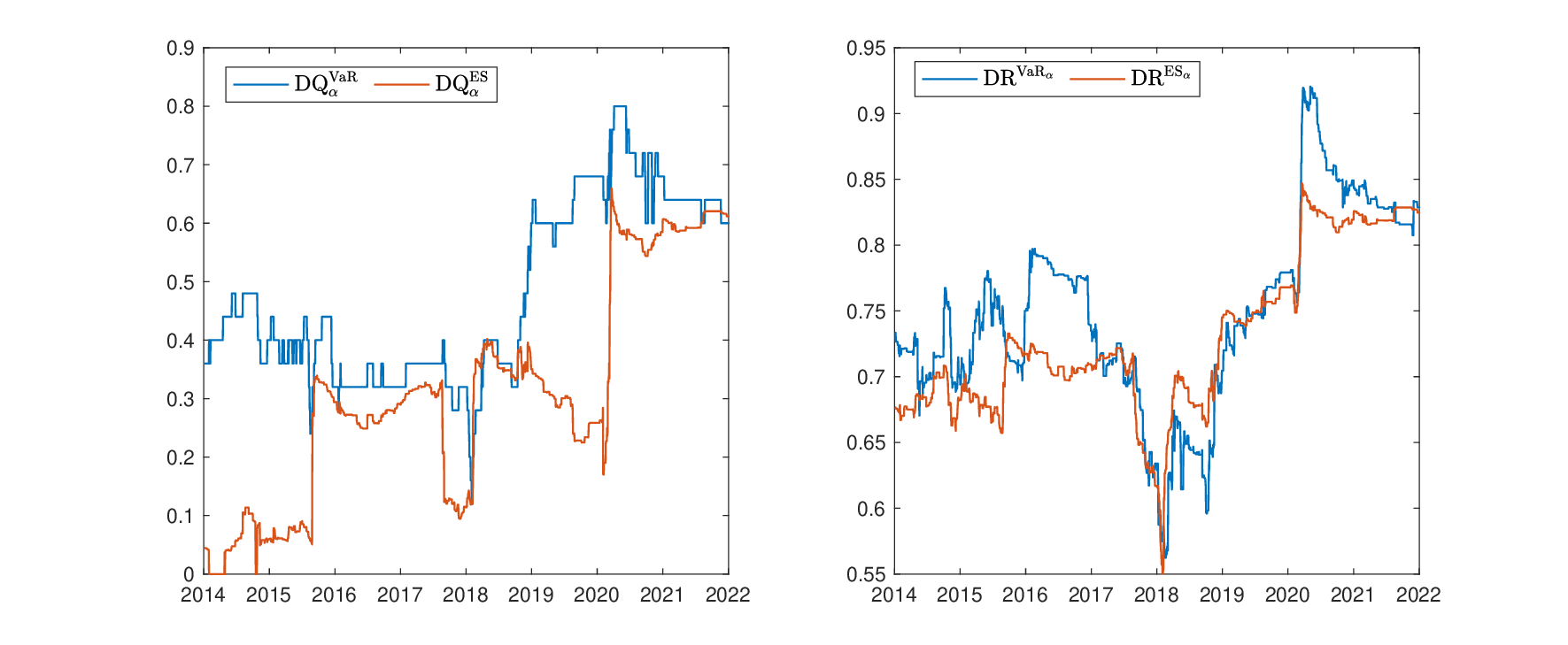}
\end{figure}

 Figure  \ref{fig:D_R_VaR}  shows that the values of DQ and DR are between 0 and 1. This corresponds to  the observation  in  Theorem \ref{prop:div_ben} that ${\rm DQ}^{\rho}_{\alpha}<1$ is equivalent to ${\rm DR}^{\rho_\alpha}<1$.
  DQ has a similar temporal pattern to DR in the above period of time, with a large jump when COVID-19 exploded, which is more visible for DQ than for DR.
  We remind the reader that DQ and DR are not meant to be compared on the same scale, and hence the fact that DQ has a larger range than DR should be taken lightly.
  We also note that the values of $\mathrm{DQ}^\VaR_{\alpha}$ are in discrete grids. This is because the empirical distribution function takes value in multiples of $1/N$ there $N$ is the sample size (500 in this experiment) and hence $\mathrm{DQ}^\VaR_{\alpha}$ takes the values $k/(N\alpha)$ for an integer $k$; see \eqref{eq:var-alter}. If  a smooth curve is preferred, then one can employ a smoothed VaR through linear interpolation. This is a  standard  technique for handling VaR;  see \citet[Section 9.2.6]{MFE15} and \citet[Remark 8 and Appendix B]{LW22}.

\subsection{DQ for different portfolios}\label{per_DQ}	 In this section,  we fix $\alpha=0.05$ and calculate the values of ${\rm DQ}^{\VaR}_{\alpha}$ and  ${\rm DQ}^{\ES}_{\alpha}$ under different portfolio compositions of stocks.
	 We consider portfolios with the following stock compositions:
	 \begin{enumerate}[(A)]
	 \item the two largest stocks  from  each of the 10 different sectors of S\&P 500;
	 \item the largest stock  from each of 5 different sectors of S\&P 500 (as in Section \ref{com:DQ_DR});
	 \item the 5 largest stocks, AAPL, MSFT, IBM, GOOGL and ORCL,  from  the  Information Technology (IT) sector;
	 \item the 5 largest stocks, BRK/B, WFC, JPM, C and BAC, from the  Financials (FINL) sector.
	 
	 \end{enumerate}
	
%	 In  Figure \ref{fig:diff},  we select  the five largest stocks: AAPL, JNJ, BRK, XOM and WMT from  different sectors and   the five largest stocks: AAPL, MSFT, GOOGL, FB and ORCL  from  the  Information Technology (IT) sector.   In Figure \ref{fig:num}, we select the ten largest stocks: AAPL, MSFT, GOOGL, FB, ORCL, V, INTC, IBM, CSCO and QCOM  from the IT sector, and divide them into two groups whose components are the top  five stocks and the total  ten  stocks, respectively. In Figure \ref{fig:sector},	  we select  the five largest stocks from the IT sector and    the five largest stocks: BRK, WFC, JPM, BAC and C from the  Financials (FINL) sector. We compare the values of  ${\rm DQ}_\alpha^{\VaR}$ and ${\rm DQ}^{\ES}_\alpha$  under these two different groups of stocks.
	 \begin{figure}[htp!]

 \caption{DQs  based on  $\VaR$ (left)  and $\ES$ (right) with $\alpha=0.05$}\label{fig:diff}
\centering
 \includegraphics[width=14.5cm]{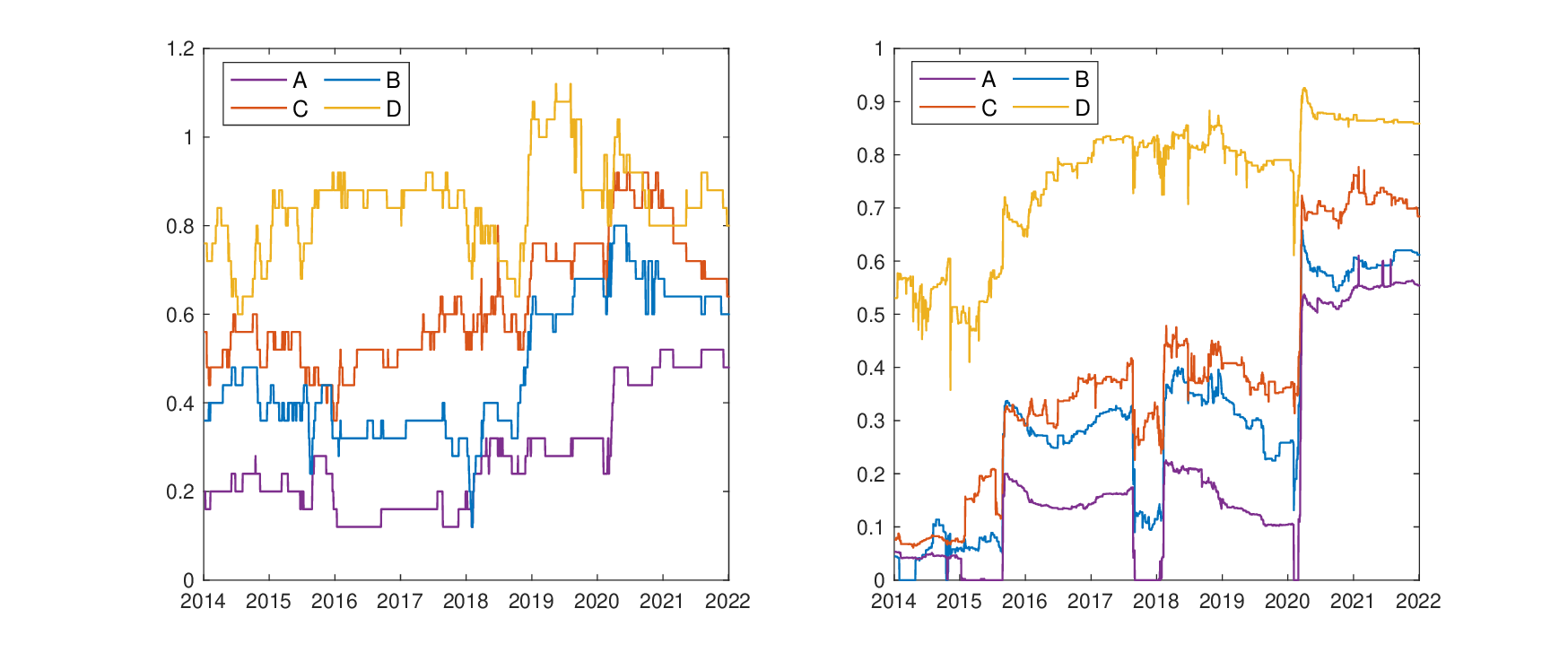}
\end{figure}

%
%	 \begin{figure}[htb!]
%
%\centering
% \includegraphics[width=15cm]{figure/compare_five_ten.eps}
% \caption{\small Values of ${\rm DQ}_{0.05}^{\VaR}$ (left) and ${\rm DQ}_{0.05}^{\ES}$ (right)  with ten stocks  (blue)  and five stocks (red) in the  IT sector}\label{fig:num}
%\end{figure}
%	
%	
%
%	 \begin{figure}[htb!]
%\centering
% \includegraphics[width=15cm]{figure/compare_IT_FINL.eps}
% \caption{\small Values of ${\rm DQ}_{0.05}^{\VaR}$ (left) and ${\rm DQ}_{0.05}^{\ES}$ (right)  with five stocks in the IT sector (blue)  and five stocks in  the FINL sector (red)}\label{fig:sector}
%\end{figure}

  We make a few observations from Figure \ref{fig:diff}. Both $\mathrm{DQ}^\VaR_\alpha$ and $\mathrm{DQ}^\ES_\alpha$ provide similar comparative results.
  The order (A)$\le$(B)$\le$(C)$ \le $(D) is consistent with our intuition.\footnote{The observations  here are consistent with those from
applying $\mathrm{DR}^{\rm SD}$ (which is also a DQ) in the same setting; see Appendix \ref{App:F}.} First, 
portfolio (A) of 20 stocks has the strongest  diversification effect among the four compositions. Second, 
 portfolio (B) across 5 sectors  has stronger diversification than (C) and (D) within one sector. Third,  portfolio  (C) of 5 stocks within the IT sector  has a stronger diversification  than portfolio (D) of 5 stocks within the FINL sector,  consistent with the fact that the stocks in  the IT sector are less correlated. Moreover, ${\rm DQ}^{\VaR}_\alpha$   for the FINL sector is larger than 1 during some period of time, which means that there is no  diversification benefit if risk is evaluated by VaR.
All DQ curves based on ES show a large up-ward jump around the COVID-19 outbreak;
such a jump also exists for curves based on VaR but it is less pronounced.

  \subsection{Optimal diversified portfolios}\label{opt_dp}
 In this section, we fix $\alpha=0.1$  and build    portfolios  via ${\rm DQ}_\alpha^{\mathrm{ VaR}}$, ${\rm DQ}_\alpha^{\ES}$, ${\rm DR}^{\rm SD}$, and the mean-variance criterion in the \cite{M52} model.\footnote{One may try other portfolio criteria other than mean-variance. For instance, \cite{LL04} found that portfolio strategies based on prospect theory perform similarly to the mean-variance strategies.}
  The optimal portfolio problems using ${\rm DR}^{\rm SD}$ and the Markowitz model are well studied in literature; see e.g. \cite{CC08}.  We compare these portfolio wealth with  the equal weighted (EW) portfolio and  the simple buy-and-hold (BH) portfolio. For an analysis on the  EW strategy, see \cite{DGU09}.%; %ee \cite{LW22} for a more recent empirical evidence) 
 %the BH strategy  is studied in e.g., \cite{CD71}.  
 
% We use the optimization methods in Section \ref{sec:opt} with tie-breaking rule in \eqref{eq:general-opt3}.

% For  variance, $$\min\limits_{\mathbf w\in\Delta_n}\rm{var}(\mathbf w^\top \mathbf X)=\min\limits_{\mathbf w\in\Delta_n}\mathbf w^\top \rm{var}(\mathbf X)\mathbf w =\min\limits_{\mathbf w\in\Delta_n} \mathbf w^\top \Sigma_{\mathbf X} \mathbf w,$$  in which   $\Sigma_{\mathbf X}$ is covariance matrix of $\mathbf X$.

% For ${\rm DR}^{\ES}$, $$\min\limits_{\mathbf w\in\Delta_n} \frac{\ES_\alpha(\mathbf w^\top \mathbf X)}{\mathbf w^\top \mathbf x_\alpha}:=D,$$ where $\mathbf x_\alpha=(\ES_\alpha(X_1),\dots,\ES_\alpha (X_n)) $.
 
 % We  consider the problem   \begin{align*} \min\limits_{\mathbf w\in \Delta_n,t>0} D, ~~\text{subject to} ~~ D \mathbf w^\top \mathbf x_\alpha\geq \frac{1}{\alpha}\mathbb{E}\left[\left(\mathbf w^\top \mathbf{X}-t\right)_{+}\right]+t \end{align*}

\begin{figure}[t]
\caption{Wealth processes for  portfolios, 40 stocks, Jan 2014 - Dec 2021}\label{sec_40} %four largest stocks from ten different sectors; 
\centering
           \includegraphics[width=14cm]{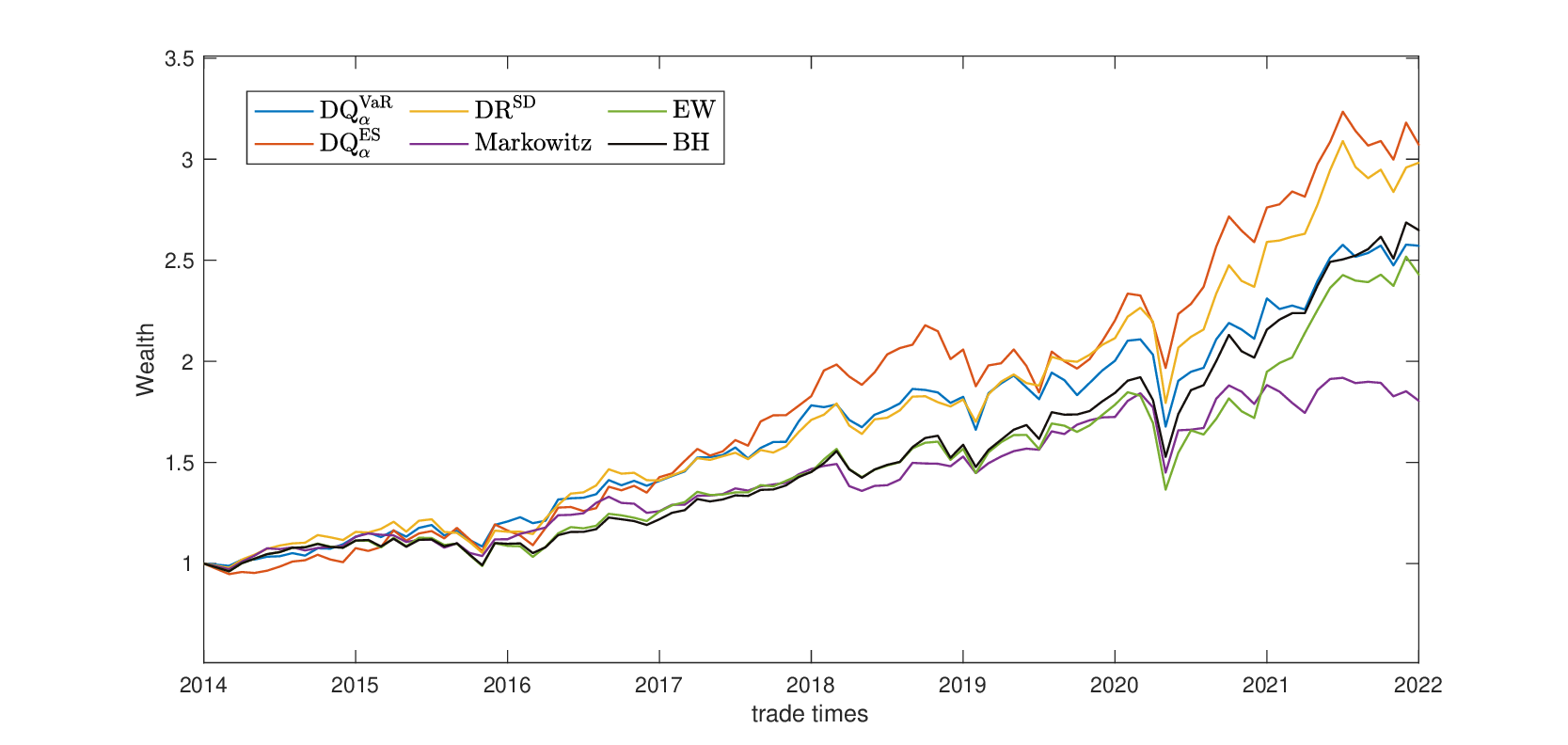}
\end{figure}

\begin{figure}[htb!]
\caption{Cumulative portfolio weights, 40 stocks, Jan 2014 - Dec 2021}\label{sec_40_weight} %four largest stocks from ten different sectors; 
\centering
           \includegraphics[width=14cm]{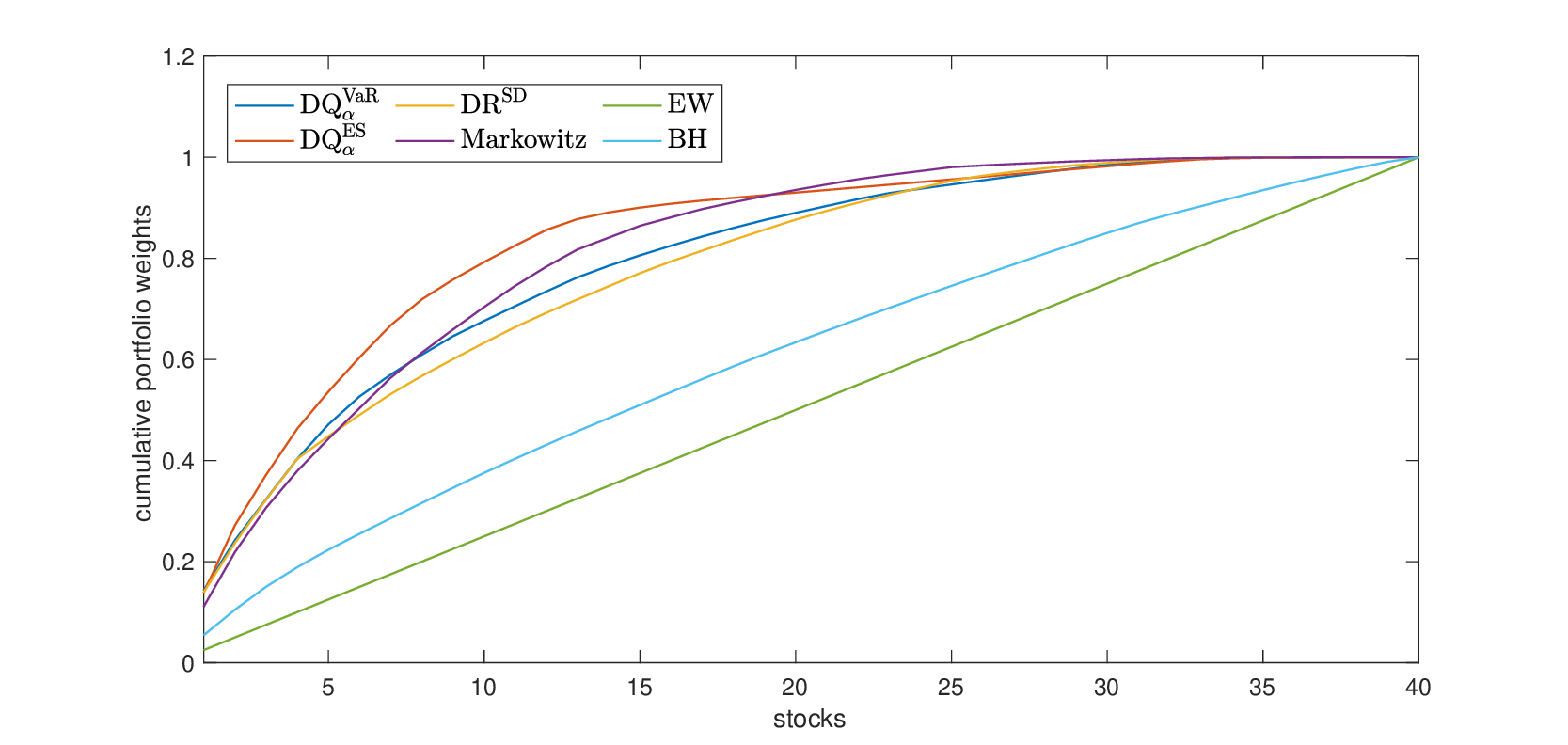}
\end{figure}

We apply  the algorithms in   Proposition \ref{thm:opt} to optimize   $\mathrm{DQ}^\VaR_\alpha$ and $\mathrm{DQ}^\ES_\alpha$, which are extremely fast.  A tie-breaking is addressed for each objective  as in \eqref{eq:general-opt3}.  Minimization of ${\rm DR}^{\rm SD}$ and the Markowitz model can be solved by existing algorithms. The initial wealth is set to $1$, and the risk-free rate  is $r=2.84\%$, which is the 10-year yield of the US treasury bill in Jan 2014.  Note that  the risk-free asset is not used to construct portfolios, but only used to calculate the Sharpe ratios.  The target annual expected return
 for the  Markowitz  portfolio is set to $10\%$.   We optimize the portfolio weights in each month with a rolling window of 500 days.  That is, in each  month, roughly 21 trading days, starting from  January 2, 2014, we use the preceding 500 trading days to compute the optimal portfolio weights using the method described above. 
 The portfolio is rebalanced every month. We choose the 4 largest stocks from each of the 10 different sectors of S\&P 500   ranked by market cap in 2012 as the portfolio compositions (40 stocks in total). The portfolio performance is reported in Figure \ref{sec_40}, and the  cumulative distribution of the sorted portfolio weights, averaged over each month, is shown in Figure \ref{sec_40_weight}.   Summary statistics,  including the annualized return (AR), the annualized volatility (AV), the Sharpe ratio (SR), and  the average trading proportion (ATP),  are reported  in Table \ref{tab_40}.\footnote{ATP is an approximation of trading costs, and it is computed as the average of $\sum_{t=1}^T| w^{t}_i-w_i^{t-}|$ over $i=1,\dots,n$, where $T=96$ is the total number of months, $w_i^{t}$ is the portfolio weight  of asset $i$ at the beginning of month $t$,
 and $w_i^{t-}$ is the portfolio weight  of asset $i$ at the end of month $t-1$, with $w_i^{1-}$ set to $w_i^1$. Note that for BH, ATP is $0$ because there is no trading, whereas for EW, ATP is positive, as rebalancing occurs at the end of each month.} 
 
 From these results, we can see that the portfolio optimization strategies based on minimizing DQ  perform  quite well,   similarly to those based on $\mathrm{DR}^{\mathrm{SD}}$, and better than the Markowitz strategy.
 Moreover, ATP and portfolio weight distribution are similar across  the  strategies based on the three diversification indices and the Markowitz strategy.  
  In contrast, the EW and BH strategies have more uniform portfolio weight distributions and smaller ATP, as anticipated. 
 We remark that it is not our intention to analyze which diversification strategy generates the highest return, which is a challenging question  that needs a separate study; also, we do not suggest   diversification should or should not be optimized in practice. The empirical results here are presented to illustrate how our proposed diversification indices work in the context of portfolio selection.  More empirical results with some other datasets and portfolio strategies are  given in Appendix \ref{App:F}, and the results show similar patterns.  % 	

\begin{table}[t]  
\def\arraystretch{1}
  \begin{center} 
    \caption{Annualized return (AR),  annualized volatility (AV), Sharpe ratio (SR), and average trading proportion (ATP)  for   different portfolio strategies from Jan 2014 to Dec 2021} \label{tab_40} 
  \begin{tabular}{c|cccccc} 
    $\%$ &${\rm DQ}^{\VaR}_\alpha$ &${\rm DQ}^{\ES}_\alpha$ &${\rm DR}^{\rm SD}$ & Markowitz & EW & BH \\    \hline
AR & 12.56    &   14.59  &14.36 & 7.93 &  11.91&12.88 \\
AV & 14.64    & 15.74 & 14.99 & 12.98 & 15.92 &14.34\\
%Real Variance & 1.7341 &   1.9825 &   2.3108  &  2.0195  &  2.1444 &   2.2541\\
SR  & 66.40     & 74.66 & 76.85&  39.22 & 56.95 &70.02\\{ATP}&19.29 & 14.75   &   15.61  &18.79 &   4.43&0 \\
 \hline \hline 
    \end{tabular}
    \end{center}%  four largest stocks from ten different sectors;
    \end{table}

\section{Concluding remarks}\label{sec:conclusion}

In this paper, we put forward six axioms  to jointly characterize a new class of  indices of diversification,
and a seventh axiom to specialize this class.   The new  diversification index DQ has favourable features both theoretically and practically, and it is  contrasted with its competitors, in particular DR.
At a high level, because of the  conceptual symmetry in Figure \ref{fig:DQDR} (see also \eqref{eq:compare-DQDR}), %that is, DR looks at quantile improvement and DQ looks at probability improvement in case of VaR,
we expect both   DQ and DR to have advantages and disadvantages in different applications, and none should fully dominate the other. Nevertheless, we find many attractive  features of DQ through the results in this paper, which suggest that DQ may be a better choice in many situations.

 We summarize these features below. Some of these features are shared by DR, but many are not.  (i) DQ  defined on a class of  MCP risk measures can be  uniquely characterized by six  intuitive  axioms (Theorem \ref{th:ax-1}).
 DQ defined on a class of coherent risk measures can be uniquely characterized by further adding the axiom of portfolio convexity (Theorem \ref{thm:quasi-conv}).
 These two results 
 lay an axiomatic foundation for using DQ as a diversification index.
(ii) DQ  further satisfies many  properties for common risk measures (Propositions \ref{prop:convex}-\ref{prop:RI}). These properties are not shared by the corresponding DR.
(iii) DQ is intuitive and interpretable with respect to dependence and common perceptions of diversification (Theorem \ref{prop:div_ben}).
%(iii) Applying DQ to variance or standard deviation, we recover the corresponding DR. Indeed, all  DRs based on non-negative $\phi$  are special cases of DQ ( Proposition \ref{pro:scx-consistent} and Proposition \ref{prop:DRgood}). 
(iv) DQ can be applied to a wide range of risk measures, such as the regulatory risk measures VaR and ES, as well as expectiles. In cases of VaR and ES, DQ has simple formulas and convenient properties (Theorem \ref{th:var} and Proposition \ref{th:var-01n}). 
% (v) DQ based on a class of convex risk measures is quasi-convex in portfolio weights (Theorem \ref{prop:convex}).
(v) Portfolio optimization of DQs based on VaR and ES can be computed very efficiently (Proposition \ref{thm:opt}). % Optimizing DQ for elliptical models and MRV models are relatively straightforward (Section \ref{sec:opt}).
%(vii) For elliptical models, DQ is able to capture heavy tail and common shock, whereas DR based on VaR, ES, variance or any other positively homogeneous risk measure only depends on the dispersion matrix (Section \ref{sec:5}).
(vi) DQ can be easily applied to real data and it produces results that are consistent with our usual perception of diversification (Section \ref{sec:emp}).

Among the class of DQ,  for most applications, 
we generally recommend the use of DQ based on ES for the following reasons: (a) it satisfies all seven axioms of intuitive appeal; (b) it has a simple optimization formula that is very convenient in portfolio optimization; (c) 
it is closely connected to financial regulation as ES is the standard risk measure of Basel IV;
(d) it has a flexible parameter $\alpha$ that allows for reflecting the sensitivity to the tail risk of the decision maker; (e) it is conceptually easy to interpret as the (usually unique) level $\beta$ of the ES family such that $\ES_\beta (\sum_{i=1}^n X_i)=\sum_{i=1}^n\ES_\alpha(X_i)$.

We also mention a few  interesting questions on DQ, which call for thorough future study. 
(i) DQ is defined through a class of risk measures. It would be interesting to formulate DQ using expected utility or behavioral decision models, to analyze the decision-theoretic implications of DQ. For instance, DQ based on entropic risk measures can be equivalently formulated using exponential utility functions.
Alternatively, one may also build DQ directly from acceptability indices (see Remark \ref{rem:literature}).
(ii)  To compute DQ, one needs to invert the decreasing function $\beta\mapsto \rho_\beta (\sum_{i=1}^n X_i)$. In the case of VaR and ES, the formula for this inversion is simple (Theorem \ref{th:var}). For more complicated classes of risk measures, this computation may be complicated and requires detailed analysis.
(iii)  For general distributions and risk measures other than VaR and ES,  finding analytical   formulas or efficient algorithms for optimal diversification  using either DQ or DR is a challenging task.
(iv) Further analysis of DQ without scale-invariance, such as those built on star-shaped risk measures (\cite{CCMTW22}), may further generalize the domain of application. 
%(iv) We mainly discussed the  desirable properties  [+], [SI], [LI] and quasi-convexity in $\mathbf w$. These properties are simple and general, and they allow for broad classes of diversification indices including DQ based on popular risk measures. More specialized properties, which may be  suitable in   different contexts,  may help to further pin down specific families of diversification indices, and they need separate studies. 
%Although it is shown in Proposition \ref{pro:non-convex} that [SI] and convexity in $\mathbf X$ are conflicting,
%it remains unclear whether there are useful diversification indices (including DQ and DR) which are convex in  $\mathbf w$ with other properties such as [+], [SI] and [LI].
%Many examples, 
% while possibly quasi-convex  in $\mathbf w$ (Theorem \ref{prop:convex}),
 %are not convex (see Example \ref{ex:DRSDnon-convex}).
%To handle large-scale optimization of DQ, efficient algorithms need to be developed. 

\textbf{Acknowledgements.}  
 We thank the Editor, an Associate Editor, three anonymous referees,  Fabio Bellini,  Ale\v s \v Cern\'y, An Chen, Jan Dhaene, Paul Embrechts, Paul Glasserman, 
Pablo Koch-Medina, Dimitrios Konstantinides,  Henry Lam, Jean-Gabriel Lauzier, Fabio Macherroni, Massimo Marinacci, Cosimo Munari,   Alexander Schied,  Qihe Tang, Stan Uryasev, Peter Wakker, and Zuoquan Xu 
for helpful comments and discussions on an earlier version of the paper.  Xia Han  is supported by the National Natural Science Foundation of China (Grant No. 12301604, 12371471). 
Liyuan Lin is supported by the Hickman Scholarship from the Society of Actuaries. 
Ruodu Wang is supported by the Natural Sciences and Engineering Research Council of Canada (CRC-2022-00141, RGPIN-2024-03728) and the Sun Life Research Fellowship.

\newpage

\appendix

\setcounter{table}{0}
\setcounter{figure}{0}
\setcounter{equation}{0}
\renewcommand{\thetable}{EC.\arabic{table}}
\renewcommand{\thefigure}{EC.\arabic{figure}}
\renewcommand{\theequation}{EC.\arabic{equation}}

\setcounter{theorem}{0}
\setcounter{proposition}{0}
\renewcommand{\thetheorem}{EC.\arabic{theorem}}
\renewcommand{\theproposition}{EC.\arabic{proposition}}
\setcounter{lemma}{0}
\renewcommand{\thelemma}{EC.\arabic{lemma}}

\setcounter{corollary}{0}
\renewcommand{\thecorollary}{EC.\arabic{corollary}}

\setcounter{remark}{0}
\renewcommand{\theremark}{EC.\arabic{remark}}
\setcounter{definition}{0}
\renewcommand{\thedefinition}{EC.\arabic{definition}}

\begin{center}
    \Large \bf
    Technical appendices 
\end{center}

\subsection*{Outline of the appendices}
We organize the   technical appendices    as follows. 
The proofs of the main results, Theorems \ref{th:ax-1}--\ref{th:var}, are presented in  Appendix \ref{App:main}. 
Additional results, discussions, and  proofs of  propositions  are presented in   Appendices \ref{App:A} (for Section \ref{sec:2}),  \ref{App:C00} (for Section \ref{sec:axiom}), 
\ref{App:C0} (for Section \ref{sec:3}), \ref{App:D}
  (for Section \ref{VaR-ES}), and  \ref{App:E} (for Section \ref{sec:opt}). Finally, in Appendix \ref{App:F},   we present other   
examples for the  optimal portfolio problem  that complement the empirical
studies in Section \ref{opt_dp}.

\section{Proofs of Theorems \ref{th:ax-1}--\ref{th:var}}\label{App:main}
 
\begin{proof}[Proof of Theorem \ref{th:ax-1}] 
For $\mathbf X\in \X^n$ and a risk measure $\phi:\X\to \R$, denote by $S(\mathbf X) = \sum_{i=1}^n X_i$ and  $  \mathbf X_\phi = (X_1-\phi(X_1),\dots,X_n-\phi(X_n))$.

We first verify the  ``if" statement. %Let $\succeq$ be a preference represented by $\rho_\alpha$.As $\rho_\alpha$ is an MCP risk measure,  $\succeq$ satisfies $\mathrm{[A1]}$-$\mathrm{[A4]}$ by Lemma \ref{pro:pre}. Then, we verify $\mathrm{DQ}^\rho_\alpha$ satisfies [+], [LI], [SI], $\mathrm{[R]}_\succeq$, $\mathrm{[C]}_\succeq$  and $\mathrm{[N]}_\succeq$.  
Using the definition of DQ and properties of MCP risk measures, it is straightforward to 
verify  $\text{[+]}$, $\text{[LI]}$, $\text{[SI]}$. Below we check the other three axioms. 
% see that $\mathrm{DQ}_\alpha^\rho$ is non-negative.  
% For all $\alpha\in I$, if  $\rho_\alpha$   satisfies $[\mathrm{CA}]$ and $[\mathrm {PH}]$, then the properties of  [LI] and [SI]  hold for  $\mathrm{DQ}_\alpha^\rho$.   To show [LI],
% for $\mathbf{c}=(c_1,\dots,c_n) \in \R^n$,   we have
% $$\begin{aligned}
% \mathrm{DQ}_\alpha^\rho(\mathbf X+\mathbf{c})&=\frac{1}{\alpha}\inf\left\{\beta \in I :  \rho_{\beta} \left(\sum_{i=1}^n  (X_i+c_i)\right) \le \sum_{i=1}^n \rho_\alpha(X_i+c_i)\right\}\\
% &=\frac{1}{\alpha}\inf\left\{\beta \in I : \rho_{\beta} \left(\sum_{i=1}^n X_i\right)+\sum_{i=1}^n c_i \le \sum_{i=1}^n  \rho_{\alpha}(X_i)+\sum_{i=1}^n c_i\right\}
% =\mathrm{DQ}_\alpha^\rho(\mathbf X).
% \end{aligned}$$
% Hence,  ${\rm DQ}^\rho_\phi$ satisfy [LI]. For $\lambda>0$, we have
% $$\begin{aligned}
% \mathrm{DQ}_\alpha^\rho(\lambda \mathbf X)&=\frac{1}{\alpha}\inf\left\{\beta \in I :  \rho_{\beta} \left(\sum_{i=1}^n \lambda X_i\right) \le \sum_{i=1}^n \rho_\alpha(\lambda X_i)\right\}\\
% &=\frac{1}{\alpha}\inf\left\{\beta \in I :  \lambda\rho_{\beta} \left(\sum_{i=1}^n X_i\right) \le \lambda \sum_{i=1}^n  \rho_\alpha(X_i)\right\} =\mathrm{DQ}_\alpha^\rho(\mathbf X).
% \end{aligned}$$
% Thus,  ${\rm DQ}_\alpha^\rho$ satisfies [SI].

 To show $\mathrm{[R]}_\phi$, for  $\mathbf X, \mathbf Y \in\X^n$ such that $\mathbf X \simeqm \mathbf Y$ and $\sum_{i=1}^n X_i \le \sum_{i=1}^n Y_i$, we have $\sum_{i=1}^n \rho_\alpha(X_i)=\sum_{i=1}^n \rho_\alpha(Y_i)$ and $\rho_\beta(\sum_{i=1}^n X_i)\leq \rho_\beta(\sum_{i=1}^n Y_i) $ for all $\beta \in I$. Hence,  
 $$\begin{aligned}
\mathrm{DQ}_\alpha^\rho( \mathbf X)&=\frac{1}{\alpha}\inf\left\{\beta \in I :  \rho_{\beta} \left(\sum_{i=1}^n  X_i\right) \le \sum_{i=1}^n \rho_\alpha(X_i)\right\}\\
&\leq\frac{1}{\alpha}\inf\left\{\beta \in I :  \rho_{\beta} \left(\sum_{i=1}^n Y_i\right) \le \sum_{i=1}^n  \rho_\alpha(Y_i)\right\} =\mathrm{DQ}_\alpha^\rho(\mathbf Y).
\end{aligned}$$  

To show $\mathrm{[N]}_\phi$, it is straightforward that  
 $\mathrm{DQ}^{\rho}_\alpha(\mathbf 0)=0.$
Let $\mathbf X=(X, \dots, X)$ for any $X \in \X$. We have 
$$\mathrm{DQ}^{\rho}_\alpha(\mathbf X)=\frac{1}{\alpha}\inf\{\beta\in I: \rho_\beta(nX)\leq n\rho_\alpha(X)\}\leq\frac{\alpha}{\alpha}=1.$$
If $\mathbf Y \succm(X, \dots, X)$ and 
$\sum_{i=1}^n Y_i\ge nX$, then  $\sum_{i=1}^n \rho_\alpha(Y_i)<n\rho_\alpha(X)\le \rho_\alpha(\sum_{i=1}^n Y_i)$. Hence, $\mathrm{DQ}^\rho_\alpha(\mathbf Y)\ge 1$.

To show $\mathrm {[C]_\phi}$, for $\mathbf X \in \X^n$, we have $a^*_{\mathbf X}=\inf\{\beta \in I: \rho_\beta(S(\mathbf X_{\rho_\alpha}))\le 0\}$.  If $a^*_{\mathbf X}=0$, it is clear that $\mathrm{DQ}^\rho_\alpha(\mathbf X)= 0$ and $\mathrm {[C]_\phi}$ holds as $\mathrm{DQ}^\rho_\alpha(\mathbf Y^k)\ge 0$ for any $\mathbf Y^k \in \X^n$. Now, we assume 
 $a^*_{\mathbf X}>0$. For any $0\le\beta< a^*_{\mathbf X}$, we have $\rho_{\beta}(S(\mathbf X_{\rho_\alpha}))>0$.
Since  $\mathbf Y^k\simeqm \mathbf X$ for each $k$ and $(S(\mathbf X)-S(\mathbf Y^k))_+ \buildrel {L^\infty} \over \longrightarrow 0$ as $k\to\infty$,    for any $\epsilon> 0$, there exists  $K$ such that $ S(\mathbf X_{\rho_\alpha})- S(\mathbf Y^k_{\rho_\alpha}) \leq\epsilon $ for  all $k\ge K$.
For any $0<\delta< a^*_{\mathbf X}$, let $0<\epsilon<\rho_{a^*_{\mathbf X}-\delta}(S(\mathbf X_{\rho_\alpha}))$. It is clear that $0<\epsilon<\rho_{\beta}(S(\mathbf X_{\rho_\alpha}))$ for all $0<\beta< a^*_{\mathbf X}-\delta$. 
Hence, for all $0<\beta< a^*_{\mathbf X}-\delta$, there exists $K$ such that   $0<\rho_\beta(S(\mathbf X_{\rho_\alpha})) -\epsilon \le \rho_\beta(S(\mathbf Y^k_{\rho_\alpha}))$ for all $k\ge K$, which implies $a^*_{\mathbf  Y^k}\ge a^*_{\mathbf X}-\delta$.
Therefore, $(\mathrm{DQ}_\alpha^\rho(\mathbf X)- \mathrm{DQ}_\alpha^\rho(\mathbf Y^k))_+\to 0$.

% To show [N], let $\mathbf X=(X, \dots, X)$ for any $X \in \X^*$. As $\rho$ is non-flat from left at $(\alpha, X)$, we have 
% $$\mathrm{DQ}^{\rho}_\alpha(\mathbf X)=\frac{1}{\alpha}\inf\{\beta\in I: \rho_\beta(nX)\leq n\rho_\alpha(X)\}=\frac{\alpha}{\alpha}=1.$$
% Moreover, it is straightforward that  
%  $\mathrm{DQ}^{\rho}_\alpha(\mathbf 0)=0.$

 Next, we show the ``only if" statement. Assume that  $D:\X^n\to \overline \R$
satisfies $\mathrm{[+]}$, $\mathrm{[LI]}$, $\mathrm{[SI]}$, $\mathrm{[R]_\phi}$,   $\mathrm{[N]_\phi}$ and $\mathrm{[C]_\phi}$.
%By Lemma \ref{pro:pre},   $\succeq$ can be  represented by an MCP  risk measure $\phi$.
Note that $\mathbf X_\phi 
\simeqm   \mathbf Y_\phi$ for all $\mathbf X,\mathbf Y\in \X^n$
since $\phi(X-\phi(X))=0$ for any $X\in \X$.
Hence,  by using $\mathrm{[R]}_\phi$,  we know that $S( \mathbf X_\phi ) =S( \mathbf Y_\phi )$ implies $D(\mathbf X_\phi ) = D(\mathbf Y_\phi ) $.
Further, by [LI], we have 
$D(\mathbf X)=D(\mathbf Y)$ if $S( \mathbf X_\phi ) =S( \mathbf Y_\phi )$.
This means that $D(\mathbf X)$ is determined by $S(\mathbf X_\phi)$.
Define the mapping
\begin{align}
\label{eq:ax-1}
R:\X\to [0,\infty],~  R(X) = \inf\{D(\mathbf X): X\le S(\mathbf X_\phi),~\mathbf X\in \X^n\},
\end{align}
with the convention $\inf \varnothing  = \infty$.
Next, we will verify several properties of $R$.
\begin{enumerate}[(a)]
\item 
$R(S(\mathbf X_\phi)) = D(\mathbf X)$ for $\mathbf X\in \X^n$. 
The inequality $R(S(\mathbf X_\phi)) \le D(\mathbf X)$ follows directly from \eqref{eq:ax-1}.
To see the opposite direction of the inequality, 
suppose $R(S(\mathbf X_\phi)) < D(\mathbf X)$. By \eqref{eq:ax-1},  there exists $\mathbf Y\in \X^n$ such that $D(\mathbf Y) < D(\mathbf X)$ and $S(\mathbf X_\phi)\le S(\mathbf Y_\phi)$. This contradicts 
  $\mathrm{[R]}_\phi$ of $D$.
\item $R(\lambda  X )=R(X)$  for all   $\lambda>0$ and $X\in \X$. This follows directly from \eqref{eq:ax-1}, [SI] of $D$ and positive homogeneity of $\phi$ which gives $ ( \lambda \mathbf X)_\phi = \lambda \mathbf X_\phi $.
\item $R( X )\le R(Y)$  for all $X,Y\in \X$ with $X\le Y$. This follows directly from \eqref{eq:ax-1}.
\item $R(0)=0$.  This follows directly from \eqref{eq:ax-1} and $D(\mathbf 0)=0$ in $\mathrm{[N]}_\phi$.
\item $\lim_{c \downarrow 0}R(S(\mathbf X_\phi)-c)=R(S(\mathbf X_\phi))$ for $\mathbf X\in \X^n$. Let $X= S(\mathbf X_\phi)$. By (c), we have $\lim_{c \downarrow 0}R(X-c)\le R(X)$. 
Assume $\lim_{c \downarrow 0}R(X-c)< R(X)$; that is, there exists $\delta>0$ such that $R(X-c)<R(X)-\delta$ for all $c >0$. Let $c_k=1/k$ for $k\in\mathbb N$.
By \eqref{eq:ax-1},  there exists  a sequence $\{\mathbf Y^{k}\}_{k \in \mathbb N}$  such that $X-c_k\le S(\mathbf Y^{k}_\phi)$ and $D(\mathbf Y^{k}_\phi)<D(\mathbf X_\phi)-\delta$. 
For $\{\mathbf Y^{k}_\phi\}_{k\in\mathbb N}$, we have $0\le (S(\mathbf X_\phi)-S(\mathbf Y^{k}_\phi))_+\le c_k$, which implies $(S(\mathbf X_\phi)-S(\mathbf Y^{k}_\phi))_+\buildrel {L^\infty} \over \longrightarrow 0 $ as $k \to \infty$.
By $\mathrm{[C]}_\phi$, we have $(D(\mathbf X_\phi)-D(\mathbf Y^{k}_\phi))_+ \to 0$; that is, for any $\delta>0$, there exists $K \in \mathbb{N}$ such that $D(\mathbf X_\phi)-\delta\le D(\mathbf Y^{k}_\phi)$ for all $k>K$, which is a contradiction. 
Therefore, we have  $\lim_{c \downarrow 0}R(S(\mathbf X_\phi)-c)=R(S(\mathbf X_\phi))$.
\end{enumerate}
%{\color{red}On the other hand, if $\phi$ is a constant risk measure, by taking $\mathbf X=(1/nX+\phi(X),\dots, 1/nX+\phi(X))$,  we have $X\le S(\mathbf X_\phi)$ and $R(X)\le D(X, \dots, X)$. Similar to (a), $R(X)< D(X, \dots, X)$ conflicts  $\mathrm{[R]}_\succeq$. Hence, we have $R(X)=D(X, \dots, X)$ from which it is easy to verify (a)-(e) for $R(X)$.}

Let $I=(0,\infty)$. 
For each $\beta  \in (0,\infty) $, 
let $\mathcal A_\beta=\{X\in \X: R(X) \le \beta\}$. 
Since $R$ is monotone, $\mathcal A_\beta$ is a decreasing set; i.e., $X\in \mathcal A_\beta$ implies $Y\in \mathcal A_\beta$ for all $Y\le X$.
Moreover, $\mathcal A_\beta$ is conic; i.e.,  $X\in \mathcal A_\beta$ implies $\lambda X\in \mathcal A_\beta$ for all $\lambda >0$.
 Moreover, we have $\mathcal A_\beta\subseteq \mathcal A_\gamma$ for $\beta\le \gamma$,
 and $\mathcal A_\beta\neq \varnothing$ since $0\in \mathcal A_0$.
 
 Let $\rho_\beta (X) = \inf\{m\in \R: X-m \in \mathcal A_\beta\}$ for $\beta \in I$.
Since $\rho_\beta$ is defined via a conic acceptance set,  $(\rho_\beta)_{\beta\in I}$ is a class of MCP risk measures; see \cite{FS16}. It is also clear that $\rho_\beta$ is decreasing in $\beta$. Note that $X\in \mathcal A_\beta$ implies   $\rho_\beta(X)\le 0$. Hence, $$ R(X)= \inf \{\beta\in I : R(X)\le \beta\} =\inf \{\beta \in I : X\in \mathcal A_\beta\} 
\ge  \inf \{\beta \in I : \rho_\beta(X)\le 0\}.$$
For $X \in \{S(\mathbf X_\phi): \mathbf X \in \X^n\}$, using (e), we have $R(X-m)\le \beta$ for all $m >0$ implies $R(X)\le \beta$. Then we have
$$ \inf \{\beta \in I : \rho_\beta(X)\le 0\}= \inf \{\beta \in I : R(X-m)\le \beta \text{ for all } m>0\}\ge \inf \{\beta \in I : R(X)\le\beta\}. $$
Therefore, $ \inf \{\beta \in I : \rho_\beta(S(\mathbf X_\phi))\le 0\}= \inf \{\beta \in I : R(S(\mathbf X_\phi))\le\beta\}=R(S(\mathbf X_\phi))$ for all $\mathbf X \in \X^n$.
Using  (a), we get, for all $\mathbf X\in \X^n$, 
$$
D(\mathbf X)= R(S(\mathbf X_\phi))
=  \inf \left \{\beta \in I : \rho_\beta\left (\sum_{i=1}^nX_i\right ) \le \sum_{i=1}^n \phi(X_i)\right\}. 
$$

Take a duplicate portfolio $\mathbf X=(X, \dots, X)$. Together with [PH] of $\rho_\beta$, $D(\mathbf{X})\le 1$  implies 
$$D(X,\dots,X)= \inf \left \{\beta \in I : \rho_\beta (X) \le   \phi(X)\right\} \le 1,$$ 
which is equivalent to
$\rho_\beta (X) \le \phi(X)$ for $\beta>1$.  
% For any $m<\phi(X)$ and $\mathbf{Y} \in \X^n$,  let $\mathbf Z=(Y_1-\phi(Y_1)+m/n, \dots, Y_n-\phi(Y_n)+m/n)$. It is clear that $S(\mathbf Z_\phi)=\sum_{i=1}^n Y_i-\phi(Y_i)$ and $(Z_1 \dots, Z_n) \succm(1/n X, \dots, 1/n X)$.
% If there is no portfolio worst than duplicate, it means that the set $\{\mathbf Y: X \le S(\mathbf Y_\phi)\}$ is empty.
For   $m<\phi(X)$, take any $\mathbf Y=(Y_1,
\dots,Y_n)$ satisfying 
$S(\mathbf Y_\phi) \ge X-m$; such $\mathbf Y$ may not exist. 
% $\sum_{i=1}^n Y_i-\sum_{i=1}^n \phi(Y_i) + m \ge X $.
Let $\mathbf Z=(Y_1-\phi(Y_1)+m/n, \dots, Y_n-\phi(Y_n)+m/n)$, yielding  $\sum_{i=1}^n Z_i=\sum_{i=1}^n Y_i-\sum_{i=1}^n \phi(Y_i)+m\ge X$ and $(Z_1 \dots, Z_n) \succm(X/n  , \dots, X/n  )$. 
 Hence, $\mathbf Z$ is worse than duplicate.  By [LI] and $\mathrm{[N]}_\phi$, we have $D(\mathbf Y)=D(\mathbf Z)\ge 1$.
 Since $D(\mathbf Y)\ge 1$ for all such $\mathbf Y$, 
by \eqref{eq:ax-1}, we have $R(X-m)\ge 1$.
%If there is no such $\mathbf Y$, then  $R(X-m)=\infty$ by \eqref{eq:ax-1}.
%In either case, we get $R(X-m)\ge 1$ for all $m<\phi(X)$.
The above observation implies $\rho_{1-\epsilon}(X)\ge \phi(X)$ for any $\epsilon>0$ 
since  $\rho_{1-\epsilon}(X)=\inf\{m \in \R: R(X-m) \le 1-\epsilon\}$.
Therefore, we get $\rho_{1-\epsilon}\ge \phi \ge \rho_{1+\epsilon}$ for all $\epsilon>0$. 
Let $\tilde \rho_1=\phi$, and $\tilde \rho_\beta=\rho_\beta$ for $\beta \ne 1$.
The class $\tilde \rho = (\tilde \rho_\beta)_{\beta \in I}$ of MCP risk measures is decreasing in $\beta$ by the above argument. 
Moreover, for any $X\in \X$, since the two decreasing curves $\beta \mapsto \rho_\beta (X) $ and $\beta \mapsto \tilde \rho_\beta (X) $  differ at only one point,  their (left) inverses coincide, and we have, for all $\mathbf X \in \X^n$,
$$ \inf \left \{\beta \in I : \rho_\beta\left (\sum_{i=1}^nX_i\right ) \le \sum_{i=1}^n \phi(X_i)\right\}=\inf \left \{\beta \in I :  \tilde  \rho_\beta\left (\sum_{i=1}^nX_i\right ) \le \sum_{i=1}^n \tilde \rho_1(X_i)\right\} , $$
which implies  $ D=\mathrm{DQ}^{\tilde \rho}_1$ on $\X^n$. 
A reparametrization via $\hat \rho_\beta=\tilde \rho_{\beta /\alpha}$ leads to
 $ D=\mathrm{DQ}^{\hat  \rho}_\alpha$ and $\hat \rho_\alpha = \phi$.
% We can show that  
% $$\begin{aligned}
% \mathrm{DQ}_{1}^\rho(\mathbf X)
% &=  \inf \left \{\beta \in I : \rho_\beta\left (\sum_{i=1}^nX_i\right ) \le \sum_{i=1}^n \rho_1(X_i)\right\}\\
% &=  \inf \left \{\beta \in I : \pi_{\beta \alpha}\left (\sum_{i=1}^nX_i\right ) \le \sum_{i=1}^n \pi_\alpha(X_i)\right\}\\
% &=\frac{1}{\alpha}  \inf \left \{\gamma \in I : \pi_{\gamma}\left (\sum_{i=1}^nX_i\right ) \le \sum_{i=1}^n \pi_\alpha(X_i)\right\}=\mathrm{DQ}_{\alpha}^\pi(\mathbf X)
% \end{aligned}$$
% and  $\pi_\alpha=\rho_1=\phi$. %which means $\phi_\alpha$ is an MCP risk measure and represents the preference $\succeq$.
%Hence, $D=\mathrm{DQ}_{\alpha}^\pi$ for some decreasing $\pi$ such that $\pi$ is decreasing in $\alpha$.
\end{proof}

The next two remarks are useful in the proof of Theorem \ref{thm:quasi-conv}.
 \begin{remark}\label{rem:r-con}
In the proof of Theorem \ref{th:ax-1}, the constructed class of risk measures $(\rho_\beta)_{\beta\in I}$ exhibits right continuity in $\beta$. This is established based on the condition   $\bigcap_{\beta>\beta^*} \mathcal A_\beta=\mathcal A_{\beta^*}$.
\end{remark} 
\begin{remark}\label{rem:nonlinear}
For a non-linear coherent risk measure $\phi$,
there exists $Y\in \X$ such that $\phi(Y)+\phi(-Y)>0$. 
Suppose otherwise. Since $\phi$ is coherent risk measure, we have $\phi(Y)+\phi(-Y)\ge 0$, and this implies $\phi(Y)+\phi(-Y)=0$ for all $Y\in \X$. We obtain $\phi(Y)\le \phi(X+Y)+\phi(- X)=\phi(X+Y)-\phi(X)$ and 
$\phi(X+Y)\le \phi(X)+ \phi(Y)$ for any $X, Y \in \X$. This implies  $\phi(X+Y)=\phi(X)+\phi(Y)$ for any  $X, Y \in \X$, contradicting the non-linearity  of  $\phi$. 
\end{remark}

\begin{proof}[Proof of Theorem \ref{thm:quasi-conv}] 
For the``if" statement, since a coherent risk measure is also MCP, it follows  that ${\rm DQ}^\rho_\alpha$ satisfies $\mathrm{[+]}$, $\mathrm{[LI]}$, $\mathrm{[SI]}$, $\mathrm{[R]_\phi}$ and $\mathrm{[N]}_\phi$, $\mathrm{[C]_\phi}$  by  Theorem \ref{th:ax-1}. Next, we show that  ${\rm DQ}^\rho_\alpha$ satisfies  $\mathrm{[PC]}$.
 
For any $\mathbf X \in \mathcal X^n$, let  $r_\beta^\mathbf X: \Delta_n\rightarrow \R $ be given by  $$r_\beta^{\mathbf X}(\mathbf w )= \rho_{\beta}\left(\sum_{i=1}^{n}w_i X_{i}\right) - \sum_{i=1}^{n} \rho_{\alpha}\left(w_iX_{i}\right)$$
%=\rho_\beta\left(\mathbf w^\top (\mathbf X-\mathbf x_\alpha^\rho)\right)$$
for $\beta \in I$. 
 From $\mathrm{[PH]}$ of  $\rho_\alpha$, we have 
 $r_\beta^{\mathbf X}(\mathbf w)= \rho_{\beta}\left(\sum_{i=1}^{n}w_i X_{i}\right) -  \mathbf w^\top\mathbf{x}_{\alpha}^{\rho}.$
Convexity  of $\rho_\beta$  implies convexity of  $\mathbf w\mapsto    r_\beta^{\mathbf X}(\mathbf w )$. %that is, 
% $r_\beta(\lambda \mathbf  w+(1-\lambda) \mathbf  v)\leq \lambda r_\beta( \mathbf w )+(1-\lambda)r_\beta( \mathbf  v)$ for any $\mathbf w,\mathbf v\in\Delta_n$ and $\lambda\in(0,1)$.
Hence, for the portfolio weight $\lambda\mathbf w+(1-\lambda)\mathbf v \in\Delta_n$, DQ based on $\rho$ at level $\alpha \in(0,1)$ is given by
 $$\begin{aligned}{\rm DQ}^{\rho}_{\alpha}((\lambda\mathbf w+(1-\lambda)\mathbf v) \odot \mathbf X)&=
  \frac{1}\alpha    \inf\left\{\beta \in I :  r_\beta^{\mathbf X}(\lambda \mathbf  w+(1-\lambda) \mathbf  v)\leq 0 \right\}\\
 &\le \frac{1}\alpha    \inf\left\{\beta \in I : \lambda r_\beta^{\mathbf X}( \mathbf w )+(1-\lambda)r_\beta^{\mathbf X}( \mathbf  v )\leq 0 \right\}\\
 &\leq \frac{1}\alpha  \max\left\{   \inf\{\beta \in I : r_\beta^{\mathbf X}( \mathbf w )\leq 0\}, \inf\{\beta \in I :r_\beta^{\mathbf X}( \mathbf  v )\leq 0 \}\right\}
 \\&=\max\left\{ {\rm DQ}^{\rho}_{\alpha}(\mathbf w \odot \mathbf X),{\rm DQ}^{\rho}_{\alpha}(\mathbf v \odot \mathbf X)\right\},\end{aligned}$$ which gives us quasi-convexity of $\mathbf w \mapsto \mathrm{DQ}^\rho_\alpha(\mathbf w \odot \mathbf X)$. By Remark \ref{rem:qc}, we have that $\mathrm{DQ}^\rho_\alpha$ satisfies [PC].
 
 For the ``only if" statement,
 we have constructed a class of MCP risk measures $\rho=
 (\rho_\beta)_{\beta \in (0, \infty)}$ with $\rho_\alpha=\phi$ in the proof of Theorem \ref{th:ax-1}.  We will further show that  $\rho$ is a class of convex risk measures using [PC].

For any $\lambda \in [0,1]$, $\mathbf w,\mathbf v\in \Delta_n$,  $\mathbf X\in \X^n$ and $\beta \in (0, \infty)$,
if $r^{\mathbf X}_\beta(\mathbf w)\le 0$ and $r^{\mathbf X}_\beta(\mathbf v)\le 0$, then we have $\mathrm{DQ}^\rho_\alpha(\mathbf w \odot \mathbf X) \le \beta$ and $\mathrm{DQ}^\rho_\alpha(\mathbf v \odot \mathbf X) \le \beta$.  By [PC], we have $\mathrm{DQ}^\rho_\alpha((\lambda\mathbf w +(1-\lambda)\mathbf v)\odot \mathbf X) \le \beta$. As discussed in Remark \ref{rem:r-con},  $\rho_\beta$ is right-continuous for any $X \in \X$. Hence, we have 
 $r^{\mathbf X}_\beta(\lambda \mathbf w+(1-\lambda) \mathbf v)\le 0$. That is, the set $\{\mathbf w \in \Delta_n: r^{\mathbf X}_\beta(\mathbf w)\le 0 \}$ is convex for any $\mathbf X \in \X^n$ and $\beta \in I$.

Let $\mathrm{Conv}\{X-\rho_\alpha(X): X \in \X\}$ be the convex hull of $\{X-\rho_\alpha(X): X \in \X\}$. %that is, $\mathrm{Conv}\{X-\rho_\alpha(X): X \in \X\}=\{\sum _{i=1}^n \lambda_i(X-\rho_\alpha(X)): \boldsymbol\lambda\in \Delta_n, X \in \X \}$.
Next, we show that $\mathrm{Conv}\{X-\rho_\alpha(X): X \in \X\}=\{X \in \X: \rho_\alpha(X)\le 0\}$.
For any $Z\in \mathrm{Conv}\{X-\rho_\alpha(X): X \in \X\}$, there exist $(\lambda_1, \dots, \lambda_n) \in \Delta_n$ and $X_1, \dots, X_n \in \X$ such that $Z=\sum_{i=1}^n \lambda_i (X_i-\rho_\alpha(X_i))$. Since $\rho_\alpha$ is convex, we have 
$$\rho_\alpha(Z)=\rho_\alpha\left(\sum_{i=1}^n \lambda_i (X_i-\rho_\alpha(X_i))\right)\le \sum_{i=1}^n \lambda_i \rho_{\alpha}(X_i-\rho_\alpha(X_i))=0.$$
Hence, $\mathrm{Conv}\{X-\rho_\alpha(X): X \in \X\}\subseteq\{X \in \X: \rho_\alpha(X)\le 0\}$.

On the other hand, since $\rho_\alpha$ is a non-linear coherent risk measure, as noted in  Remark \ref{rem:nonlinear}, there exists $Y_0$ such that $\rho_\alpha(Y_0)+\rho_\alpha(-Y_0)>0$. Let $Y_1=1-Y_0$.   For any $Z\in \X$ with $ \rho_\alpha(Z)\le 0$, we can find $\theta>0$ such that 
$$Z=(1-2\lambda)(X-\rho_\alpha(X))+\lambda(\theta Y_0-\rho_\alpha(\theta Y_0))+\lambda(\theta Y_1-\rho_\alpha(\theta Y_1))$$
with $\lambda=-{\rho_\alpha(Z)}/{(\theta\rho_\alpha(Y_0)+\theta\rho_\alpha(-Y_0))}$ and $X={1}/(1-2\lambda)Z$. It is clear that $\lambda \in [0,1/2]$ holds if $\theta$ is sufficiently large. Hence, $Z \in \mathrm{Conv}\{X-\rho_\alpha(X): X \in \X\}$. This implies $\mathrm{Conv}\{X-\rho_\alpha(X): X \in \X\}\supseteq\{X \in \X: \rho_\alpha(X)\le 0\}$.
%Moreover, since $\rho_\alpha$ is [PH], if  $Z \in \mathrm{Conv}\{X-\rho_\alpha(X): X \in \X\}$, we have  $\theta Z \in \mathrm{Conv}\{X-\rho_\alpha(X): X \in \X\}$ for any $\theta>0$. Therefore, for any $Z\in \X$ with $\rho_\alpha(Z)<-(\rho_\alpha(Y)+\rho_\alpha(-Y))/2$, it can be written as the convex combination of $\theta X-\rho_\alpha(\theta X)$, $\theta Y-\rho_\alpha(\theta Y)$ and $\theta Y'-\rho_\alpha(\theta Y')$ for some $\theta>1$. Hence, for any $Z \in \X$ with $\rho_\alpha(Z) \le 0$, we have $Z \in \mathrm{Conv}\{X-\rho_\alpha(X): X \in \X\}$; that is $\mathrm{Conv}\{X-\rho_\alpha(X): X \in \X\}=\{X \in \X: \rho_\alpha(X)\le 0\}$.

Therefore, for any $ X,  Y\in \X$ with $\rho_\alpha( X) \le 0$ and $\rho_\alpha(Y) \le 0$, we can find  $\mathbf w, \mathbf v\in \Delta^n$ and  $\mathbf X \in \X^n$ with $n \ge 4$ such that $ X=\mathbf w^\top (\mathbf X-\mathbf x_\alpha^\rho)$ and  $ Y=\mathbf v^\top (\mathbf X-\mathbf x_\alpha^\rho)$.  Since $r_\beta(\mathbf w )=\rho_\beta\left(\mathbf w^\top (\mathbf X-\mathbf x_\alpha^\rho)\right)$, 
 we have $\rho_\beta( X)=r_\beta^{\mathbf X}(\mathbf w)$, $\rho_\beta( Y)=r_\beta^{\mathbf X}(\mathbf v)$ and $\rho_\beta(\lambda  X+(1-\lambda) Y)=r_\beta^{\mathbf X}(\lambda\mathbf w+(1-\lambda)\mathbf v)$. 
If $\rho_\beta( X)\le 0$ and $\rho_\beta( Y)\le 0$, we have $r^{\mathbf X}_\beta(\lambda \mathbf w+(1-\lambda)\mathbf v)\le 0$ for any $\lambda \in [0,1]$; that is $\rho_\beta(\lambda  X+(1-\lambda)  Y)\le 0$. Hence, $\rho_\beta$ is quasi-convex. Since $\rho_\beta$ is MCP, we further have $\rho_\beta$ is coherent.
\end{proof}

\begin{proof}[Proof of Theorem \ref{prop:div_ben}]$\text{(i)}$
As  $\sum_{i=1}^{n} X_{i}\le\sum_{i=i}^{n}\rho_\alpha(X_i)$ a.s.~and $\rho_0\leq\esssup$, it is clear that $\alpha^*=0$ in \eqref{def:alpha*}, which implies $\mathrm{DQ}_\alpha^\rho(\mathbf{X})=0$. Conversely, if $\mathrm{DQ}_\alpha^\rho(\mathbf{X})=0$, then $\alpha^*=0$.
By definition of $\rho_0$ and $\mathrm{DQ}^\rho_\alpha$,
this implies   $\rho_0(\sum_{i=1}^{n} X_{i})\le\sum_{i=1}^{n}  \rho_\alpha(X_i)$,
and hence
$ \sum_{i=1}^{n} X_{i} \le \sum_{i=1}^{n}  \rho_\alpha(X_i)$ a.s.

$\text{(ii)}$
We first show the ``only if" statement. 
 As $\rho$ is left continuous and non-flat from the left at $(\alpha,\sum_{i=1}^n X_i$) and 
 $\sum_{i=1}^n \rho_\alpha(X_i)-\rho_\alpha \left(\sum_{i=1}^n X_i\right) >0,$ there exists $\delta>0$ such that
 $$ \rho_\beta \left(\sum_{i=1}^n X_i\right)-\rho_\alpha \left(\sum_{i=1}^n X_i\right)  <  \sum_{i=1}^n \rho_\alpha(X_i)-\rho_\alpha \left(\sum_{i=1}^n X_i\right)  $$ for all $\beta \in (\alpha-\delta,\alpha)$.   
Hence, we have $\alpha^*\le \alpha-\delta<\alpha$, which leads to ${\rm DQ}^{\rho}_\alpha(\mathbf X)<1$. 

Next, we  show the ``if" statement. 
As ${\rm DQ}^{\rho}_\alpha(\mathbf X)<1$, we have $\alpha>\alpha^*$.
By \eqref{def:alpha*}, there exists $\beta \in (\alpha^*,\alpha)$ such that
$$\sum_{i=1}^n \rho_\alpha(X_i)\ge \rho_{\beta}\left(\sum_{i=1}^n X_i\right).$$
Because $\rho$ is non-flat from the left at $(\alpha,\sum_{i=1}^n X_i)$, we have
$$\sum_{i=1}^n \rho_\alpha(X_i)\ge\rho_{\beta}\left(\sum_{i=1}^n X_i\right)>\rho_\alpha\left(\sum_{i=1}^n X_i\right).$$

$\text{(iii)}$ If $\rho_\alpha$ satisfies {$\mathrm{[PH]}$}, for $\mathbf{X}=(\lambda_1 X,\dots,\lambda_nX)$ where $\lambda_1,\dots,\lambda_n\ge 0$, we have
$$\alpha^*= \inf\left\{\beta \in I :  \rho_{\beta}\left(\sum_{i=1}^n\lambda_i X\right) \le \sum_{i=1}^n \lambda_i\rho_{\alpha}(X) \right\}.$$
It is clear that $\rho_{\alpha}\left(\sum_{i=1}^n\lambda_i X\right)=(\sum_{i=1}^n\lambda_i)\rho_{\alpha}(X)$.
Together with the non-flat condition and $\rho_{\beta}\left(\sum_{i=1}^n\lambda_i X\right)>\sum_{i=1}^n\lambda_i\rho_{\alpha}(X)$ for all $\beta <\alpha$, we have  $\alpha^*=\alpha$,
and thus ${\rm DQ}^{\rho}_\alpha(\mathbf X)=1$.

$\text{(iv)}$ If  $\rho_\alpha$ is comonotonic-additive and $\mathbf X$ is comonotonic, then  $$\alpha^*= \inf\left\{\beta \in I :  \rho_{\beta}\left(\sum_{i=1}^nX_i\right) \le \sum_{i=1}^n \rho_{\alpha}(X_i)=\rho_{\alpha}\left(\sum_{i=1}^nX_i\right) \right\},$$ which, together with the non-flat condition, implies that $\alpha^*=\alpha$, and thus ${\rm DQ}^{\rho}_\alpha(\mathbf X)=1$.
\end{proof}

\begin{proof}[Proof of Theorem \ref{th:var}] 
We first show \eqref{eq:var-alter}.
For any $X\in \X$, $t\in \R$ and $\alpha\in (0,1)$,  by Lemma 1 of \cite{GJW22}, 
$\p(X>t)\le \alpha$ if and only if 
$\VaR_\alpha(X)\le t$. 
Hence, 
\begin{align*}
\p\left(\sum_{i=1}^n X_i> \sum_{i=1}^n \VaR_{\alpha}(X_i)\right)  &  =
 \inf
\left\{\beta \in (0,1) : \p\left(\sum_{i=1}^n X_i> \sum_{i=1}^n \VaR_{\alpha}(X_i)\right)\le \beta \right\} 
\\& =   \inf
\left\{\beta \in (0,1) :  \VaR_{\beta} \left(\sum_{i=1}^n X_i\right) \le \sum_{i=1}^n \VaR_{\alpha}(X_i) \right\} ,
\end{align*} 
and \eqref{eq:var-alter} follows.
The formula \eqref{eq:es-alter}  for $\mathrm{DQ}^\ES_\alpha$ follows from a similar argument to \eqref{eq:var-alter} by noting that $Y $ is a random variable with  $\VaR_\alpha(Y)=\ES_\alpha(\sum_{i=1}^n X_i) $.

Next, we show the last statement of the theorem.
If  $\p(\sum_{i=1}^n X_i > \sum_{i=1}^n \ES_\alpha(X_i))=0$,
 then $\mathrm{DQ}^\ES_\alpha(X)=0$ by Theorem \ref{prop:div_ben} (i).

 Below, we assume   $\p(\sum_{i=1}^n X_i > \sum_{i=1}^n \ES_\alpha(X_i))>0$. 
The formula  \eqref{eq:es-alter3}  is very similar to  Proposition 2.2 of \cite{MU18}, where we additionally show that the minimizer to \eqref{eq:es-alter3} is not $0$.
Here we present a self-contained proof based on the well-known formula of ES (\cite{RU02}),
$$ 
\mathrm{ES}_{\beta}(X)=\min _{t \in \mathbb{R}}\left\{t+\frac{1}{\beta} \mathbb{E}\left[(X-t)_{+}\right]\right\}, \mbox{~~ for $X \in \mathcal{X}$ and $\beta \in(0,1)$}.
$$ 
Using this formula, we obtain, by writing $X_i'=X_i-\ES_\alpha(X_i)$ for $i\in [n]$,
\begin{align*}
\mathrm{DQ}^{\ES}_\alpha (\mathbf{X})& = \frac{1}{\alpha} \inf\left\{\beta \in (0,1) :  \ES_{\beta} \left(\sum_{i=1}^n X_i\right) - \sum_{i=1}^n \ES_{\alpha}(X_i) \le0 \right\}\\&= \frac{1}{\alpha}\inf\left\{\beta \in (0,1) : \min _{t \in \mathbb{R}} \left\{t+ \frac{1}{\beta}\mathbb{E}\left[ \left(\sum_{i=1}^n   X_i' -t\right)_{+}\right]\right\}\leq0\right\}
\\&= \frac{1}{\alpha}\inf\left\{\beta \in (0,1) :   \frac{1}{\beta}\mathbb{E}\left[ \left(\sum_{i=1}^n   X_i' -t\right)_{+}\right]\leq -t  \mbox{~for some } t\in \R \right\} 
\\ &= \frac{1}{\alpha}\inf\left\{\beta \in (0,1) :   \mathbb{E}\left[ \left(r \sum_{i=1}^n  X_i' +1\right)_{+}\right]\le  \beta \mbox{~for some } r\in (0,\infty) \right\} 
\\&= \frac{1}{\alpha}\inf_{r\in (0,\infty)} \mathbb{E}\left[ \left(r \sum_{i=1}^n  X_i'  +1\right)_{+}\right].
\end{align*} 
%Hence,
%$$\mathrm{DQ}^{\ES}_\alpha (\mathbf{X})=\inf_{t\in \R, \beta \in (0,1)} \beta 
%~~~\mbox{ s.t.}~~~ t+\frac{1}{\beta}\mathbb{E}\left[ \left(\sum_{i=1}^n \left( X_i -  \ES_{\alpha}(X_i)\right)-t\right)_{+}\right]\leq 0.$$
%It is clear that $t>0$ can not satisfy the constraint. Hence, we can limit $t$ in $(-\infty, 0]$.
%By replacing $t$ with $-1/r$, we can get 
%$$\mathrm{DQ}^{\ES}_\alpha (\mathbf{X})=\inf_{r\in (0,+\infty], \beta \in (0,1)} \beta 
%~~~\mbox{ s.t.}~~~ \mathbb{E}\left[ \left(r\sum_{i=1}^n \left( X_i -  \ES_{\alpha}(X_i)\right)+1\right)_{+}\right]\leq \beta,$$
%which implies
%$$\mathrm{DQ}^{\ES}_\alpha (\mathbf{X})=\min_{r\in (0,+\infty]} \left\{\mathbb{E}\left[ \left(r\sum_{i=1}^n \left( X_i -  \ES_{\alpha}(X_i)\right)+1\right)_{+}\right],1\right\}.$$
Let $f:[0,\infty)\to [0,\infty),~ r\mapsto  \mathbb{E}[ \left(r\sum_{i=1}^n   X_i'  +1\right)_{+}]$. It is clear that $f(0)=1$.
Moreover, 
$$f(r)\ge r \E\left[\left( X_i'  \right)_+\right]\to \infty~~~~\mbox{as $r\to\infty$}.$$
By Theorem \ref{th:ax-1} (iii),  we have $\mathrm{DQ}^\ES_\alpha(\mathbf X) \le 1$, and hence
$\inf_{r \in (0,\infty)} f (r) \le \alpha<1 $.
 The continuity of $f$ yields 
$\inf_{r\in (0,\infty)} f (r) = \min_{r\in (0,\infty)} f(r), $
and thus   \eqref{eq:es-alter3}  holds.
%If all $X_1,\dots,X_n$ are constants, then $f(r)=1$ for all $r\in (0,\infty)$ and \eqref{eq:es-alter3} holds.
%Next, suppose that at least one of  $X_1,\dots,X_n$ is not a constant. 
%We can compute 
%$$f'(r)=\E\left[\sum_{i=1}^n \left( X_i -  \ES_{\alpha}(X_i)\right) \id_{\left\{r\sum_{i=1}^n \left( X_i -  \ES_{\alpha}(X_i)\right)+1>0\right\}}\right].$$
%As $r \downarrow 0$, we have  $\p(  r\sum_{i=1}^n \left( X_i  -  \ES_{\alpha}(X_i)\right)+1>0) \to 1$ and $$\lim_{r\downarrow 0} f'(r) \to \E\left[\sum_{i=1}^n \left( X_i -  \ES_{\alpha}(X_i)\right) \right]=\sum_{i=1}^n\left( \E[ X_i] -  \ES_{\alpha}(X_i)\right) < 0,$$
%where the last inequality uses   the fact that at least one of $X_1,\dots,X_n$ is not a constant, which is implied by $\p(\sum_{i=1}^n X_i>\sum_{i=1}^n \ES_\alpha(X_i))>0$.
%Moreover, $$\lim_{r\to \infty} f'(r) =  \E\left [\sum_{i=1}^n \left( X_i -  \ES_{\alpha}(X_i)\right) \id_{\{\sum_{i=1}^n \left( X_i -  \ES_{\alpha}(X_i)\right)>0\}}\right]> 0. $$ 
%The above two facts and  the continuity of $f$ yield 
%that 
%$\inf_{r\in (0,\infty)} f (r) = \min_{r\in (0,\infty)} f(r), $
%and thus   \eqref{eq:es-alter3}  holds. 
\end{proof}

%\begin{proof}[Proof of Proposition \ref{prop:convex}] Fixed $\mathbf X \in \mathcal X^n$.  Let  $r_\beta: \Delta_n\rightarrow \R $ be given by  $r_\beta(\mathbf w )= \rho_{\beta}\left(\sum_{i=1}^{n}w_i X_{i}\right) - \sum_{i=1}^{n} \rho_{\alpha}\left(w_iX_{i}\right)$ for $\beta \in I$.  From $\mathrm{[PH]}$ of  $\rho_\alpha$, we have $r_\beta(\mathbf w)= \rho_{\beta}\left(\sum_{i=1}^{n}w_i X_{i}\right) -  \mathbf w^\top\mathbf{x}_{\alpha}^{\rho}.$ Convexity  of $\rho_\beta$  implies convexity of  $\mathbf w\mapsto    r_\beta(\mathbf w )$. Hence, for the portfolio weight $\lambda\mathbf w+(1-\lambda)\mathbf v \in\Delta_n$, DQ based on $\rho$ at level $\alpha \in(0,1)$ is given by  $$\begin{aligned}{\rm DQ}^{\rho}_{\alpha}((\lambda\mathbf w+(1-\lambda)\mathbf v) \odot \mathbf X)&= \frac{1}\alpha    \inf\left\{\beta \in I :  r_\beta(\lambda \mathbf  w+(1-\lambda) \mathbf  v)\leq 0 \right\}\\&\le \frac{1}\alpha    \inf\left\{\beta \in I : \lambda r_\beta( \mathbf w )+(1-\lambda)r_\beta( \mathbf  v )\leq 0 \right\}\\ &\leq \frac{1}\alpha  \max\left\{   \inf\{\beta \in I : r_\beta( \mathbf w )\leq 0\}, \inf\{\beta \in I :r_\beta( \mathbf  v )\leq 0 \}\right\} \\&=\max\left\{ {\rm DQ}^{\rho}_{\alpha}(\mathbf w \odot \mathbf X),{\rm DQ}^{\rho}_{\alpha}(\mathbf v \odot \mathbf X)\right\},\end{aligned}$$ which gives us quasi-convexity.\end{proof}

\section{Additional results for Section \ref{sec:2}}\label{App:A}

In this appendix, we   present an   impossibility result showing a conflicting nature of the three natural properties {\rm[+]}, {\rm[LI]} and {\rm[SI]} for some diversification indices defined via risk measures.
% We first give an example of DR based on a centered   version of risk measures.
% \begin{example}\label{ex:div-index2} For a given risk measure $\phi$, another commonly used  diversification index is defined as (see, e.g., \cite{BDI08})
%      $$\frac{\phi\left(\sum_{i=1}^n X_i\right)-\E[\sum_{i=1}^n X_i]}{\sum_{i=1}^n (\phi(X_i)-\E[X_i])};$$
%      this corresponds to the diversification ratio ${\rm DR}^{\phi-\E}$ where   $\phi-\E$ is called a risk-based capital by \cite{BDI08}.
%   The   index  ${\rm DR}^{\phi-\E}$
%   is location-scale invariant if $\phi$ satisfies $\mathrm{[CA]}$  and  $\mathrm{[PH]}_\gamma$ with $\gamma\in\R$  (e.g., $\phi=\VaR$) as soon as the denominator is not $0$;
%   it does not satisfy [+] in general, unless $\phi \ge \E$ on $\X$.  If $\phi$ is a law-invariant coherent risk measure, then $\phi-\E$  is a generalized deviation measure of \cite{RUZ06}.
%   Note that   ${\rm DR}^{\phi-\E}$   is not $\phi$-determined; see \eqref{eq:tau-based} below. \end{example}
As mentioned  in Section \ref{sec:2},  the most  commonly  used diversification indices  depend on $\mathbf X$ through its values assessed by some risk measure $\phi$.
That is, given a risk measures $\phi$ and a portfolio $\mathbf{X}$, the diversification index can be written as  \begin{align}\label{eq:tau-based} D(\mathbf{X})=R\left(\phi\left(\sum_{i=1}^n X_i\right), \phi(X_i),\dots,\phi(X_n)\right) \mbox{~for some function $R:\R^{n+1}\to \overline{\R}$}.
\end{align} 
We will say that $D$ is \emph{$\phi$-determined} if \eqref{eq:tau-based} holds. 
Often, one may further choose $R$ so that  $D(\mathbf X)$  decreases in $\phi(\sum_{i=1}^n X_i) $
and increases in $\phi ( X_i)$ for each $i\in [n]$, for a proper interpretation of measuring diversification.

%Given an MCP risk measure $\phi$,    there is a profound conflict  between the properties {\rm[+]}, {\rm[LI]} and {\rm[SI]}  for any $\phi$-determined diversification index.  
We show that    a  diversification index  based on an MCP risk measure, such as VaR or ES  satisfying  all three properties [+], [LI] and [SI]   can take at most $3$ different values. In this case, we will say that the diversification index $D$ is \emph{degenerate}. In fact, this result can be extended  to   more general  properties  $\mathrm{[PH]}_\gamma$  and $\mathrm{[CA]}_m$ with $\gamma\in\R$ and $m\in \R$ of the risk measure $\phi$, with definitions  given at the beginning of Section \ref{sec:3}.

 \begin{proposition}\label{thm:div-index}
Fix $n\ge 1$.
Suppose that a risk measure $\phi$ satisfies  {$\mathrm{[PH]}_\gamma$} and  $[\rm CA]_m$ with $\gamma\in\R$ and $m\ne 0$. A   diversification index   $D$
is $\phi$-determined and satisfies {\rm[+]}, {\rm[LI]} and {\rm[SI]}   if and only if  for all $\mathbf{X}\in \X^n$,
    \begin{equation}\label{eq:D}D(\mathbf{X})=C_1\id_{\{d<0\}}+C_2\id_{\{d=0\}}+C_3\id_{\{d>0\}},\end{equation}
where $d=\mathrm{DB}^\phi(\mathbf X) =\sum_{i=1}^n \phi(X_i)-\phi\left(\sum_{i=1}^n X_i\right)$  for some  $C_1,C_2,C_3\in \R_+\cup\{\infty\}$.
\end{proposition}
 We first present a lemma to prepare for the proof of Proposition \ref{thm:div-index}.
\begin{lemma}\label{lem:div-ratio}
A function $R:\R^{n+1} \to \overline \R$ satisfies, for all   $x_0 \in \R$,  $\mathbf{x}=(x_1,\dots,x_n) \in \R^n$, $\mathbf{c}=(c_1,\dots,c_n) \in \R^n$ and $\lambda>0$,
(i) $R\left(x_0+\sum_{i=1}^n c_i,\mathbf{x}+\mathbf{c}\right)=R(x_0,\mathbf{x})$ and (ii) $R(\lambda x_0,\lambda \mathbf{x})=R( x_0,\mathbf{x})$, if and only if there exist $C_1,C_2,C_3\in \overline\R$ such that
   \begin{equation}\label{eq:R}
   R(x_0,\mathbf{x})=C_1\id_{\{r<0\}}+C_2\id_{\{r=0\}}+C_3\id_{\{r>0\}},
   \end{equation}
 where  $r=\sum_{i=1}^n x_i-x_0$, for all $x_0\in \R$ and  $\mathbf{x} \in\R^n$.  
\end{lemma}
\begin{proof} 
First, we show that $R$ in \eqref{eq:R} satisfies (i) and (ii).
Assume  $r<0$. For any  $\mathbf{c} \in \R^n$ and $\lambda>0$, it is clear that  $x_0+\sum_{i=1}^n c_i<\sum_{i=1}^n (x_i+c_i)$ and  $\lambda x_0<\sum_{i=1}^n \lambda x_i$. Therefore, (i) and (ii) are satisfied. The cases of  $r=0$ and $r>0$ follow by the same argument.

Next, we verify the ``only if" part. Given $\mathbf{x}= (x_1,\dots,x_n)$ and $\mathbf{y}=(y_1,\dots,y_n)$ satisfying $\sum_{i=1}^n x_i=\sum_{i=1}^n y_i$,
let $\mathbf{c}=\mathbf{y}-\mathbf{x}$. For any $x_0 \in \R$, we have $\sum_{i=1}^n c_i=\sum_{i=1}^n (y_i-x_i)=0$. Therefore, $$R(x_0,\mathbf{x})=R\left(x_0+\sum_{i=1}^n c_i,\mathbf{x}+\mathbf{c}\right)=R(x_0,\mathbf{y}).$$
Thus, the value of $R(x_0,\mathbf{x})$ only depends on $x_0$ and $\sum_{i=1}^n x_i$. Let $\tilde R: \R^2 \to \overline \R$ be a function such that  $\tilde R\left(x_0,\sum_{i=1}^n x_i\right)=R(x_0,\mathbf{x})$. From the properties of $R$,  $\tilde R$ satisfies
$\tilde R(a+c,b+c)=\tilde R(a,b)$ for any $c \in \R$, and
 $\tilde R(\lambda a,\lambda b)=R(a,b)$ for any $\lambda>0$. Hence, we have
$$\tilde R(a,b)=\tilde R (a-b,0)=\tilde R(1,0) ~\text{ for } ~a>b,$$
$$\tilde R(a,b)=\tilde R (0,b-a)=\tilde R(0,1) ~\text{ for }~ a<b,$$
and
$$\tilde R(a,b)=\tilde R (a-a,b-a)=\tilde R(0,0) ~\text{ for }~ a=b.$$
Let $C_1=\tilde R(1,0)$, $C_2=\tilde R(0,0)$ and $C_3=\tilde R(0,1)$. We have $R(x_0,\mathbf{x})=\tilde R(x_0,\sum_{i=1}^n x_i)$, which has the form in \eqref{eq:R}.
\end{proof}

\begin{proof}[Proof of Proposition \ref{thm:div-index}]  
Let us first prove sufficiency.  By definition, $D$ satisfies [+] and $D$ is $\phi$-determined.  Next, we prove $D$ satisfies [LI] and [SI]. Similarly to Lemma \ref{lem:div-ratio}, we only prove the case $d<0$. It is straightforward  that  $$\phi\left(\sum_{i=1}^n \lambda X_i\right )=\lambda^\gamma \phi\left(\sum_{i=1}^n X_i \right)<\lambda^\gamma \sum_{i=1}^n \phi(X_i)=\sum_{i=1}^n \phi(\lambda X_i),$$
and
$$\phi\left(\sum_{i=1}^n (X_i+c_i)\right)=\phi\left(\sum_{i=1}^n X_i\right)+m\sum_{i=1}^n c_i<\sum_{i=1}^n \left(\phi(X_i)+m c_i\right)=\sum_{i=1}^n \phi(X_i+c_i).$$
Thus, we have $D(\lambda \mathbf{X})=C_1$ and $D( \mathbf{X}+\mathbf{c})=C_1$, which completes the proof of sufficiency. 

Next, we  show the necessity. Define the set $$\mathcal{A}= \left\{\left(\phi\left(\sum_{i=1}^n X_i\right),\phi(X_1),\dots,\phi(X_n)\right): (X_1,\dots,X_n) \in \X^n \right\}.$$
Note that $\phi$ satisfies $\mathrm{[PH]}_\gamma$ with $\gamma \ne 0$ since $\mathrm{[CA]}_m$ for $m\ne 0$ implies $\rho(2)\ne \rho(1)$, which in turn implies $\gamma\ne 0$.
We always write $\mathbf x=(x_1,\dots,x_n)$ and $\mathbf c=(c_1,\dots,c_n)$.
Consider the two operations $(x_0,\mathbf x)\mapsto (x_0+\sum_{i=1}^n c_i,\mathbf x+\mathbf c) $ for some $\mathbf c\in \R^n$
and $(x_0,\mathbf x)\mapsto (\lambda x_0,\lambda \mathbf x)$ for some $\lambda >0$. 
% By [CA] and [PH], if $(x_0,\mathbf x) \in \mathcal A$, then $(x_0+\sum_{i=1}^n c_i,\mathbf x+\mathbf c) \in \mathcal A$ and $(\lambda x_0,\lambda \mathbf x) \in \mathcal A $ for all $\mathbf c \in \R^{n}$ and $\lambda>0$. 
%If $(x_0,\mathbf x) \in \R^{n+1}\setminus \mathcal A$, then $(x_0+\sum_{i=1}^n c_i,\mathbf x+\mathbf c) \in \R^{n+1}\setminus \mathcal A$ and $(\lambda x_0,\lambda \mathbf x)  \in \R^{n+1}\setminus \mathcal A$ for all $\mathbf c \in \R^{n}$ and $\lambda>0$.
%Since the above operations of location and scale change are invertible, we get 
 % that both $\mathcal A$ 
%and  $\R^{n+1}\setminus \mathcal A$ are  closed under these operations.
Let  $r(x_0,\mathbf x)=\sum_{i=1}^n x_i-x_0$.
By using $\mathrm{[CA]}_m$ and $\mathrm{[PH]}_\gamma$ of $\phi$, we have that (see also the proof of Lemma \ref{lem:div-ratio}) 
  the regions $\mathcal A_+:=\{(x_0,\mathbf x):r(x_0,\mathbf x)>0\}$,  
$\mathcal A_0:=\{(x_0,\mathbf x):r(x_0,\mathbf x)=0\}$
and
$\mathcal A_{-}:=\{(x_0,\mathbf x):r(x_0,\mathbf x)<0\}$
are closed under the above two operations,
and each of them is connected via the above two operations.
 Therefore, $\mathcal A$ is the union of some of $\mathcal A_+$, $\mathcal A_0$ and $\mathcal A_-$.

We define a function $R: \R^{n+1}\to \overline{\R}$.
For $(x_0,\mathbf x)\in \mathcal{A}$, 
 let  $R(x_0,\mathbf x)=D(X_1,\dots,X_n)$, 
where  $(X_1,\dots,X_n)$ is any random vector such that $x_0=\phi(\sum_{i=1}^n X_i)$ and  $\mathbf{x} =(\phi(X_1),\dots,\phi(X_n))$. 
The choice of $(X_1,\dots,X_n)$ is irrelevant since $D$ is $\phi$-determined. 
 For $(x_0,\mathbf x)\in \R^{n+1} \setminus \mathcal{A}$, let $R(x_0,\mathbf x)=0$.
We  will verify that  $R$ satisfies conditions (i) and (ii) in Lemma \ref{lem:div-ratio}.

For $(x_0,\mathbf x)\in \mathcal{A}$, there exists $\mathbf{X}=(X_1,\dots,X_n)\in \X^n$ such that $x_0=\phi(\sum_{i=1}^n X_i)$ and $\mathbf{x} =(\phi(X_1),\dots,\phi(X_n))$.
 For any $\mathbf{c}\in \R^n$, using  $\mathrm{[CA]}_m$ with $m\ne 0$ of $\phi$ and [LI] of $D$,  we obtain  
 $$\begin{aligned}
R\left(x_0+\sum_{i=1}^n c_i,\mathbf{x}+\mathbf{c}\right)&=R\left(\phi\left(\sum_{i=1}^n X_i\right)+\sum_{i=1}^n c_i,\phi(X_1)+c_1, \dots, \phi(X_n)+c_n\right)\\
& = R\left(\phi\left(\sum_{i=1}^n \left(X_i+\frac{c_i}{m}\right)\right),\phi\left(X_1+\frac{c_1}{m}\right), \dots, \phi\left(X_n+\frac{c_n}{m}\right)\right)\\
&=D\left(\mathbf{X}+\mathbf{c}/m\right)=D(\mathbf{X})=R\left(x_0,\mathbf x\right).
\end{aligned}$$
Using  $\mathrm{[PH]}_\gamma$ with $\gamma\ne 0$ of $\phi$ and [SI] of $D$, for any $\lambda>0$, we obtain
$$\begin{aligned}
R(\lambda x_0,\lambda\mathbf{x})& =R\left(\lambda\phi\left(\sum_{i=1}^n X_i\right),\lambda\phi(X_1), \dots, \lambda\phi(X_n)\right)
\\ &=
R\left(\phi\left(\sum_{i=1}^n \lambda^{1/\gamma} X_i\right),\phi(\lambda^{1/\gamma} X_1), \dots, \phi(\lambda^{1/\gamma} X_n)\right)\\
&=D(\lambda^{1/\gamma}\mathbf{X})=D(\mathbf{X})=R\left( x_0,\mathbf x\right).
\end{aligned}$$
Hence,  $R$ satisfies (i) and (ii) in Lemma \ref{lem:div-ratio} on $ \mathcal A$. 
By definition,   $R$ satisfies (i) and (ii) also on $\R^{n+1}\setminus \mathcal A$.
Since $\mathcal A$ and $\R^{n+1}\setminus \mathcal A$ are both closed under the two operations,  we know that $R$ satisfies (i) and (ii) on $\R^{n+1}$. 

Using Lemma \ref{lem:div-ratio},  we have $R$ has the representation \eqref{eq:R}, which gives 
 $$
  D(\mathbf{X})=C_1\id_{\{d<0\}}+C_2\id_{\{d=0\}}+C_3\id_{\{d>0\}}$$
with $d=\sum_{i=1}^n \phi(X_i)-\phi\left(\sum_{i=1}^n X_i\right)$  and  $C_1,C_2,C_3\in \overline \R$ for all $\mathbf{X}\in X^n$.
As $D$ satisfying [+], we have $C_1,C_2,C_3\in \R_+\cup \{\infty\}$.  
\end{proof}

 \section{Additional  results and proofs for Section 
\ref{sec:axiom}}
\label{App:C00}

\subsection{Existence of worse-than-duplicate portfolios}
\label{app:worse}
We discuss the existence of worse-than-duplicate portfolios. First, note that  if a vector $\mathbf X^{\rm wd}=(X_1^{\rm wd},\dots,X_n^{\rm wd})$ is worse than a duplicate portfolio $\mathbf X^{\rm du}=(X, \dots, X)$ under a given MCP risk measure $\phi$, then we have
$$\phi\left(\sum_{i=1}^n X_i^{\rm wd}\right)\ge \phi(nX)=n\phi(X)>\sum_{i=1}^n \phi(X_i^{\rm wd})$$
and thus $\phi$ violates subadditivity with $\mathbf X^{\rm wd}$. Therefore, a necessary condition for the existence of a vector that is worse than a duplicate under a MCP risk measure $\phi$ is that $\phi$ violates subadditivity. 

We further provide a necessary and sufficient condition for the existence or  non-existence of duplicate portfolios.
\begin{lemma}\label{lem:1}
For a monotone risk measure $\phi$,  there exists a worse-than-duplicate  portfolio   if and only if  
there exist 
 $X,X_1,\dots,X_n\in \X$ with $X_1+\dots+X_n = n X$ 
 such that $\phi (X)> \phi (X_i)$ for $i\in [n]$. 
\end{lemma}
\begin{proof}
This follows directly by monotonicity.
\end{proof}

A risk measure $\phi:\X\to\R$ is \emph{scale-continuous} if the mapping $\lambda\mapsto \phi(\lambda X)$ on $(0,1)$ is continuous for every $X$. This condition is very weak; for instance it is weaker than continuity  on any $L^p$-space $\X$.

\begin{proposition}\label{pro:no-worse}
 For a  monotone risk measure $\phi$  scale-continuous  on $\X=L^\infty$,   there exists no worse-than-duplicate  portfolio   if and only if  $\phi$ is quasi-convex.
\end{proposition} 
\begin{proof}
If $\phi$ is quasi-convex, then for any $X_1,\dots,X_n$, 
\begin{equation}\label{eq:1}
 \phi \left (\frac {X_1}n+\dots+\frac{X_n}n\right ) \le \max\{ \phi(X_1),\dots,\phi(X_n)\}.
\end{equation}
By Lemma \ref{lem:1}, there exists no worse-than-duplicate portfolio. 
Conversely, if  exists no worse-than-duplicate portfolio,
 then \eqref{eq:1} holds for all $X_1,\dots,X_n$. It suffices to verify that this implies quasi-convexity of $\phi$.
That is, we need to show that for $\lambda \in (0,1)$ and $X_1,X_2$, 
 \begin{equation}\label{eq:3} 
\phi(\lambda X_1 + (1-\lambda) X_2) \le \max\{ \phi(X_1),\phi(X_2)\}.
\end{equation} 
First, suppose that $\lambda =p/n^q$ where $p,q\in \N$.
Repeatedly applying \eqref{eq:1} $q$ times, we get, for all $Y_1,\dots,Y_{m}$ where $m=n^q$,
 \begin{equation}\label{eq:4} 
 \phi \left (\sum_{i=1}^{m} \frac {Y_i}{m} \right ) \le \max\{ \phi(Y_1),\dots,\phi(Y_{m})\}.
\end{equation} 
Letting $Y_i=X_1$ for $i\le p$ and $Y_i=X_2$ for $i>p$ in \eqref{eq:4}, we get  \eqref{eq:3}.
Next, consider a general $\lambda \in (0,1)$. 
Let $X_1'=  \lambda X_1/t$ 
and $X_2'= (1-\lambda)  X_2/(1-t) $,
where  $t=p /n^q \in (0,1)$   for some $p,q\in \N$.
Using \eqref{eq:4}, we get 
 \begin{equation}\label{eq:5} 
\phi(\lambda X_1 + (1-\lambda) X_2) = \phi(t X'_1 + (1-t) X'_2) \le \max\{ \phi(X'_1),\phi(X'_2)\}.
\end{equation}  
Sending $t\to \lambda$ and using continuity we obtain the desired result.
\end{proof}

\subsection{Examples and proofs related to portfolio convexity}\label{app:convex} 
 
In the first example, we show that  
convexity or quasi-convexity of $\mathbf X \mapsto D(\mathbf X)$  should not hold for a diversification index $D$. 
\begin{example}[Quasi-convexity on $\X^n$ is not desirable]\label{ex:non-convex}

Let $(X,Y)\in \X^2$ represent any diversified portfolio (e.g., with iid normal components), and assume that $Z:=(X+Y)/2$ is not a constant.
Since
the portfolio $(Z,Z)$ relies only on one asset and has no diversification benefit, for a good diversification index $D$ 
we naturally want $D(Z,Z)$ to be larger than both $D(X,Y)$ and $D(Y,X)$; recall that $D(Z,Z)=1$ in the setting of Theorem \ref{prop:div_ben} (iii). %This rules out quasi-convexity of $D$, and
This argument shows that it is unnatural to require  $D $ to be convex or quasi-convex on $\X^2$; the case of $\X^n$ is similar. 
%Let $X$ and $Y$ be iid standard normal random variables. 
%Both $(X+Y, X-Y)$  and $(X-Y, X+Y)$ have iid normal components and   should be intuitively considered as having some diversification benefit, but their linear combination $(X,X)$  relies only on one asset $X$ and has no diversification benefit.  
%Hence, for a diversification index $D$, we want $D(X,X)>\max\{D(X+Y,X-Y),D(X-Y,X+Y)\}$;
%it would be unnatural to require  $D $ to be convex or quasi-convex on $\X^n$.
Indeed, if a real-valued $D$ satisfies  $\mathrm{[SI]}$ and convexity on $\X^n$, then it is a constant; this is shown in the proposition below.
\end{example}
%Despite that quasi-convexity of $D$ is unnatural on $\X^n$,  quasi-convexity may hold for $\mathbf w \mapsto  D(\mathbf w \odot \mathbf X)$ for each given $\mathbf X$.  We term this property \emph{portfolio quasi-convexity}.}\com{I want to move the blue part to the appendix}

%The following result shows that there is a conflict between convexity and   [SI]. This result is mentioned in Example \ref{ex:non-convex}.
\begin{proposition}\label{pro:non-convex}
A mapping $D: \mathcal X^n \to \R$ satisfies {$\mathrm{[SI]}$} and convexity if and only if $D(\mathbf X)=c$ for all $\mathbf X \in \mathcal X$ and some constant $c\in \R$.
\end{proposition}

\begin{proof}

If $D$ is a constant for all $\mathbf X \in \mathcal X^n$, it is clear that $D$ satisfies  [SI] and convexity. Next we will show the ``only if'' part.  Let  $d_0=D(\mathbf 0) \in \R$.
\begin{enumerate}[(i)]
\item  If $d_0\ge D(\mathbf X)$ for all $\mathbf X \in \mathcal X^n$ and there exists $\mathbf X_0$ such that $D(\mathbf X_0)<d_0$, then 
$$D\left(\frac{1}{2}\mathbf X_0+\frac{1}{2}(-\mathbf X_0) \right)=D(\mathbf 0)>\frac{1}{2} D(\mathbf X_0)+\frac{1}{2} D(-\mathbf X_0) ,$$
which contradicts the convexity of $D$.
\item If there exists $\mathbf X_0$ such that $d_0<D(\mathbf X_0)$, then, by   [SI] of $D$,
$$D\left(\frac12  \mathbf 0+\frac12 \mathbf X_0\right)=D(\mathbf X_0)>\frac12 D(\mathbf 0)+\frac12 D(\mathbf X_0),$$ which  contradicts the convexity of $D$. 
\end{enumerate} 
By (i) and (ii), we can conclude that $D$ only takes the value $d_0$.
\end{proof}

From the proof of Proposition \ref{pro:non-convex},  we see that the conflict between convexity and [SI]  holds for real-valued mappings on any closed convex cone, not necessarily on $\X^n$.

In the second example, we see that convexity of $\mathbf w\mapsto D(\mathbf w\odot \mathbf X)$ is not desirable either for a good diversification index. 
 \begin{example}[Convexity in $\mathbf w$ is not desirable]
\label{ex:DRSDnon-convex} 
Let 
  $Z$ be standard normal and $\epsilon>0$ be a small constant.
Consider a portfolio vector $\mathbf X=((1-\epsilon) Z,-\epsilon Z)$.  Let $\mathbf w=(1,0)$
and $\mathbf v= (\epsilon,1-\epsilon)$.
Note that  $\mathbf w \odot \mathbf X=(1-\epsilon)  ( Z,0)$
and $\mathbf v \odot \mathbf X =  (\epsilon-\epsilon^2)   ( Z,- Z) $.
The portfolio $\mathbf w \odot \mathbf X$ is not diversified since it has only one non-zero component, and the portfolio  $\mathbf v \odot \mathbf X$ is perfectly hedged since the sum of its components is $0$. Hence, for a good diversification index  $D$, it should hold that $D (\mathbf w \odot \mathbf X)=1$
and $D (\mathbf v \odot \mathbf X)=0$;   Theorem \ref{prop:div_ben} confirms this. On the other hand, 
the portfolio $$\left(\frac{1}{2}\mathbf w+ \frac{1}{2}\mathbf v\right) \odot \mathbf X = \frac 12 \left( (1-\epsilon  ^2) Z, -(\epsilon-\epsilon^2)  Z \right)$$
is not well diversified  since its second component is very small compared to its first component. 
Intuitively, for $\epsilon\approx 0$, we expect $D( (\mathbf w/2+ \mathbf v/2) \odot \mathbf X ) \approx 1>  D (\mathbf w \odot \mathbf X)/2 +  D (\mathbf v \odot \mathbf X)/2$. This shows that  $\mathbf w\mapsto   D(\mathbf w\odot \mathbf X)$ is not convex. One can verify that this is indeed true if $D$ is DQ or DR based on commonly used risk measures such as SD, VaR $(\alpha<1/2)$ and ES.

%Recall from Proposition \ref{pro:scx-consistent} that  $\mathrm{DR}^{\mathrm{SD}} = \mathrm{DQ}^\rho_\beta $ for any $\beta \in (0,\infty)$ where $ \rho=(\mathrm{SD}/\alpha)_{\alpha \in(0,\infty)}$, and hence $\mathrm{DR}^{\mathrm{SD}}(\mathbf w \odot \mathbf X)$ is quasi-convex in $\mathbf w$.
%By taking $\mathbf X\sim \mathrm{N}(\mathbf 0,\Sigma)$ where $\Sigma=((1,0)^\top,(0,100)^\top)$, we can see that $\mathrm{DR}^\mathrm{SD}(\lambda (1,0)\odot \mathbf X+(1-\lambda) (0,1)\odot \mathbf X) = \sqrt{\lambda^2+100(1-\lambda)^2}/(10-9\lambda)$ for $\lambda\in (0,1)$, which is not convex in $\lambda$, implying that $\mathbf w\mapsto  \mathrm{DR}^{\mathrm{SD}}(\mathbf w\odot \mathbf X)$ is not convex.
\end{example}

% \begin{proposition}\label{prop:convex} Let  $\rho=\left(\rho_{\alpha}\right)_{\alpha \in I}$ be a class of convex risk measures  satisfying  $\mathrm{[PH]}$ and decreasing in $\alpha$. For every   $\mathbf{X}\in \X^n$ and $\alpha\in I$, $\mathrm{DQ}^{\rho}_\alpha$ satisfies $\mathrm{[PC]}.$ 
% \end{proposition} 
% \begin{proposition}\label{prop:non_deg}
% Assume that $n \ge 3$ and $\rho$ is a class of non-linear coherent risk measures. If  there exists $X \in \X$ such that $\rho_\beta(X)$ is a strictly decreasing function of $\beta$, then $\{{\rm DQ}^\rho_\alpha (\mathbf X): \mathbf X \in \X^n\}=[0,1]$.
% \end{proposition}

\begin{proof}[Proof of Proposition \ref{prop:convex}]
 Since the proof of Theorem \ref{thm:quasi-conv} solely relies on convexity and positive homogeneity of $\rho_\alpha$ to show [PC], it is clear that $\mathrm{DQ}^{\rho}_\alpha$ satisfies [PC].

The subadditivity of $\rho_\alpha$ implies that $\mathrm {DQ}^{\rho}_\alpha$ takes value in $[0,1]$. Consequently,  $\{{\rm DQ}^\rho_\alpha (\mathbf X): \mathbf X \in \X^n\}\subseteq [0,1]$. We only need to show  $[0,1]\subseteq\{{\rm DQ}^\rho_\alpha (\mathbf X): \mathbf X \in \X^n\}$.

Since $\rho_\alpha$ is non-linear and sub-linear, there exists $Y$ such that $\rho_\alpha(Y)+\rho_\alpha(-Y)>0$ following the argument of Remark \ref{rem:nonlinear}.
Consider $\mathbf X^\theta =(X,\theta Y, -\theta Y, 0, \dots, 0)$ with  $\theta\ge 0$. We have
$$\mathrm{DQ}^\rho_\alpha(\mathbf X^\theta)=\frac{1}{\alpha}\inf\{\beta\in I: \rho_\beta(X)\le \rho_\alpha(X)+\theta \rho_\alpha(Y)+\theta\rho_\alpha(-Y)\}.$$
It is clear that $\mathrm{DQ}^\rho_\alpha(\mathbf X^{0})=1$,  and  there  exists $\tilde \theta$ such that $\mathrm{DQ}^\rho_\alpha(\mathbf X^{\tilde \theta})=0$   with $\rho_\alpha(X)+\tilde\theta \rho_\alpha(Y)+\tilde \theta\rho_\alpha(-Y)>\rho_0(X)$. Since $\beta\mapsto \rho_\beta(X)$ is strictly decreasing, its generalized inverse is continuous and we  can conclude    
$\{\mathrm{DQ}^{\rho}_\alpha(\mathbf X^\beta): \beta \in [0, \tilde \theta]\}=[0,1]$. Hence, $[0,1]\subseteq\{{\rm DQ}^\rho_\alpha (\mathbf X): \mathbf X \in \X^n\}$. 
\end{proof}

\subsection{Constructing DQ from a single risk measure}
\label{sec:43}

In this section, we discuss how to construct DQ from only a single risk measure $\phi$.
For commonly used risk measures like VaR and ES, a natural family $\rho$ with $\rho_\alpha=\phi$ exists. If in some applications one needs to use a different $\phi$ which does not belong to an existing family, we will need to construct a family of risk measures for $\phi$. 

First, suppose that $\phi$ is MCP.
A simple approach is to take  $\rho_\alpha=(1-\alpha)\esssup+\alpha\phi$ for $\alpha \in (0,1)$. Clearly, $\rho_1=\phi$.
As $\phi$ is  MCP, we have $\phi(X)\le \phi(\esssup(X))=\esssup(X)$ for all $X \in L^\infty$. Hence,   $\rho$ is a decreasing class of MCP risk measures. Therefore, $\mathrm{DQ}^\rho_1$ satisfies the six axioms in Theorem \ref{th:ax-1}. 
Moreover, by checking the definition, this DQ has an explicit formula
$$
\mathrm{DQ}^{\rho}_1(\mathbf X)=\left(\frac{ \esssup\left(\sum_{i=1}^n X_i\right)-\sum_{i=1}^n \phi(X_i)}{\esssup\left(\sum_{i=1}^n X_i\right)-\phi\left(\sum_{i=1}^n X_i\right)}\right)_+.
$$
If $\sum_{i=1}^n X_i\le \sum_{i=1}^n \phi(X_i)$, we have $\esssup(\sum_{i=1}^n X_i)\le \sum_{i=1}^n \phi(X_i)$ and $\mathrm{DQ}^\rho_1(\mathbf X)=0$; this is also reflected by Theorem \ref{prop:div_ben} when $\esssup(\sum_{i=1}^n X_i)>\phi(\sum_{i=1}^n X_i)$.  
% For any given $X\in L^\infty$, $\alpha \mapsto \rho_\alpha(X)$ is a linear function. That is,   $\rho$ is left continuous and non-flat from the left with $\rho_0=\esssup$. By  Theorem \ref{prop:div_ben} (i), we have $\mathrm{DQ}^\rho_\alpha(\mathbf X)=0$ if and only if there is no insolvency risk with pooled individual capital, i.e. $\sum_{i=1}^n X_i\le \sum_{i=1}^n \phi(X_i)$.

For any arbitrary risk measure $\phi$, we can always define the decreasing family $\{\phi_+/\alpha:\alpha \in I\}$ for constructing DQ; here $\phi_+$ is the positive part of $\phi$.  
This approach leads to DQs that are also DRs.

% For a single non-negative risk measure $\phi$ and its corresponding DR, we   can construct a class   $\rho =(\rho_{\alpha})_{\alpha \in I}$  such that $\mathrm{DQ}^{\rho}_{\alpha}=\mathrm{DR}^{\phi}$ which shows that DR is a subclass of DQ. 
\begin{proposition}\label{prop:equiv}
    For a given $\phi: \mathcal{X} \rightarrow \mathbb{R}_{+}$, let $\rho=(\phi / \alpha)_{\alpha \in(0, \infty)}$. For $\alpha \in(0, \infty)$, we have $\mathrm{DQ}_\alpha^\rho=\mathrm{DR}^\phi$. The same holds if $\rho=(b \mathbb{E}+c \phi / \alpha)_{\alpha \in(0, \infty)}$ for some $b \in \mathbb{R}$ and $c>0$ and $\mathcal{X}=L^1$.
\end{proposition} 
\begin{proof}
First, we compute $\alpha^*$ by the definition of ${\rm DQ}^{\rho}_{\alpha}$. For any $\mathbf{X} \in (L^1)^n$,
\begin{align*}
\alpha^*&=\inf\left\{\beta \in (0,\infty): b\E\left[\sum_{i=1}^n X_i\right] +\frac{c}{\beta}\phi \left(\sum_{i=1}^n X_i\right) \le b \sum_{i=1}^n\E\left[ X_i\right] +\sum_{i=1}^n \frac{c}{\alpha}\phi(X_i) \right\}\\
&= \inf\left\{\beta \in  (0,\infty) : \frac{\phi \left(\sum_{i=1}^n X_i\right)}{\beta} \le \frac{\sum_{i=1}^n\phi(X_i) }{\alpha} \right\}.
\end{align*}
If  $\phi(\sum_{i=1}^n X_i)=0$ and $\sum_{i=1}^n \phi(X_i)=0$, then
$\alpha^*=0$.  If $\phi(\sum_{i=1}^n X_i)>0$ and $\sum_{i=1}^n \phi(X_i)=0$, then
$\alpha^*=\infty$ because the set on which the infimum is taken is empty.  If  $\phi(\sum_{i=1}^n X_i)>0$ and $\sum_{i=1}^n \phi(X_i)>0$, then $\alpha^*=\alpha\phi(\sum_{i=1}^n X_i)/\sum_{i=1}^n \phi(X_i)$.
Hence, ${\rm DQ}^{\rho}_{\alpha}(\mathbf{X})={\rm DR}^{\phi}(\mathbf{X})$ holds for all $\mathbf{X}\in (L^1)^n$. 
By the same argument, for $\rho=(\phi/\alpha)_{\alpha \in (0,\infty)}$,  we get $\mathrm{DQ}^\rho_\alpha(\mathbf{X})=\mathrm{DR}^{\phi}(\mathbf{X})$ for all $\mathbf{X}\in \X^n$.
\end{proof}
As a result of Proposition \ref{prop:equiv},
DQ built on the family $\rho$ of the mean-SD functions given by $\rho_\alpha(X)= \E[X]+ \mathrm{SD}(X)/\alpha $ is precisely $\mathrm{DR}^{\mathrm{SD}}$.

\subsection{Axiomatization of DQ using preferences}\label{app:Preference} 
 %We  identify the axioms satisfied by MCP risk measures with a  relation of preference.     

 The axioms $\mathrm{[R]_\phi}$, $\mathrm{[N]}_\phi$ and $\mathrm{[C]_\phi }$ are formulated based on an exogenously specified risk measure $\phi$, usually by financial regulation. 
 This choice can also be endogenized in the context of internal decision making. 
 % We present in Section \ref{app:Preference} an axiomatization of DQ without specifying a risk measure via a few additional axioms on the preference of a decision maker. 
 In this section, we provide an axiomatization of DQ as in Theorem \ref{th:ax-1} without specifying a risk measure $\phi$. 
We first define the preference of a decision maker over risks. 
A preference relation  $\succeq$ is defined by  a non-trivial total preorder\footnote{A preorder is a binary relation  on $\mathcal{X}$, which is reflexive and transitive. A binary relation $\succeq$ is reflexive if $X \succeq X$ for all $X \in \mathcal{X}$, and transitive if $X \succeq Y$ and $Y \succeq Z$ imply $X \succeq Z$. A non-trivial total preorder is a preorder that in addition is complete, that is, $X \succeq Y$ or $Y \succeq X$ for all $X, Y \in \mathcal{X}$, and there exist at least two alternatives $X$, $Y$  such that $X$ is preferred over $Y$ strictly.\label{foot3}}   on   $\X$. As usual,   $\succ $ and $ \simeq $  correspond to the antisymmetric and equivalence relations, respectively. 
%$ \succ:=\{\succeq \text{and}  \}$ and $\simeq$ stand for  the (strictly) antisymmetric and equivalence relation, respectively.
On the preference $\succeq$ of risk, the relation  $X\succeq Y$  means the agent prefers $X$ to $Y$ for any $X,Y\in\X$. We will use the following  axioms.  
\begin{itemize}
    \item[{[A1]}]$X\le Y~\Longrightarrow~X\succeq Y$.
    \item[{[A2]}] $X\succeq Y~\Longrightarrow~X+c\succeq Y+c$ for any $c \in \R$.
    \item[{[A3]}] $X\succeq Y~\Longrightarrow~ \lambda X\succeq \lambda Y$ for any $\lambda >0$.

    \item[{[A4]}]  For any $X \in \X$, there exists $c \in \R$ such that $X \simeq c$.
\end{itemize}
%   A preference is usually considered associated with a  risk measure.
The four axioms are rather standard and we only briefly explain them. 
  The  axiom [A1]  means that  the agent  always prefers a smaller loss. The axioms [A2] and [A3] mean that  if the agent   prefers one  random loss over another, then this is preserved under any strictly increasing   linear transformations.  The  axiom [A4] implies that  any random losses  can be equally favourable  as a constant loss which is commonly referred to as a   certainty equivalence.

A numerical representation of a preference  $\succeq$  is a mapping $\phi: \X \rightarrow\R$, such that
 $X \succeq Y \Longleftrightarrow \phi(X)\le \phi(Y)$   
for all $X,Y\in \X$.  
 %It is well known  that a preference  $\succeq$  admits a risk measure representation if   it is reflexive, transitive and separable. 
 In other words, $\succeq$ is the preference of an agent favouring less risk evaluated via  $\phi$. There is a simple relationship between preferences satisfying [A1]-[A4] and MCP risk measures. 
\begin{lemma}\label{pro:pre}
A preference  satisfies $\mathrm{[A1]}$--$\mathrm{[A4]}$ if and only if it can be represented by  an  MCP  risk measure  $\phi$.
\end{lemma}
 %We first show the ``if" part. Let $\succeq$  be represented by    an  MCP risk measure $\phi$ through  $X \succeq Y \Longleftrightarrow \phi(X)\le \phi(Y)$.% and $\X \succeq Y$ if and only if  $\phi(X)\ge \phi(Y)$. \begin{enumerate}[(a)]  \item If $X\le Y$, we have $\phi(X)\le \phi(Y)$ by [M] of $\phi$. Hence,   $X\succeq Y$ and  [A1] holds.  \item If $X\succeq Y$, we have $\phi(X) \le \phi(Y)$. By [CA], we have $\phi(X+c)\le \phi(Y+c)$ for any $c \in \R$. Hence, $X+c\succeq Y+c$ and [A2] holds.  \item If $X\succeq Y$, we have $\phi(X) \le \phi(Y)$. By [PH], we have $\phi(\lambda X)\le \phi(\lambda Y)$ for any $\lambda>0$. Hence, $\lambda X\succeq \lambda Y$ and [A3] holds.  \item If $X \succeq Y$ and $Y \succeq Z$, we have $\phi(X)\le \phi(Y)$ and $\phi(Y) \le \phi(Z)$. It follows that  $\phi(X)\le \phi(Z)$, which implies $X \succeq Z$. Hence, [A4] holds   \item As $\phi$ is an MCP risk measure, we have $\phi(c)=c$ for any $c \in \R$. It is clear that $X \sim \phi(X)$ for any $X \in \X$ as $\phi(X)=\phi(\phi(X))$. If   there exists $c_0\neq\phi(X)$ such that $X \sim c_0$, then we have $\phi(X)=\phi(c_0)=c_0$, which is a contradiction.   Hence, [A5] holds. \end{enumerate}
 \begin{proof}
 The ``if" statement is straightforward to check, and we will show the ``only if" statement. The preference $\succeq$ can be represented by a risk measure  $\phi$ through  $X \succeq Y \Longleftrightarrow \phi(X)\le \phi(Y)$ for all $X,Y\in \X$ since $\succeq$ is separable by [A1] and [A4]; see \cite{D54} and \cite{DK13}.
 If $\phi(0)= \phi(1)$, then by using [A1]--[A3], the preference $\succeq$ is trivial, contradicting our assumption on $\succeq$. Hence, using [A1], $\phi(0)< \phi(1)$, we can further let $\phi(0)=0$ and $\phi(1)=1$. It is then straightforward to verify that  $\phi$ is MCP from [A1]-[A3].
 %we first assume there exist $a,b\in \R$ such that $a \neq b$ and $b \simeq a$.  By [A2], we have $a+c \simeq b+c$ for all $c\in \R$.  Hence,  for all $a,b \in R$, we have  $a\simeq b$. By [A4], for any $X, Y \in \X$, there exist $c_X, c_Y \in \R$ such that $X \simeq c_X\simeq c_Y \simeq Y$. Hence, we can simply choose $\phi(X)=c$ for some constant $c \in \R$.  Now, we assume that if $b \simeq a$, then $a=b$.
%Next, we show that $\phi$ is an MCP risk measure.
%\begin{enumerate}[(a)] \item If $X\le Y$, we have $X\succeq Y$ by [A1]. Hence, $\phi(X) \le \phi(Y)$, which implies $\phi$ satisfies [M]. \item By the definition of $\phi$, we have $X \simeq \phi(X)$. By [A2], we have $X+c \simeq \phi(X)+c$. Hence, $\phi(X+c)=\phi(X)+c$, which implies $\phi$ satisfies [C]. \item As $X \simeq \phi(X)$,  we have $\lambda X \simeq \lambda \phi(X)$ by [A3]. Hence, $\phi(\lambda X)=\lambda\phi(X)$, which implies $\phi$ satisfies [PH].  \end{enumerate}}
\end{proof}
%  In Lemma \ref{pro:pre}, we show that a preference satisfies [A1]--[A4] if and only if it can be represented by an MCP risk measure. Such a result states the one-to-one correspondence between risk orders and risk measures.  The idea of expressing the numerical representation of a total preorder by means of a family of risk measures was used in other works; see for example \cite{BDS12} and \cite{DK13}. 

 %We   present a lemma to prepare for the proof of Theorem \ref{th:ax-1}.

Similarly to Section \ref{sec:axiom}, but with the preference $\succeq$ replacing the risk measure $\phi$,
we denote by $
\mathbf X \simeqm   \mathbf Y$  if 
$X_i \simeq  Y_i$ for each $i\in [n]$,
by $
\mathbf X  \succeqm  \mathbf Y$ if   
$X_i  \succeq Y_i$ for each $i\in [n]$, and by $\mathbf X   \succm\mathbf Y$ if  $X_i  \succ Y_i$ for each $i\in [n]$.
With this new formulation  and everything else unchanged, the axioms of rationality,    normalization and continuity are now denoted by $\mathrm{[R]}_\succeq$, $\mathrm{[N]}_\succeq$ and $[\mathrm{C}]_\succeq$.  
% In the following result, we characterize DQ by Axioms $\mathrm{[+]}$, $\mathrm{[LI]}$,  $\mathrm{[SI]}$, $\mathrm{[R]}_\succeq$,  $\mathrm{[N]}_\succeq$ and $\mathrm{[C]_\succeq}$ where  $\succeq$  satisfies the axioms $\mathrm{[A1]}$--$\mathrm{[A4]}$. 

 \begin{proposition}\label{prop:preference}
 A diversification index $D:\X^n\to \overline \R$
satisfies $\mathrm{[+]}$, $\mathrm{[LI]}$, $\mathrm{[SI]}$, $\mathrm{[R]_\succeq}$, $\mathrm{[N]}_\succeq$ and $\mathrm{[C]_\succeq}$ for some preference $\succeq$ satisfying $\mathrm{[A1]}$--$\mathrm{[A4]}$
   if and only if $D$ is $\mathrm{DQ}_\alpha^\rho$  for some  decreasing  families  $\rho $ of  MCP risk measures. %such that $\rho$ is non-flat from left at $(\alpha, X)$ for any $X \in \X^*$
   Moreover, in both directions of the above equivalence, it can be required that $\rho_\alpha$ represents $\succeq$.
\end{proposition}
\begin{proof}
The proof  follows from Theorem \ref{th:ax-1}   by noting that  Lemma \ref{pro:pre} allows us to convert between a preference $\succeq$ satisfying [A1]-[A4] and an MCP risk measure $\phi$.
\end{proof}

Theorem \ref{thm:quasi-conv}
also admits a formulation via preferences  similar to Proposition \ref{prop:preference}.

% \begin{remark}
% The order $\sum_{i=1}^n X_i\le \sum_{i=1}^n Y_i$ in $\mathrm{[R]}_\succeq$ cannot be replaced by $\sum_{i=1}^n X_i\succeq \sum_{i=1}^n Y_i$ as $\sum_{i=1}^n X_i\succeq \sum_{i=1}^n Y_i$ implies the value of $D(\mathbf X)$ is determined by $\phi(\sum_{i=1}^n X_i)-\sum_{i=1}^n \phi(X_i)$ where $\phi$ is an MCP risk measure representing $\succeq$.
% \end{remark}

\subsection{Uniqueness of the risk measure family representing a DQ}
 Proposition \ref{prop:uni_rho} below shows that the choice of the risk measure family is unique up to strictly increasing transformation of the parameter
  if the ordering structure on portfolio diversification is specified by a given ordering relation $\succeq$ on $\X^n$ that can be numerically represented by a DQ.

\begin{proposition}\label{prop:uni_rho}

Let $n \ge 3$, $I=[0,1]$, $\alpha\in I$
and 
 $\phi$ is a positively homogeneous  risk measure with  $ \phi( Y)+\phi(- Y) >0$ for some $Y \in \X$.
  Suppose that a weak order $\succeq$ is numerically represented by both $\mathrm{DQ}^\rho_\alpha$
and 
$\mathrm{DQ}^\tau_\alpha$ such that $\{\mathrm{DQ}^\rho_\alpha(\mathbf X): \mathbf X\in \X^n \}= \{\mathrm{DQ}^\tau_\alpha(\mathbf X): \mathbf X\in \X^n \}=[0,1]$,
where   $\rho=(\rho_\beta)_{\beta \in I}$  and  $\tau=(\tau_\beta)_{\beta \in I}$
are continuous decreasing families of   risk measures satisfying $\phi=\rho_\alpha=\tau_\alpha$.  
Then, there exists a
 strictly increasing $f:(0,\alpha)\to (0,\alpha)$ such that  
$\tau_{\beta}=\rho_{f(\beta)}$ for all $\beta \in (0,\alpha)$.
\end{proposition}
\begin{proof}
Since $\succeq$ is represented by both $\mathrm{DQ}^\rho_\alpha$
and 
$\mathrm{DQ}^\tau_\alpha$, there exists a strictly increasing function $g:[0,1]\to [0,1]$ such that 
$\mathrm{DQ}^\rho_\alpha = g(\mathrm{DQ}^\tau_\alpha)$.
Let $f(\beta)=g(\beta/\alpha)$ for $\beta \in (0,\alpha)$.

Assume that there exists $\beta^* \in (0,\alpha)$ such that $\tau_{\beta^*}(X)> \rho_{f(\beta^*)}(X)$. 
By positive homogeneity 
of $\phi$ and $\phi(Y)+\phi(-Y)>0$, there exists $\epsilon>0$ such that 
$\tau_{\beta^*}(X)>\phi(X)+\phi(\epsilon Y)+\phi(-\epsilon Y)> \rho_{f(\beta^*)}(X)$.
 Let $\mathbf X=(X, \epsilon Y, -\epsilon Y,0,\dots,0)$.
Since  $\beta \mapsto \rho_\beta(X)$ and $\beta \mapsto  \tau_\beta(X)$ are continuous, we have 
 $$g(\mathrm{DQ}^{\tau}_\alpha( \mathbf X))=f\left(\inf\left\{\beta \in I: \tau_\beta(X) \le \phi(X)+\phi(\epsilon Y)+\phi(-\epsilon Y)\right\}\right)>f(\beta^*)$$
and    
 \begin{align*}
 \mathrm{DQ}^{\rho}_\alpha (\mathbf X) 
&=\inf\left\{\beta \in I: \rho_\beta(X) \le \phi(X)+\phi(\epsilon Y)+\phi(-\epsilon Y)\right\} \le f\left(\beta^*\right),
 \end{align*}
 which contradicts $\mathrm{DQ}^\rho_\alpha(\mathbf X) = g(\mathrm{DQ}^\tau_\alpha(\mathbf X))$.
\end{proof}

\section{Additional  results and proofs   for Section \ref{sec:3}}
\label{App:C0}

 In this section, we present  additional results, proofs, and discussions  supplementing   Sections  \ref{sec:44} and \ref{sec:45}.

\subsection{Worst-case and best-case dependence for DQ (Section \ref{sec:44})}
\label{App:dep}	
	We assume that two random vectors $\mathbf X$ and $\mathbf Y$ have the same marginal distributions, and we study the effect of the dependence structure.  
	We will assume that  a tuple of distributions  $\mathbf F=( F_1,\dots,F_n)$ is given and each component has a finite mean. 
	Let $$\mathcal{Y}_\mathbf{F}=\left\{(X_1,
	\dots, X_n): X_i \sim F_i \text{ for each } i=1,
	\dots,n\right\}.$$
	For $\mathbf X, \mathbf Y\in \mathcal Y_{\mathbf F}$, we say that $\mathbf X$ is smaller than $\mathbf Y$ in sum-convex order, denoted by $\mathbf X \le_{\rm{scx}} \mathbf Y$, if  $\sum_{i=1}^n X_i \ge_{\rm{SSD}} \sum_{i=1}^n Y_i$; see
	\cite{CR06}. We refer to \cite{SS07} for a general treatment of multivariate stochastic orders.  
	With arbitrary dependence structures,
	the best-case value and worst-case value of ${\rm DQ}^{\rho}_{\alpha}$ are given by
	$$\inf_{\mathbf X \in \mathcal Y_\mathbf{F}}{\rm DQ}^{\rho}_{\alpha}(\mathbf X )~~~\text{ and }~~~ \sup_{\mathbf X \in \mathcal Y_\mathbf{F}}{\rm DQ}^{\rho}_{\alpha}(\mathbf X ).$$
	For some mapping on $\X^n$,  finding the best-case and worst-case values and structures over $\mathcal{Y}_\mathbf{F}$ is   known as a problem of risk aggregation under dependence uncertainty; see \cite{BJW14} and \cite{EWW15}.
	
	If $\rho=(\rho_{\alpha})_{\alpha \in I}$ is a class of  {SSD-consistent}  risk measures such as ES, then, by Proposition \ref{pro:scx-consistent}, ${\rm DQ}^{\rho}_{\alpha}$ is consistent with the sum-convex order on $ \mathcal Y_\mathbf{F}$. This leads to the following observations on the corresponding dependence structures.
	\begin{enumerate}[(i)]
		\item It is well-known (e.g., \cite{R13})  that
		the $\le_{\rm scx} $-largest element of $\mathcal Y_{\mathbf F}$ is comonotonic, and thus a comonotonic random vector has the largest ${\rm DQ}^{\rho}_{\alpha}$ in this case. Note that such $\rho$ does not include VaR. Indeed, as we have seen from Proposition \ref{th:var-01n}, $\mathrm{DQ}^\VaR_\alpha(\mathbf X)=1$ for comonotonic $\mathbf X$ under mild conditions, which is not equal to its largest value $n$.
		
		\item
		In  case $n=2$, the $\le_{\rm scx} $-smallest element of $\mathcal Y_{\mathbf F}$ is  counter-comonotonic,
		and thus  a comonotonic random vector has the smallest ${\rm DQ}^{\rho}_{\alpha}$.
		\item For $n \ge 3$,  the $\le_{\rm scx} $-smallest elements of $\mathcal Y_{\mathbf F}$ are generally hard to obtain.  If  each pair $(X_i,X_j)$ is counter-monotonic for $i\ne j$, then $\mathbf X$
		is  a $\le_{\rm scx} $-smallest element of $\mathcal Y_{\mathbf F}$. Pairwise counter-monotonicity puts very strong restrictions on the marginal distributions. For instance, it rules out all continuous marginal distributions; see \cite{PW15}.
		
		\item  If a joint mix, i.e., a random vector with a constant component-wise sum, exists in $\mathcal Y_{\mathbf F}$, then any joint mix is  a $\le_{\rm scx} $-smallest element of $\mathcal Y_{\mathbf F}$ by Jensen's inequality. See \cite{PW15} and \cite{WW16} for results on the existence of joint mixes.
		In case a joint mix does not exist, the $\le_{\rm scx} $-smallest elements are obtained by
		\cite{BJW14} and
		\cite{JHW16} under some conditions on the marginal distributions such as monotonic densities.
	\end{enumerate}
	In optimization problems over dependence structures  (see e.g., \cite{R13} and \cite{EWW15}), the above observations yield guidelines on where to look for the optimizing structures.

\subsection{Proofs and related discussions on RI and RC (Section \ref{sec:45})}

Here we present the proof of Proposition \ref{prop:RI} and an additional result (Proposition \ref{prop:RI2}) on the properties RI and RC.

\begin{proof}[Proof of Proposition \ref{prop:RI}]
\begin{enumerate}[(i)] 
\item 
%Since $\rho_\alpha$ satisfies $\mathrm{[CA]}_m$ with a common $m\in \R$, for all $\alpha \in I$, we have $$\rho_\alpha(X+c)=\rho(X)+mc=\rho_\alpha(X)+\rho_\alpha(c)-\rho_\alpha(0)=\rho_\alpha(X)+\rho_\alpha(c).$$ 
For any $n \in \mathbb N$, $\mathbf X\in (L^p)^n$ and $c\in \R$,  by $\mathrm{[CA]}_m$ of $(\rho_\alpha)_{\alpha \in I}$,
$$\begin{aligned}{\rm DQ}_\alpha^\rho( \mathbf X,c )
&=\frac{1}{\alpha}\inf\left\{\beta \in I :  \rho_{\beta} \left(\sum_{i=1}^n X_i+ c\right) \le \sum_{i=1}^n \rho_{\alpha}(X_i) +\rho_{\alpha}(c)\right\}\\
&=\frac{1}{\alpha}\inf\left\{\beta \in I :  \rho_{\beta} \left(\sum_{i=1}^n X_i\right)+mc \le \sum_{i=1}^n \rho_{\alpha}(X_i) + mc  \right\}={\rm DQ}_\alpha^\rho( \mathbf X ),
\end{aligned}$$  
and hence  ${\rm DQ}^{\rho}_\alpha$ satisfies {\rm[RI]}.

\item For any $n \in \mathbb N$ and $\mathbf X\in (L^p)^n$,  by $\mathrm{[PH]}$ of $(\rho_\alpha)_{\alpha \in I}$,
$$\begin{aligned}{\rm DQ}_\alpha^\rho( \mathbf X,\mathbf X )
&=\frac{1}{\alpha}\inf\left\{\beta \in I :  \rho_{\beta} \left(2\sum_{i=1}^n X_i\right) \le 2 \sum_{i=1}^n \rho_{\alpha}(X_i) \right\}
%\\
%&=\frac{1}{\alpha}\inf\left\{\beta \in I :  \rho_{\beta} \left(\sum_{i=1}^n X_i\right)\le \sum_{i=1}^n \rho_{\alpha}(X_i) \right\}
={\rm DQ}_\alpha^\rho( \mathbf X ),
\end{aligned}$$  
and hence  ${\rm DQ}^{\rho}_\alpha$ satisfies {\rm[RC]}. \qedhere
\end{enumerate}
\end{proof}

\begin{proposition}\label{prop:RI2}
 Let   $\phi:L^p\to \R $ be a continuous and law-invariant risk measure.  
 %$\phi$ is a risk measure.
 \begin{enumerate}[(i)]
   %  \item If $\rho_\alpha$ satisfies $\mathrm{[CA]_m}$  with $m\in\R$ and $\rho_{\alpha}(0)=0$ then ${\rm DQ}^{\rho}_\alpha$ satisfies {\rm[RI]}.
     \item Suppose that $\mathrm{DR}^\phi$ is not degenerate for some  input dimension. Then    ${\rm DR}^{\phi}$ satisfies {\rm[RI]} and {\rm [+]}  if and only if  $\phi$ satisfies  $\mathrm{[CA]}_0$, $[\pm]$ and $\phi(0)=0$.
    %  \item If $  \rho_\alpha  $ satisfies  $\mathrm{[PH]}$, then  ${\rm DQ}^{\rho}_\alpha$ satisfies $\mathrm{[RC]}$.
   \item     If $\phi$ satisfies $\mathrm{[PH]}$, then  ${\rm DR}^{\phi}$ satisfies $\mathrm{[RC]}$.
 \end{enumerate}
\end{proposition}
\begin{proof}
\begin{enumerate}[(i)] 
\item 
We first show the ``if" part.  If $\phi$ satisfies ${\rm [CA]}_0$ and $\phi(0)=0$, then $\phi(c)=\phi(0)=0$ for all $c\in \R$.
For any $n \in \mathbb N$, $\mathbf X\in (L^p)^n$ and $c\in \R$, 
$$\begin{aligned}{\rm DR}^{\phi}( \mathbf X,c )
=\frac{ \phi \left(\sum_{i=1}^n X_i+c\right)}{ \sum_{i=1}^n \phi( X_i)+\phi(c)}
=\frac{ \phi \left(\sum_{i=1}^n X_i\right)}{ \sum_{i=1}^n \phi( X_i)}=\mathrm{DR}^{\phi}(\mathbf X)\end{aligned}.$$
 Thus, ${\rm DR}^{\phi}$ satisfies {\rm[RI]}. 
 
For the ``only if" part,  we first assume $\phi(0)\neq0$. 
 Since ${\rm DR}^{\phi}$ satisfies {\rm[RI]}, for all $n \in \mathbb N$, $c\in\R$ and $\mathbf X= {\bf 0}\in \R^n$, we have  $$\begin{aligned}{\rm DR}^{\phi}( \mathbf X,c )=\frac{ \phi \left( c\right)}{n \phi( 0)+\phi(c)}={\rm DR}^{\phi}( \mathbf X )=\frac{ \phi \left( 0\right)}{ n\phi( 0)}=\frac{1}{n}.\end{aligned}$$ The above equality  means that $\phi(c)=n\phi(0)/(n-1)$ holds for any $n \in \mathbb N$ and $c\in\R$, and thus  we have $\phi(0)=0$, which violates the assumption $\phi(0)\neq0$. Hence,  $\phi(0)=0$.

If there exists $c_1\in \R$ such that $\phi(c_1)\ne 0$, then by [RI] and 
$\phi(0)=0$, we have 
$$
 {\rm DR}^{\phi}( c_1,0,0,\dots,0,c )=\frac{\phi(c_1+c)}{\phi(c_1)+\phi(c)}= {\rm DR}^{\phi}( c_1,0,0,\dots,0 )=\frac{\phi(c_1)}{\phi(c_1)} = 1 ,
$$ 
 and thus  $\phi(c_1+c)=\phi(c_1)+\phi(c)$ as long as $\phi(c_1) $ or $\phi(c)$ is not zero. If both of $\phi(c_1) $ and $\phi(c)$ are $0$, then $\phi(c_1+c)=0$.
To sum up, $\phi$ is additive on $\R$. Since $\phi$ is also continuous on $\R$, 
we know that $\phi$ is linear, that is, $\phi(c)=\beta c$ for some $\beta \in \R$. 

Suppose that 
there exists $X$ such that $\phi(X)\ne 0$; otherwise there is nothing to show. 
Using [RI] and $\phi(0)=0$, we have, for $c\in \R$, $${\rm DR}^{\phi}(   X,0,0,\dots,0,c )= \frac{\phi(X+c)}{\phi(X)+\phi(c)}=\mathrm{DR}^\phi(X,0,\dots,0)=1,$$
which implies $\phi(X+c)=\phi(X)+\phi(c)=\phi(X)+\beta c$.
%Therefore, $\phi$ satisfies $\mathrm{[CA]}_\beta$. 

Using the fact that  $\mathrm{DR}^\phi$ is not degenerate for some dimension $n$, there exists $\mathbf X=(X_1,\dots,X_n)$ such that $\mathrm{DR}^\phi(\mathbf X) \in \R\setminus \{0,1\}$.
Note that $\phi(\sum_{i=1}^n X_i) \neq 0$ and $\sum_{i=1}^n \phi(X_i)\neq 0$. Hence, 
$$
\mathrm{DR}^\phi(\mathbf X,1) = \frac{\phi(\sum_{i=1}^n X_i +1 ) }{\sum_{i=1}^n \phi( X_i) +\phi(1)  }
= \frac{\phi(\sum_{i=1}^n X_i )+\beta  }{\sum_{i=1}^n \phi( X_i)  +\beta  } =\mathrm{DR}^\phi(\mathbf X)= \frac{\phi(\sum_{i=1}^n X_i )  }{\sum_{i=1}^n \phi( X_i)    } . 
$$ 
This implies $\beta =0$, $\phi(c)=0$ for all $c\in \R$ and $\phi(X+c)=\phi(X)$ for all $X\in L^p$ such that $\phi(X) \neq 0$.
For any $X \in L^p$ such that $\phi(X)=0$ and $c\in\R$, we have
$$\mathrm{DR}^\phi(X,0,\dots,0,c)=\frac{\phi(X+c)}{\phi(X)+\phi(c)}=\frac{\phi(X+c)}{\phi(X)}=\frac{\phi(X)}{\phi(X)}=\mathrm{DR}^\phi(X,0,\dots,0),$$
which implies $\phi(X+c)=0=\phi(X)$.
Therefore, $\phi$ satisfies $\mathrm{[CA]}_0$.

%The proof for $[\pm]$ follows from the   same arguments in the proof (Step 4) of Proposition \ref{prop:DRgood}. 

Finally, we show  $\phi$ is either non-negative or non-positive by considering the following three cases.
\begin{enumerate}[(a)]
\item Assume that there exists $X \in L^{p}$ such that $\phi(X)+\phi(-X)>0$.
If there exists $Y \in L^{p}$ such that $\phi(Y)<0$, then by continuity of $\phi$ and $\phi(0)=0$, there exists $m>0$ such that $0<-\phi(mY)<\phi(X)+\phi(-X)$. We have 
$$\mathrm{DR}^{\phi}(mY,X,-X,0,\dots,0)=\frac{\phi(m Y)}{\phi(m Y)+ \left(\phi(X)+\phi(-X)\right)}< 0,$$
which   contradicts the fact that $\mathrm{DR}^{\phi}$ is non-negative.
Hence, $\phi(Y)\ge 0$ for all $Y \in L^{\infty}$.
\item By the same argument, if there exists $X \in L^{p}$ such that $\phi(X)+\phi(-X)<0$, then $\phi(Y)\le 0$ for all $Y \in L^{\infty}$.
\item Assume $\phi(X)+\phi(-X)=0$ for all $X \in L^{\infty}$.
Suppose that there exists $Y \in L^{\infty}$ such that $\phi(Y) < 0$.
Using Lemma 1 of \cite{WW20} again, there exist $Z,Z'\in L^\infty$ satisfying $Z \laweq Z' $ and $Z-Z'\laweq Y-\E[Y]$.  For $\mathbf Z=(Z,-Z',0,\dots,0)$, using the law invariance of $\phi$, we have
    $$\mathrm{DR}^{\phi}(\mathbf Z)=\frac{\phi\left(Z-Z'\right)}{\phi(Z)+\phi(-Z')} =\frac{\phi\left(Y-\E[Y]\right)}{\phi(Z)+\phi(-Z')}=\frac{\phi\left(Y\right)}{\phi(Z)+\phi(-Z)}=\frac{\phi(Y)}{0}= - \infty,$$
    which  contradicts $\mathrm{DR}^{\phi}(\mathbf Z) \ge 0$. Hence, $\phi(X) \ge 0$ for all $X\in L^\infty$. Together with $\phi(X)+\phi(-X)=0$, we get $\phi(X)=0$. To extend this to $L^p$, we simply use continuity. For  $X \in L^p$, let $Y_M=(X\wedge M)\vee (-M)$. Hence, $Y_M \in L^{\infty}$ and $Y_M  \stackrel{L^p}{\longrightarrow} X $ as $M \to \infty$. As a result, we have $\phi(X)=\lim_{M\to \infty }\phi(Y_M)=0$.
\end{enumerate}
In conclusion, we have $\phi(Y)\ge 0$ or $\phi(Y)\le 0$ for all $X \in L^{p}$.
Case (c) is not possible because it contradicts that $\mathrm{DR}^{\phi}$ is not degenerate. 
Cases (a) and (b) are possible, corresponding to, for instance, (a) $\phi=\mathrm{SD}$; (b) $\phi=-\mathrm{SD}$.

\item   If $\phi$ satisfies ${\rm [PH]}$, then 
for any $n \in \mathbb N$ and $\mathbf X\in (L^p)^n$, 
$$\begin{aligned}{\rm DR}^{\phi}( \mathbf X, \mathbf X )
=\frac{ \phi \left(2\sum_{i=1}^n X_i\right)}{2 \sum_{i=1}^n \phi( X_i)}
=\frac{ \phi \left(\sum_{i=1}^n X_i\right)}{ \sum_{i=1}^n \phi( X_i)}=\mathrm{DR}^{\phi}(\mathbf X)\end{aligned}.$$
 Hence, ${\rm DR}^{\phi}$ satisfies {\rm[RC]}. \qedhere
\end{enumerate}
\end{proof}

In Proposition \ref{prop:RI2}, we show that  if [RI] is assumed, then the only option for DR is to use a non-negative $\phi$ (we can use $-\phi$ if $\phi$ is non-positive) such as $\var$ or $\mathrm{SD}$.  By Proposition \ref{prop:equiv}, all such DRs belong to the class of DQs.  
%The result in Proposition \ref{prop:RI2} has a similar implication to Proposition \ref{prop:DRgood} with [RI] replacing [LI] and [SI]. 
%If [RI] is considered as desirable, then DQ becomes useful compared to DR as it offers more choices, and in particular, it works for any classes $\rho$ of monetary risk measures (\cite{FS16}) with $\rho_\alpha(0)=0$ including VaR and ES. Both   DQ and DR are  replication consistent for  risk measures satisfying $\mathrm{[PH]}$.

\section{Additional results and proofs for Section \ref{VaR-ES}}\label{App:D}

In this section, we present  the proof for Proposition \ref{th:var-01n} and an additional numerical result to complement those in Section \ref{sec:normal-t}.
\begin{proof}[Proof of Proposition \ref{th:var-01n}]
This statement on ES follows from Proposition \ref{prop:convex}; for the one on VaR, see Theorem 1 (i) of \cite{HLW22}. 
\end{proof}
  
%One  omitted empirical result in Section  \ref{sec:normal-t} is presented as follows. 

\begin{figure}[htb!]
\caption{$D(\mathbf Y')/D(\mathbf Y)$ based on VaR  and ES for $\alpha\in (0,0.1]$ with fixed  $n=10$} \label{fig:ratio} \centering\includegraphics[width=15cm]{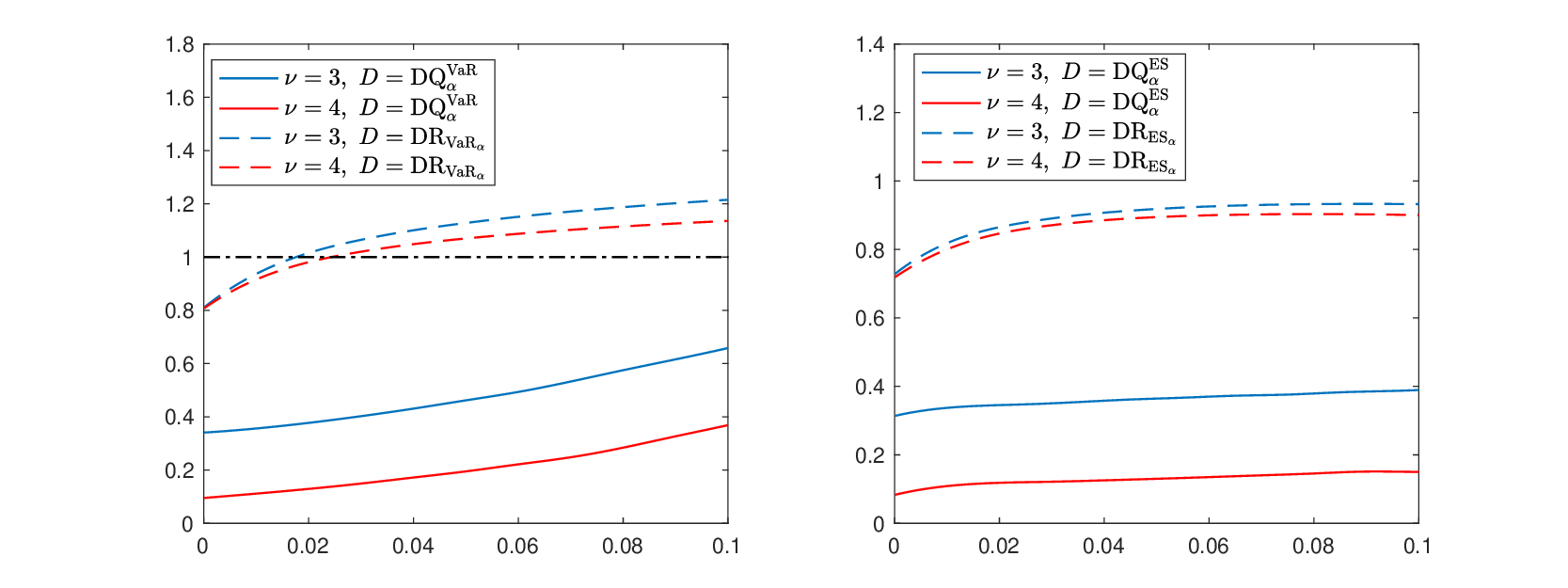} \end{figure}
We   look at the models $\mathbf{Y}'$ and $\mathbf{Y}$ in the setting of Tables \ref{tab:Ind1} and \ref{tab:Ind2}. In Figure \ref{fig:ratio}, we  observe that the values of $D(\mathbf Y')/D(\mathbf Y)$ for $D={\rm DQ^{\VaR}_\alpha}$ or ${\rm DQ^{\ES}_\alpha}$  are always smaller than $1$ for  $\alpha\in(0,0.1]$, while the values of $D(\mathbf Y')/D(\mathbf Y)$  for $D={\rm DR^{{\VaR}_\alpha}}$     are only   smaller than $1$ when  $\alpha$ is relatively small.
We always observe that, if 
the desired relation $D(\mathbf Y')/D(\mathbf Y)<1$
holds for $D=\mathrm{DR}^{\VaR_\alpha}$
or 
$\mathrm{DR}^{\ES_\alpha}$
then it holds for $D=\mathrm{DQ}^\VaR_\alpha$
or 
$\mathrm{DQ}^\ES_\alpha$, but the converse does not hold.
This means that if the iid model is preferred to the common shock model by DR, then it is also preferred by DQ, but in many situations, it is only preferred by DQ not by DR. 
Similarly to Tables \ref{tab:Ind1} and \ref{tab:Ind2},   the iid normal model shows a  stronger diversification  according to
DQ, and this is not the case for DR.

\section{Proofs for Section \ref{sec:opt}}\label{App:E}

%In this section, we first  show one property of $\mathrm{DQ}_{\alpha}^{\rho}(\mathbf w \odot\mathbf{X})$ and then give the proofs of results in Section \ref{sec:opt}. 

\begin{proof}[Proof of Proposition \ref{thm:opt}]
For the case of $\mathrm{DQ}^{\VaR}_{\alpha}(\mathbf X)$,  \eqref{eq:var-alter} in Theorem \ref{th:var}  gives  that to minimize $\mathrm{DQ}^{\VaR}_{\alpha}(\mathbf X)$ is equivalent to minimize 
\begin{align*}
 \p\left(\mathbf w ^\top  \mathbf X  > \mathbf w^\top \mathbf x^{\VaR}_{\alpha}\right)= \p\left(\mathbf w ^\top \left( \mathbf X   - \mathbf x^{\VaR}_{\alpha}\right)>0 \right)~~~~ \mbox{over $\mathbf w\in \Delta_n$}.
 \end{align*}

Next, we discuss the case of $\mathrm {DQ}^{\ES}_\alpha(\mathbf X)$. Let $f(\mathbf v)=\E[(\mathbf v^\top (\mathbf X-\mathbf x^{\ES}_\alpha)+1)_+]$ for $\mathbf v\in \R_+^n$.
It is clear that $f$ is convex. 
%For any $a \in (0,1)$, we have
%\begin{align*}
%f(a\mathbf v_1+(1-a)\mathbf v_2)&=\E[((a\mathbf v_1+(1-a)\mathbf v_2)^\top (\mathbf X-\mathbf x^{\ES}_\alpha)+1)_+]\\
%&\le a\E[(\mathbf v_1^\top (\mathbf X-\mathbf x^{\ES}_\alpha)+1)_+]+(1-a)\E[(\mathbf v_2^\top (\mathbf X-\mathbf x^{\ES}_\alpha)+1)_+]\\
%&= af(\mathbf v_1)+(1-a)f(\mathbf v_2).
%\end{align*}
%Hence, $f(\mathbf v)$ is convex.
Furthermore, for any $i\in [n]$, we have, for almost every $\mathbf v\in \R^n_+$,
  \begin{align*}
  \frac{\partial f}{\partial v_i}(\mathbf v)
  &=\E\left[(X_i-\ES_\alpha(X_i))\id_{\{\mathbf v^\top (\mathbf X-\mathbf x^{\ES}_\alpha)+1>0\}}\right]\\
  &=\E\left[(X_i-\ES_\alpha(X_i))\id_{\left\{\{\mathbf v^\top (\mathbf X-\mathbf x^{\ES}_\alpha)+1>0 \}\cap \{X_i-\ES_\alpha(X_i)>0\}\right\}}\right]\\
  &\quad +\E\left[(X_i-\ES_\alpha(X_i))\id_{\left\{\{\mathbf v^\top (\mathbf X-\mathbf x^{\ES}_\alpha)+1>0 \}\cap \{X_i-\ES_\alpha(X_i)<0\}\right\}}\right].
  \end{align*}
%Let $I_1=\E[(X_i-\ES_\alpha(X_i))\id_{\left\{\{(\mathbf v^\top (\mathbf X-\mathbf x^{\ES}_\alpha)+1)>0 \}\cap \{X_i-\ES_\alpha(X_i)>0\}\right\}}]$ and $I_2=\E[(X_i-\ES_\alpha(X_i))\id_{\left\{\{(\mathbf v^\top (\mathbf X-\mathbf x^{\ES}_\alpha)+1)>0 \}\cap \{X_i-\ES_\alpha(X_i)<0\}\right\}}]$.
The set $\{(\mathbf v^\top \mathbf X-\mathbf x^{\ES}_\alpha)+1>0 \}\cap \{X_i-\ES_\alpha(X_i)>0\}$ increases in $v_i$ and the set $\{(\mathbf v^\top \mathbf X-\mathbf x^{\ES}_\alpha)+1>0 \}\cap \{X_i-\ES_\alpha(X_i)<0\}$ decreases in  $v_i$. Hence, $ v_i\mapsto \partial f/\partial v_i (\mathbf v)$ is increasing.
Furthermore,  $ \partial f/\partial v_i(\mathbf v)\to \E[(X_i-\ES_\alpha(X_i))\id_{\{X_i-\ES_\alpha(X_i)>0\}}]>0$ as $v_i \to \infty$. 
Also, 
$\partial f /\partial v_i(\mathbf v)\to \E[X_i-\ES_\alpha(X_i)]<0$ as $\mathbf v\downarrow \mathbf 0$ component-wise.
Hence, there  exists a minimizer $\mathbf v^*$ of the problem $\min_{\mathbf v \in \R_+^n\setminus\{\mathbf 0\}} \E[(\mathbf v^\top (\mathbf X-\mathbf x^{\ES}_\alpha)+1)_+]$.

Let $A=\{\mathbf v \in \R_+^n\setminus\{\mathbf 0\}: \p(\mathbf v(\mathbf X-\mathbf x_\alpha^\ES)>0)>0\}$ and $B=\{\mathbf v \in \R_+^n\setminus\{\mathbf 0\}: \p(\mathbf v(\mathbf X-\mathbf x_\alpha^\ES)>0)=0\}$. If $B$ is empty, it is clear that $\min_{\mathbf w\in \Delta_n} \mathrm{DQ}^{\ES}_\alpha(\mathbf w \odot \mathbf X)=\min_{\mathbf v \in \R_+^n\setminus\{\mathbf 0\}} \E[(\mathbf v^\top (\mathbf X-\mathbf x^{\ES}_\alpha)+1)_+]$ by Theorem \ref{th:var}.

 If $B$ is not empty, assume $\mathbf v^* \in A$.
For any $\mathbf v_A \in A$, $\mathbf v_B \in B$ and $k>0$, we have
$$\E\left[\left((\mathbf v_A+k\mathbf v_B)^\top(\mathbf X-\mathbf x^{\ES}_\alpha)+1\right)_+\right]\le \E\left[\left(\mathbf v_A^\top(\mathbf X-\mathbf x^{\ES}_\alpha)+1\right)_+\right].$$
This implies
$f(\mathbf v^*+k\mathbf v_B)= f(\mathbf v^*)$
for all $k>0$, which  contradicts $\partial f/\partial v_i (\mathbf v)>0$ as $v_i \to \infty$. Hence, we have $\mathbf v^* \in B$.
For $\mathbf w^*=\mathbf v^*/\Vert \mathbf v^*\Vert$, we have $\p((\mathbf w^*)^\top (\mathbf X-\mathbf x^{\ES}_\alpha)>0)=0$ and $\mathrm{DQ}^{\ES}_\alpha (\mathbf w^* \odot \mathbf X)=0$ by Theorem \ref{th:var}, which means that $\mathbf w^*$ is the minimizer of the problem $\min_{\mathbf w\in \Delta_n} \mathrm{DQ}^{\ES}_\alpha(\mathbf w \odot \mathbf X)$.
\end{proof}

\section{Additional empirical results for Section \ref{sec:emp}}\label{App:F}

In this section, we present  some omitted empirical results to complement those in Sections \ref{per_DQ} and  \ref{opt_dp}.  
\subsection{DR based on SD and variance}
\begin{figure}[htb!]
 \caption{DRs  based on  SD (left)  and var (right)}\label{fig:diff2}
\centering
 \includegraphics[width=15cm]{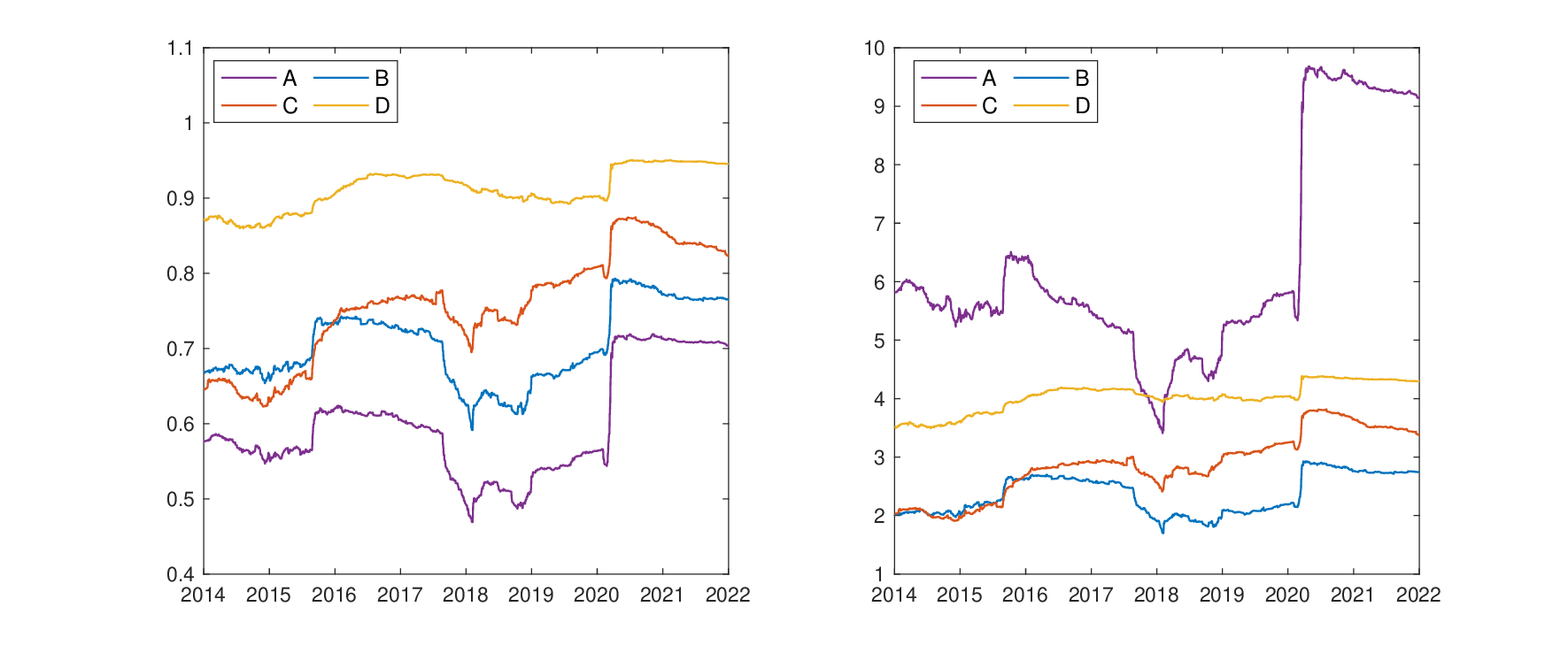}
\end{figure}

In Section \ref{per_DQ},   the values of DQs based on VaR and ES are reported under different portfolio compositions of stocks during the period from 2014 to 2022.  Using the same stock compositions in (A)-(D), we calculate the values of DRs based on SD and var (recall that they are also DQs), to see how they perform. The results are reported in   Figure \ref{fig:diff2}. 

We can see that the same intuitive order  (A)$\le$(B)$\le$(C)$ \le $(D) as in Figure \ref{fig:diff} in Section \ref{per_DQ}  holds for $\mathrm {DR^{SD}}$, showing some consistency between DQs based on VaR and ES and $\mathrm {DR^{SD}}$. The values of $\mathrm {DR^{SD}}$  are between $0$ and $1$. On the other hand,  the values of $\mathrm {DR^{var}}$ are all larger than $1$, and portfolio
(A) of 20 stocks has the weakest diversification effect according to  $\mathrm {DR^{var}}$ among the four compositions. This is not in line with our intuition, but is to  be expected since variance has a different scaling effect than SD, and more correlated stocks lead to a larger value of  $\mathrm{DR^{var}}$ in general. For example, $\mathrm{DR^{var}}$ equals  $1$ even for an iid normal model of arbitrarily large dimension (which is often considered as quite well-diversified), and $\mathrm{DR^{var}}$ equals  $n$ if the portfolio has one single asset. These observations show that   $\mathrm{DR^{var}}$ is difficult to interpret if it is used to measure diversification across dimensions.

%We present  some omitted empirical results to complement the empirical results 

\subsection{Portfolio performance using other datasets}
\label{EC:72}
In Section \ref{opt_dp},  we used the period from January 3, 2012,  to December 31, 2021, to build up the portfolios.
Next, we consider two different datasets from Section \ref{opt_dp}, first using the period 2002-2011 and second using 20 instead of 40 stocks in 2012-2021, to see how the results vary. 
%For the first comparison, we  choose the top  two   stocks from each sector in footnote \ref{foot3}  to build a portfolio.   The results including another two portfolios built by ${\rm DR}^{\mathrm {VaR_\alpha}}$ and ${\rm DR}^{\mathrm {ES_\alpha}}$  are reported in Figure \ref{sec_20} and Table \ref{tab_sector_20}, and they show similar observations to the those in Section \ref{opt_dp}.  % 

\begin{figure}[htb!]
\caption{Wealth processes for portfolios,  40 stocks, Jan 2004 - Dec 2011}\label{sec_40_02}
\centering
           \includegraphics[width=15cm]{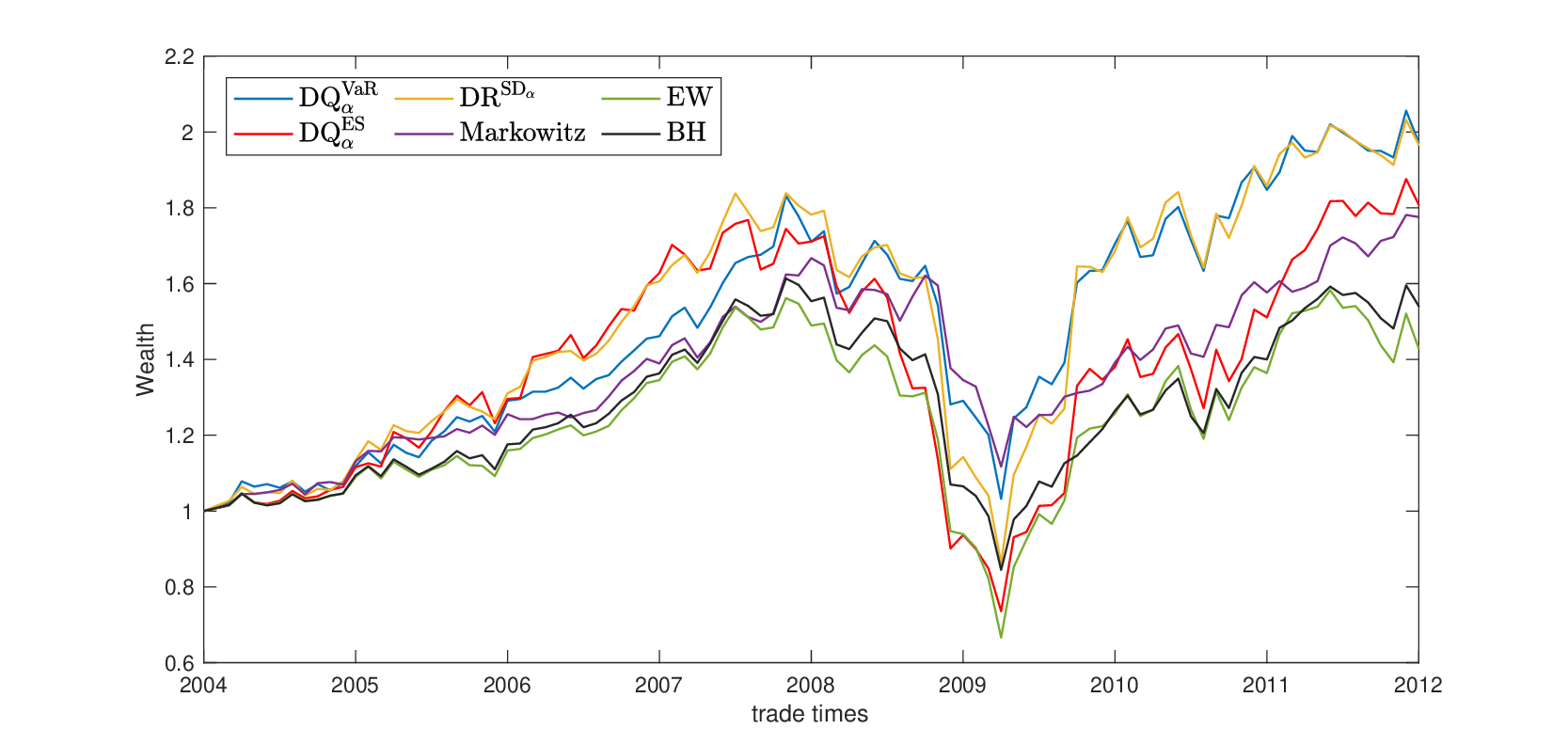}

\end{figure}
\begin{figure}[htb!]
\caption{Cumulative portfolio weights,  40 stocks, Jan 2004 - Dec 2011}\label{sec_40_02_cu}
\centering
           \includegraphics[width=15cm]{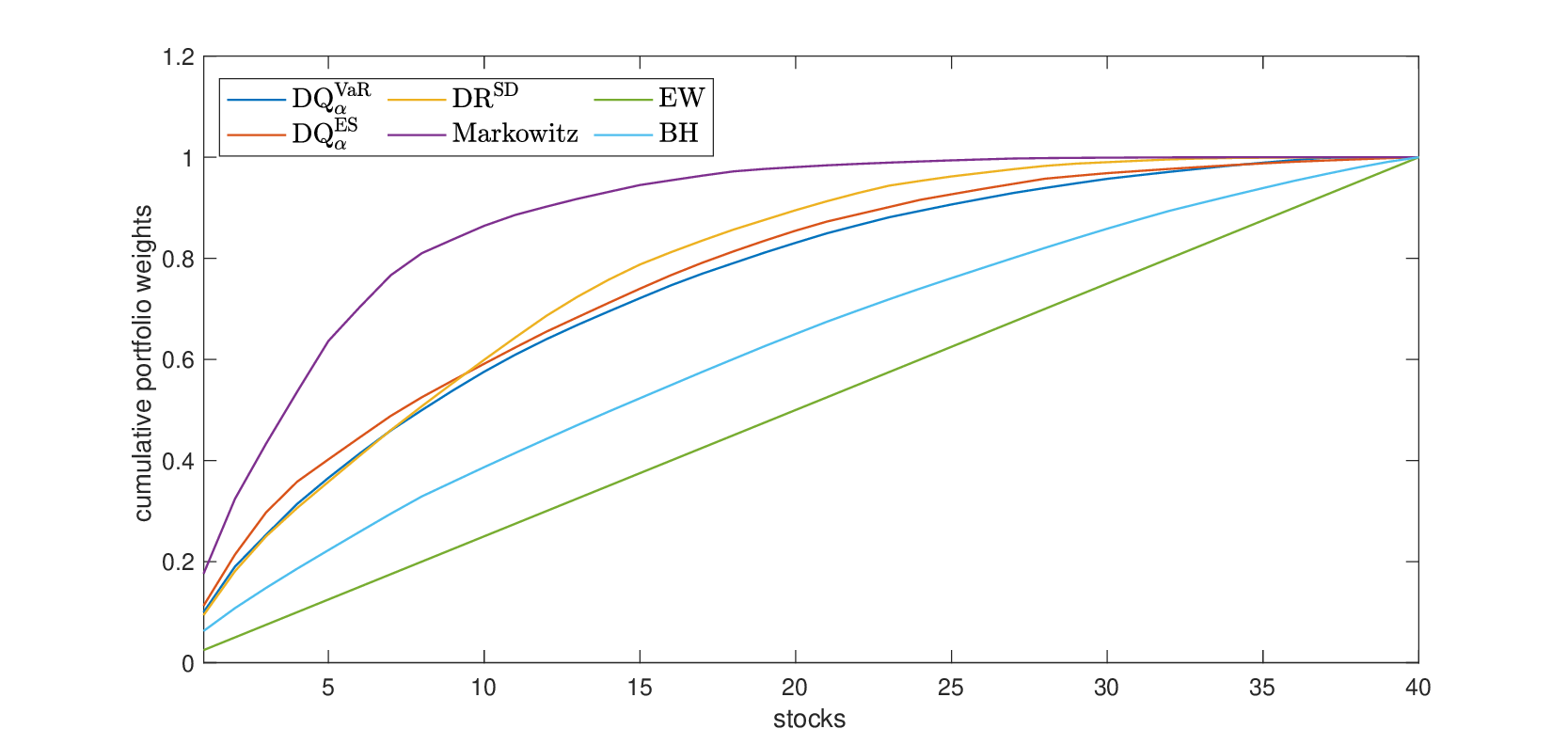}
\end{figure}

\begin{table}[htb!]  
\def\arraystretch{1}
  \begin{center} 
    \caption{Annualized return (AR),  annualized volatility (AV), Sharpe ratio (SR)  and  average trading proportion (ATP)  for   different portfolio strategies from Jan 2004 to Dec 2011} \label{tab_40_02} 
  \begin{tabular}{c|cccccc} 
    $\%$ &${\rm DQ}^{\VaR}_\alpha$ &${\rm DQ}^{\ES}_\alpha$ &${\rm DR}^{\rm SD}$ & Markowitz & EW & BH \\    \hline
AR & 9.46    &   8.13  &9.10 & 7.98 &  5.30&6.23 \\
AV & 16.65    & 21.45 & 20.92 & 11.98 & 20.15 &15.53\\
%Real Variance & 1.7341 &   1.9825 &   2.3108  &  2.0195  &  2.1444 &   2.2541\\
SR  & 30.48    & 17.47 & 22.58&  30.06& 4.57 &11.94\\{ATP}&37.23 & 28.59   &   20.24  &24.56 &   5.04&0\\ 
 \hline \hline 
    \end{tabular}
    \end{center} %  four largest stocks from ten different sectors;
    \end{table}

 For the first experiment, we choose the four largest stocks  from each of the 10 different sectors of S\&P 500   ranked by market cap in 2002 as the portfolio compositions and use the period from January 3, 2002,  to December 31, 2011, to build up the portfolio.   We use the risk-free rate $r = 4.38\%$ to calculate portfolio statistics, and the target annual expected return for the Markowitz portfolio is set to $5\%$ due to infeasibility of setting $10\%$.
  The results   are reported in Figures \ref{sec_40_02}, \ref{sec_40_02_cu} and Table \ref{tab_40_02}.

\begin{figure}[htb!]
\caption{Wealth processes for portfolios, 20 stocks, Jan 2014 - Dec 2021}\label{sec_20}
\centering
           \includegraphics[width=14cm]{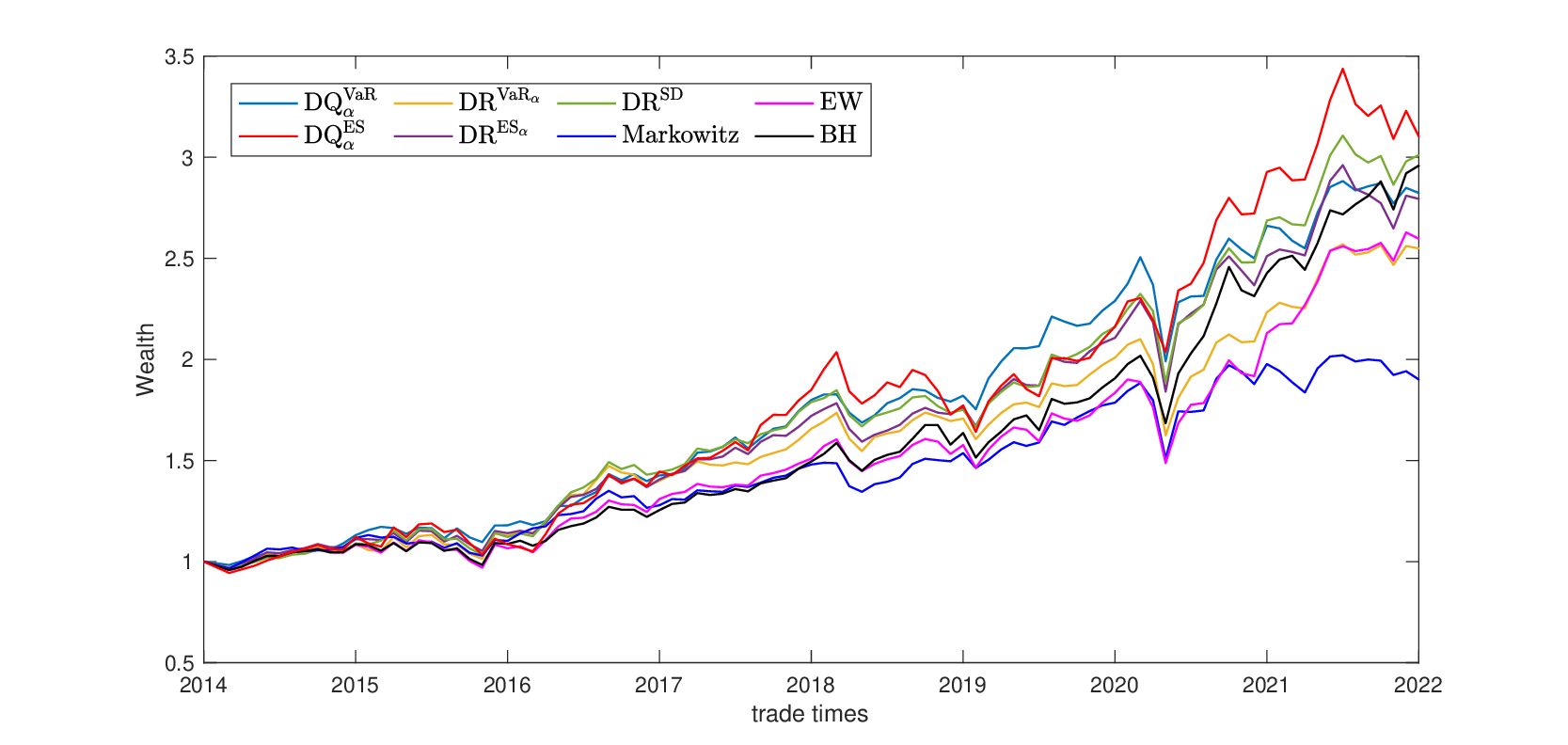}
\end{figure}
\begin{figure}[htb!]
\caption{Cumulative portfolio weights,  20 stocks, Jan 2014 - Dec 2021}\label{sec_20_cu}
\centering
           \includegraphics[width=15cm]{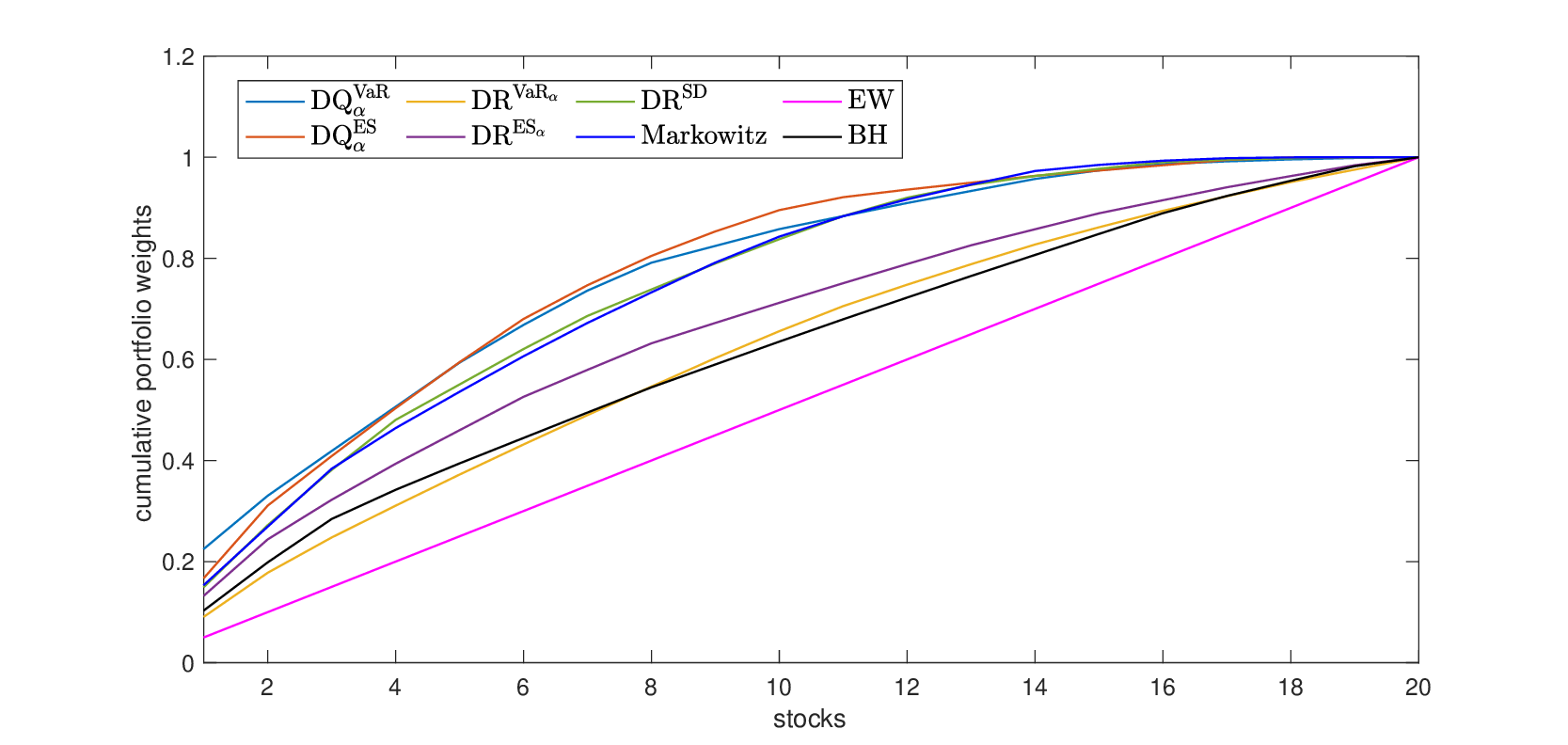}
\end{figure}

      \begin{table}[htb!]  
\def\arraystretch{1}
  \begin{center} 
    \caption{Annualized return (AR),  annualized volatility (AV), Sharpe ratio (SR) and    average trading proportion (ATP) for   different portfolio strategies from Jan 2014 to Dec 2021} \label{tab_sector_20} 
  \begin{tabular}{c|cccccccc} 
    $\%$ &${\rm DQ}^{\VaR}_\alpha$ &${\rm DQ}^{\ES}_\alpha$& ${\rm DR}^{\VaR_\alpha}$& ${\rm DR}^{\ES_\alpha}$  &${\rm DR}^{\rm SD}$ & Markowitz & EW & BH \\    \hline
AR & 13.54    &   14.79 & 12.77 & 13.85&14.37 & 8.59 &  12.74&14.22 \\
AV & 13.43 & 15.90 &14.41&14.53 & 14.29 & 12.74 & 14.68 &13.96\\
%Real Variance & 1.7341 &   1.9825 &   2.3108  &  2.0195  &  2.1444 &   2.2541\\
SR  & 79.69     & 75.17& 68.89& 75.79 & 80.67&  45.14 & 67.40 &81.54\\{ATP}&16.07 & 19.24   &   64.77  &57.56 &11.81 &15.19 &   4.45&0\\
 \hline \hline 
    \end{tabular}
    \end{center} % The two largest stocks from 10 different sectors
    \end{table}

      For the second experiment, we  choose the  top two  stocks  from  each  sector   to build the portfolios, and all other parameters are the same as in Section \ref{opt_dp}. The results, including two other portfolios built by ${\rm DR}^{\mathrm {VaR_\alpha}}$ and ${\rm DR}^{\mathrm {ES_\alpha}}$,  are reported  in Figures \ref{sec_20} and \ref{sec_20_cu} and Table \ref{tab_sector_20}.  
      Since we do not find an efficient algorithm for computing  ${\rm DR}^{\mathrm {VaR_\alpha}}$ and ${\rm DR}^{\mathrm {ES_\alpha}}$, we use the preceding 500 trading days to compute the optimal portfolio weights using the random sampling method,  which is relatively slow and not very stable. (If the previous month has an optimal weight $\mathbf w^*_{t-1}$, then $10^5$ new weights are sampled from $\lambda \mathbf w^*_{t-1}+(1-\lambda)\Delta_n$, where $\lambda$ is chosen as $0.9$. Tie-breaking is done by picking the one that is closest to $\mathbf w^*_{t-1}$. We set $\mathbf w^*_{0}=(1/n,\dots,1/n)$.) The results show similar observations to  those in Section \ref{opt_dp}.
      The additional observation   from  Table \ref{tab_sector_20} is that DR strategies have much larger ATP than the others, but this may be partially caused by our random sampling algorithms to optimize DR.
 
 \subsection{Portfolio performance under  other risk assessments criteria}\label{App:G}
\begin{figure}[htb!]
\caption{Wealth processes for portfolios, 40 stocks, Jan 2014 - Dec 2021}\label{sec_40_VaR_ES_var}
\centering
           \includegraphics[width=15cm]{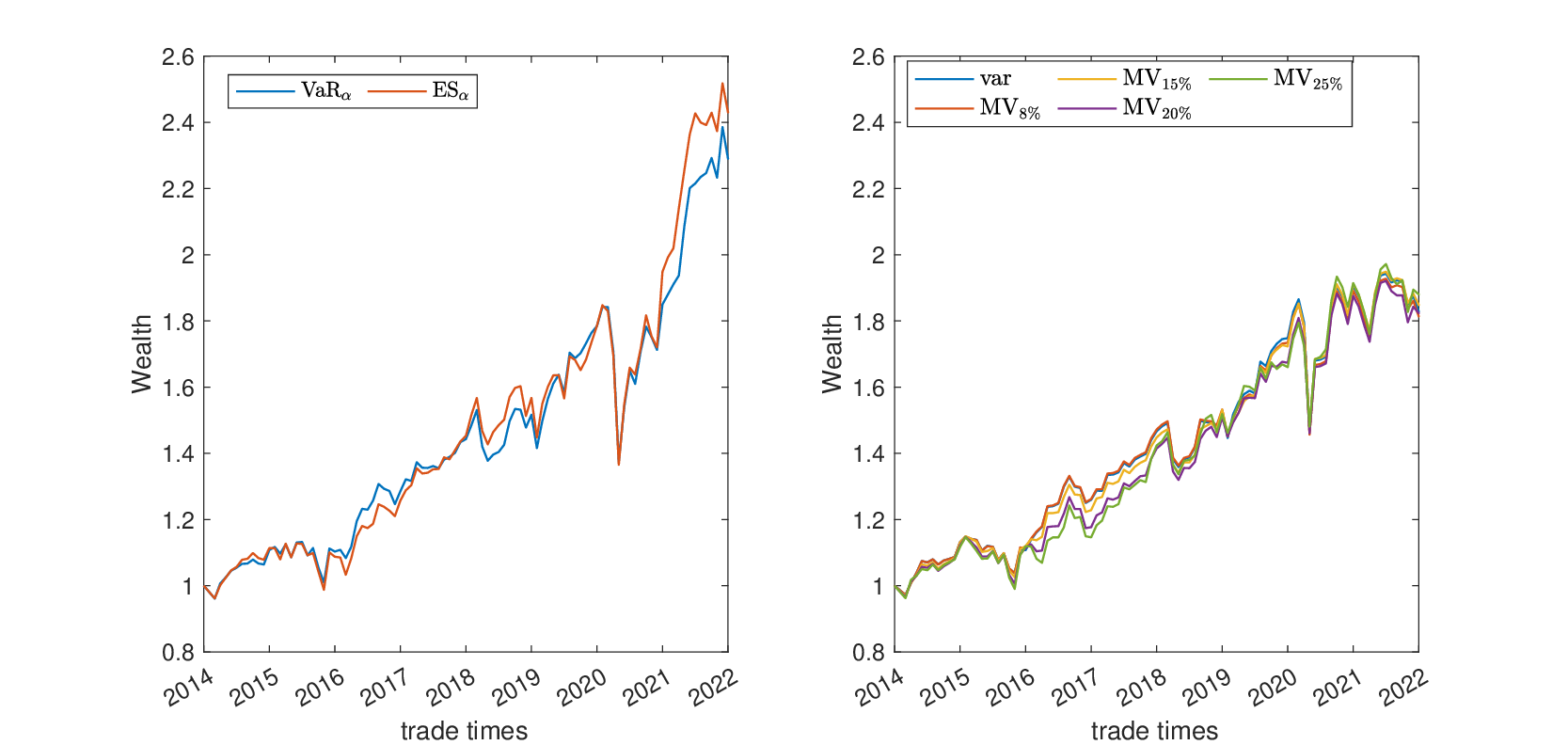}
\end{figure}

\begin{figure}[htb!]
\caption{Wealth processes for portfolios, 20 stocks, Jan 2014 - Dec 2021}\label{sec_20_VaR_ES_var}
\centering
           \includegraphics[width=15cm]{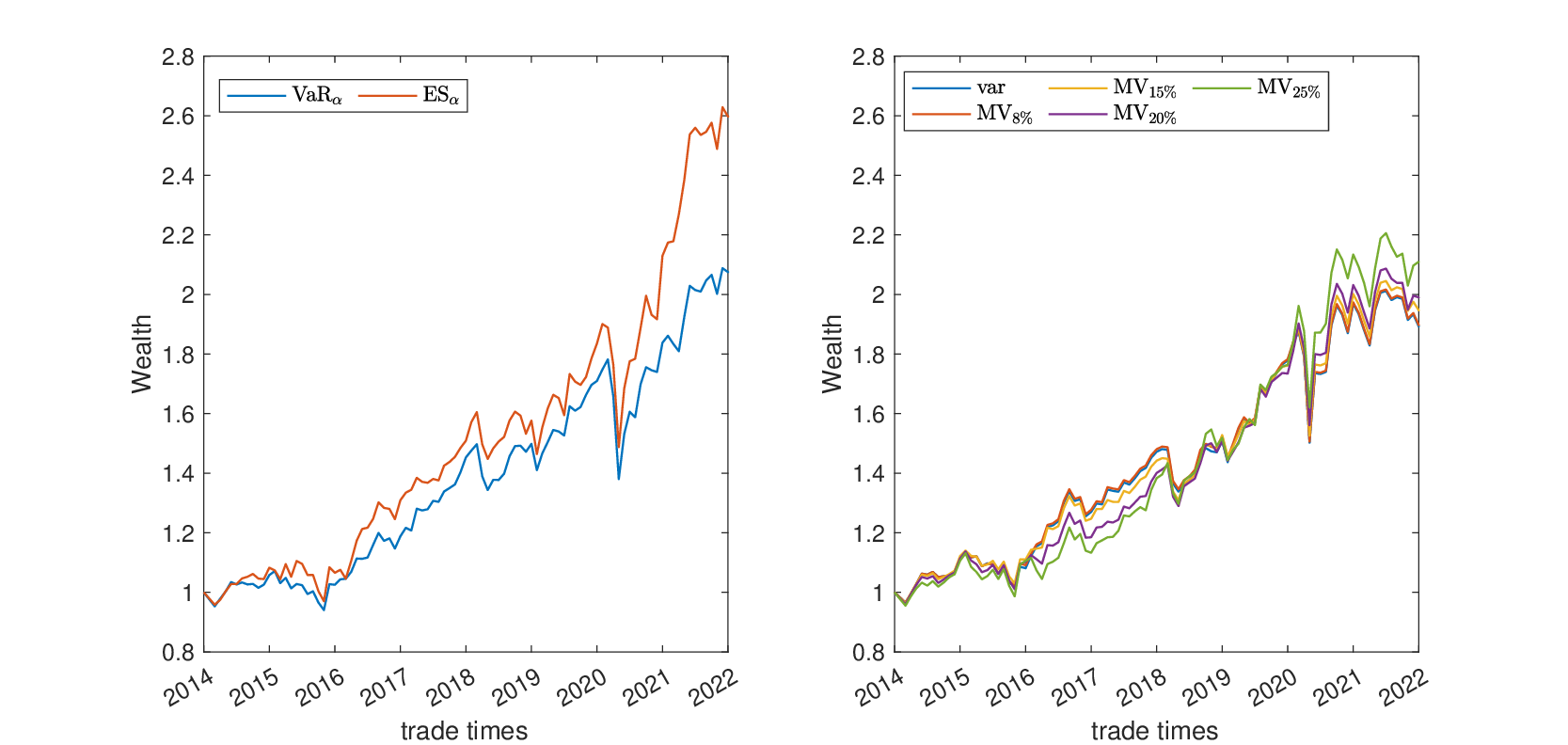}
\end{figure}

In this section, we analyze empirical results by minimizing a few other risk assessment criteria, the variance and VaR and ES with $\alpha=0.1$, of the total portfolio return. Portfolios are constructed using data spanning from January 3, 2012, to December 31, 2021 as in the setting of Section \ref{opt_dp}.  We first select the same set of 40 stocks    in Section \ref{opt_dp}, and then  we explore portfolios built from 20 stocks in Section \ref{EC:72}. In addition, we consider the Markowitz  mean-variance   approach  in Section  \ref{opt_dp}  with different levels of target annual expected return. The results are reported in Figures \ref{sec_20_VaR_ES_var} and \ref{sec_40_VaR_ES_var} and Tables \ref{tab_40_VaR} and \ref{tab_20_VaR}.

\begin{table}[H]  
\def\arraystretch{1}
  \begin{center} 
    \caption{Annualized return (AR),  annualized volatility (AV), Sharpe ratio (SR), and average trading proportion (ATP)  for   different portfolio strategies from Jan 2014 to Dec 2021 with 40 stocks, where $\mathrm{MV}_{r}$ represents the Markowitz model with target return level $r$} \label{tab_40_VaR} 
  \begin{tabular}{c|ccccccc} 
    $\%$ &$\VaR_\alpha$ &$\ES_\alpha$ &var & $\mathrm{MV}_{8\%}$  &$\mathrm{MV}_{15\%}$  & $\mathrm{MV}_{20\%}$ & $\mathrm{MV}_{25\%}$ \\    \hline
AR & 11.15    &   11.90  &8.09 & 7.99 &  8.24&8.09&8.49 \\
AV & 14.90    & 15.91 & 12.99 & 12.97& 12.89 & 12.79 &13.11\\
%Real Variance & 1.7341 &   1.9825 &   2.3108  &  2.0195  &  2.1444 &   2.2541\\
SR  & 55.81     & 56.96 & 40.40& 39.68 & 41.92 &41.04&43.10 \\{ATP}&77.88 & 4.43  &   16.79  &17.33&   25.06&33.34&40.57 \\
 \hline \hline 
    \end{tabular}
    \end{center}%
    \end{table}

\begin{table}[H]  
\def\arraystretch{1}
  \begin{center} 
    \caption{Annualized return (AR),  annualized volatility (AV), Sharpe ratio (SR), and average trading proportion (ATP)  for   different portfolio strategies from Jan 2014 to Dec 2021 with 20 stocks, where $\mathrm{MV}_{r}$ represents the Markowitz model with target return level $r$} \label{tab_20_VaR} 
  \begin{tabular}{c|ccccccc} 
    $\%$ &$\VaR_\alpha$ &$\ES_\alpha$ &var & $\mathrm{MV}_{8\%}$  &$\mathrm{MV}_{15\%}$  & $\mathrm{MV}_{20\%}$ & $\mathrm{MV}_{25\%}$ \\    \hline
AR & 10.03    &   12.74  &8.53 & 8.56 &  8.89&9.18&9.23 \\
AV & 12.99    & 14.68 & 13.20 & 13.43& 13.41 & 13.43 &14.44\\
%Real Variance & 1.7341 &   1.9825 &   2.3108  &  2.0195  &  2.1444 &   2.2541\\
SR  & 55.34     & 67.40 &   44.69 &44.91 & 47.47 &49.13 &51.35\\{ATP}&66.17 & 4.45  &   13.07  &13.88&   22.30&30.26&38.34 \\
 \hline \hline 
    \end{tabular}
    \end{center}%
    \end{table}
    
  Comparing  Figure \ref{sec_40_VaR_ES_var} with Figure \ref{sec_40},     we observe that the outcomes for DQ are  comparable to those achieved with VaR and ES. 
 Moreover,  minimizing variance as an objective corresponds  to a Markowitz model  without expected return constraint. In our experiments, we find that under our parameter settings, targeting a lower annual expected return, such as $5\%$, imposes no effective constraint; the resulting portfolio outcomes mirror those achieved through variance minimization. However, aiming for a higher return, such as $30\%$, turns out to be infeasible. Additionally, Figure \ref{sec_40_VaR_ES_var} illustrates that the Markowitz strategies perform similarly across different levels of target return. Comparing Figure \ref{sec_20_VaR_ES_var} with Figure \ref{sec_20} reveals similar findings.   Similarly to the cases of DR strategies in the previous section, we can see from Tables \ref{tab_40_VaR} and \ref{tab_20_VaR}   that VaR strategies  result in higher ATP values compared to other strategies. This  may be partly due to the random sampling methods used for VaR optimization.

\end{document}